%% file: ms.tex
\pdfoutput=1

\RequirePackage[l2tabu,orthodox]{nag}

\documentclass[11pt,letterpaper,reqno,oneside]{article}

\input{preamble.tex}

\title{\scshape Slow persuasion%
\thanks{We are grateful to 
Giacomo Calzolari,
Eddie Dekel, 
Piero Gottardi, 
Alessandro Pavan and 
Bruno Strulovici 
for guidance and comments,
and to
three anonymous referees,
Marina Halac (as co-editor),
Gregorio Curello,
Théo Durandard,
Piotr Dworczak,
Jeff Ely,
Alex Frug,
George Georgiadis,
Duarte Gonçalves,
Yingni Guo,
Navin Kartik,
Frédéric Koessler,
Elliot Lipnowski,
Ludmila Matysková,
Konstantin Milbradt,
Marco Ottaviani,
Eduardo Perez-Richet,
Francisco Poggi,
Sven Rady,
Marciano Siniscalchi,
Alex Smolin,
Pierpaolo Soravia,
Quitzé Valenzuela-Stookey
and audiences at
EUI,
Northwestern,
the 2019 Barcelona GSE Summer Forum
and the 2019 Summer School of the Econometric Society
for comments and suggestions.
A previous version of this paper was titled `Strategic research funding'.}}

\author{Matteo Escudé \\ LUISS \and Ludvig Sinander \\ University of Oxford}

\date{4 April 2022}

\makeatletter
	\AtBeginDocument{ \hypersetup{ 
		pdftitle = {Slow persuasion}, 
		pdfauthor = {Matteo Escudé and Ludvig Sinander} 
		} }
\makeatother

\begin{document}

\maketitle

\begin{abstract}
	What are the value and form of optimal persuasion
	when information can be generated only slowly?
	We study this question in a dynamic model
	in which a `sender' provides public information over time subject to a graduality constraint,
	and a decision-maker takes an action in each period.
	Using a novel `viscosity' dynamic programming principle,
	we characterise the sender's equilibrium value function and information provision.
	We show that the graduality constraint inhibits information provision relative to unconstrained persuasion.
	The gap can be substantial,
	but closes as the constraint slackens.
	Contrary to unconstrained persuasion,
	less-than-full information may be provided
	even if players have aligned preferences but different prior beliefs.
\end{abstract}

\section{Introduction}
\label{sec:intro}

Information production takes time.
Experts seeking to influence behaviour
can provide information no faster than it can be produced.
In this paper, we study the impact of such graduality constraints on strategic communication.

Consider, for example, a firm undertaking a risky project.
Management produces information about the soundness of the investment with a view to persuading investors to back the enterprise.
This typically cannot be achieved instantaneously,
as evidence of profitability accumulates only gradually over the months
as the project is taken from conception to market.

Alternatively, consider a funding body that wishes to influence policy-making (or individual behaviour) by commissioning research.
Consensus on scientific questions is built over dozens of studies,
each of which takes time to conduct and be peer-reviewed.
This constrains how rapidly information can be produced.

How much can an expert benefit from strategic information provision in such an environment? How does the graduality constraint affect the amount of information provided?
In this paper, we answer these questions by characterising the value of graduality-constrained information provision
and by delineating how graduality shapes the extent of information production.

The value that an expert can derive from strategic information provision
is a key concern of the literature on Bayesian persuasion.
In the well-known model of \textcite{KamenicaGentzkow2011},
an uninformed `sender' can costlessly produce public information about the unknown state of the world by designing a public signal,
and a (symmetrically uninformed) decision-maker observes the signal's realisation and takes an action.

Our \emph{slow persuasion} model augments this setting with a graduality constraint.
Since graduality is inherently dynamic, so is our model.
In each period, the sender can produce at most a \emph{small} amount of public information about the unknown (and fixed) state of the world.
The players' shared belief about the state evolves over time as evidence accumulates.
Informed by this belief, the decision-maker takes an action in each period,
which together with the state determines (flow) payoffs.
This defines a two-player stochastic game in continuous time, which we study by describing the set of Markov perfect equilibria with the common belief as state variable.
We characterise equilibrium payoffs and behaviour (in particular, information provision).

The persuasion literature is distinguished from the earlier cheap-talk literature
by its \emph{commitment assumption:} once the signal has been chosen, its realisation cannot be concealed from the decision-maker.
Thus the two players remain symmetrically informed throughout their interaction,
which eliminates communication frictions
and so permits a sharper focus on the value of (constrained) information provision.
We maintain the commitment assumption within each period: whatever information is produced becomes public.
(We do not assume commitment \emph{over time.}
But given our focus on Markov perfect equilibria, nothing would change if we did.)

Our first result characterises the value of slow persuasion.
The sender's \emph{value function} describes, at each public belief, her expected discounted continuation payoff in equilibrium.
\Cref{proposition:value_fn} asserts that the value function
is a version of the concave envelope (the value in unconstrained persuasion)
that accounts for graduality and impatience.
In particular, the value is strictly convex and below the concave envelope
whenever the latter is affine (and exceeds the sender's flow payoff).

To see why, consider `splitting' the current public belief across two posteriors by producing information.
In unconstrained persuasion,
the split can be effected immediately,
so that its value equals the average of payoffs at the two posteriors.
When persuasion is slow, it takes time for the belief to reach (one of) the target posteriors, which has two effects.
First, the sender impatiently discounts the payoffs she anticipates once one of the posteriors has been reached.
Secondly, until the target posteriors are reached, the actions chosen by the decision-maker will be those she considers optimal at the beliefs prevailing in the interim.

The value of slow persuasion increases
when the graduality constraint is slackened (allowing faster information production)
and is well-approximated by the concave envelope for a sufficiently slack constraint (\Cref{corollary:value_fn_limit}).
Outside of this special case, the value and concave envelope may differ substantially.

We next study behaviour.
Using our characterisation of the value, we obtain a description of the sender's equilibrium information-provision strategy.
To assess the impact of the graduality constraint, we derive comparative statics (\Cref{proposition:beliefs_comp_stat}): as the constraint tightens,
the sender optimally provides Blackwell-less information over the course of the relationship.
(She \emph{could} provide the same aggregate information, albeit more slowly; we prove that she prefers not to.)
This holds no matter what the players' preferences.

In slow persuasion, Blackwell-less information is provided overall
than in the unconstrained benchmark (\Cref{proposition:beliefs_persuasion}).
In general, the gap can be large.
However, in the special case of fast information arrival,
long-run information provision is well-approximated by the prediction of the unconstrained-persuasion model (also \Cref{proposition:beliefs_persuasion}).%
	\footnote{Formally, long-run beliefs (the set of posteriors to which the belief martingale converges) are close to those chosen in the unconstrained problem.}
The seemingly similar `slow' limit lacks this continuity property:
a sufficiently \emph{tight} constraint does \emph{not} generally lead to negligible aggregate information provision.

We conclude by studying the case
in which the conflict of interest between the players
arises not from different preferences over actions conditional on the state,
but rather from their having different prior beliefs.
This case is salient
in organisations, where preferences may already have been aligned using contracts,
and for policy questions such as climate-change mitigation, where disagreement is primarily over the extent of anthropogenic climate change, rather than over the correct policy response to a given amount of warming.

In unconstrained persuasion, a purely belief-based conflict of this sort amounts to no conflict at all: full information is provided, and ex-post optimal actions are taken \parencite[see e.g.][]{AlonsoCamara2016}.
We show by contrast that when persuasion is constrained to be slow,
only partial information is produced over the long run if the constraint is tight enough (\Cref{proposition:disagree_nonfull}).
It follows in particular that
instituting contracts to align preferences conditional on the state
is in general insufficient to encourage information production.

Clearly prior disagreement harms the sender when persuasion is slow,
since then only partial evidence is available in the medium term,
leading players' priors to colour their posterior beliefs and (thus) their views on what actions are best.
We show that this welfare effect is monotone:
the greater the prior difference, the lower is the sender's value (\Cref{proposition:disagree_value}).

The key to our results is \Cref{proposition:value_fn} (the value characterisation),
from which the remaining propositions are derived.
To prove it, we follow the natural route of studying the value
as a solution of the Hamilton–Jacobi–Bellman (HJB) equation of the sender's best-reply problem.
This poses a technical challenge: the value function may have kinks, in which case it fails to solve the HJB equation (a second-order differential equation) in the classical sense.
Such kinks are an unavoidable by-product of two features of the sender's problem:
her ability to freeze the public belief by halting information provision,
and the discontinuities in her flow payoff that occur whenever the decision-maker switches from one action to another.

To overcome this hurdle, we extend existing results from the mathematics literature
to prove a novel dynamic programming principle (\Cref{theorem:viscosity}):
the value function is a \emph{viscosity solution} of the HJB equation.
This permits us to use the powerful theory of viscosity solutions of differential equations to characterise the value function in \Cref{proposition:value_fn}. We view this as a methodological contribution, and believe that our viscosity approach will prove useful for the study of other stochastic models in continuous time.

\subsection{Related literature}
\label{sec:intro:related}

This paper belongs to the Bayesian persuasion literature initiated by \textcite{KamenicaGentzkow2011,RayoSegal2010,AumannMaschler1995} (see \textcite{Kamenica2019,BergemannMorris2019} for surveys).
We contribute in particular to the growing strand of that literature which examines the impact of constraints on (or costs of) information production
\parencite[e.g.][]{GentzkowKamenica2014,LetreustTomala2019,DovalSkreta2021}.%
	\footnote{Some of this work is surveyed by \textcite{KamenicaKimZapechelnyuk2021}.
	\textcite{CalzolariPavan2006_RAND,CalzolariPavan2006_JET,PavanCalzolari2009,Rosar2017,GeorgiadisSzentes2020,Dworczak2020,DovalSkreta2021,BoleslavskyKim2021} study persuasion problems subject to \emph{incentive} constraints.}
Whereas these papers consider static settings, we study a dynamic constraint: graduality.

We also contribute to the larger strand on dynamic persuasion.
In some important papers on this topic \parencite{Ely2017,RenaultSolanVieille2017,Ball2019}, the state of the world evolves over time.
We instead consider a fixed state, but impose a graduality constraint.

Several recent papers study dynamic persuasion models with a fixed state
in which the decision-maker chooses when to take a game-ending action
\parencite{Au2015,ElySzydlowski2020,OrlovSkrzypaczZryumov2020,BizzottoRudigerVigier2021,Smolin2021,CheKimMierendorff2021}.
Our decision-maker instead selects freely among her actions in each period.
This is an important difference: whereas commitment over time has no value for the sender in our setting, it is strictly valuable in these models.
The last two papers feature graduality constraints; the remainder do not.

More closely related are \textcite{HenryOttaviani2019} and \textcite[§6]{SiegelStrulovici2020}, who study models of graduality-constrained information provision about a fixed state
in which the \emph{sender} chooses when to stop irreversibly, whereupon the decision-maker takes an action.%
	\footnote{\textcite{BrocasCarillo2007} study a similar setting
	in discrete time, with no discounting. See also \textcite[§4 and §5]{FershtmanPavan2022}.}
Our decision-maker instead acts in every period, earning the sender a flow payoff.
Until she stops, these papers' sender incurs a flow cost $c>0$, absent in our model.
Both papers focus on welfare-improving institutional design, whereas we study equilibrium behaviour in a fixed game.
Siegel and Strulovici do not have results analogous to any of ours, but their environment exhibits similar properties to the special case of our model with only two or three actions.

Henry and Ottaviani further assume that there are only two actions and that the sender's payoff is independent of the state.
In this special case, the dependence of information provision on the speed at which evidence accumulates that we emphasise (\Cref{proposition:beliefs_comp_stat}) is entirely absent if (as in our model) $c=0$. It becomes important as soon as the sender's payoff is allowed to be state-dependent or there are more than two actions
(see our two- and three-action examples on \cpageref{sec:value:example,suppl:hjb_details:3act}).
The authors characterise the sender's value function, and show that information provision is close to that in unconstrained persuasion when the cost $c$ is small,
results analogous to a specialisation of our \Cref{proposition:value_fn,proposition:beliefs_persuasion}.%
	\footnote{Henry and Ottaviani also show that when $c$ is small \emph{and} the sender is patient, her value is close to the concave envelope. This is analogous to our \Cref{corollary:value_fn_limit}.}
Their two-action environment is tractable enough to avoid the need for viscosity methods.

Our material on purely belief-based conflicts of interest
contributes to the literature on strategic communication with heterogeneous beliefs \parencite[e.g.][]{CheKartik2009,VandenSteen2009}.
In unconstrained persuasion, full information is provided when the conflict of interest is rooted in differing beliefs alone \parencite[e.g.][]{AlonsoCamara2016}.
We show that slow persuasion overturns this conclusion.

Viscosity solutions of differential equations were introduced by \textcite{CrandallLions1983}. We give a brief exposition and some references in \cref{suppl:viscosity_intro}. Viscosity methods have begun to be used in economic theory \parencite[e.g.][]{NikandrovaPancs2018,KeVillas-Boas2019,KellerRady2020,Zhong2019,BarillaGoncalves2022}.
Our sender's best-reply problem is non-standard
due the discontinuities in her flow payoff
and her ability to freeze the state variable (the public belief),
precluding off-the-shelf use of standard results.

Our technical contribution is related to \textcite{KuvalekarLipnowski2020},
who also adapt arguments from the viscosity literature to deal with discontinuities in flow payoffs.
Their goal is to establish smooth pasting for their model,
a property which fails in our setting (as the value may have kinks).
We instead derive a dynamic programming principle.

\subsection{Roadmap}
\label{sec:intro:roadmap}

We formulate the model in the next section, and pin down the decision-maker's equilibrium behaviour in §\ref{sec:myopic}.
We then study the sender's best-reply problem,
characterising her value function in §\ref{sec:value} and her equilibrium information provision in §\ref{sec:beliefs}.
We conclude in §\ref{sec:disagree} by considering the case in which the conflict of interest arises from differing beliefs alone.

\section{Model}
\label{sec:model}

Our model is a stochastic game in continuous time.
The players are a \emph{decision-maker} and a \emph{sender,}
who take respective actions $a_t$ and $\lambda_t$ in each period $t \in \R_+$. Flow payoffs depend on the decision-maker's action $a_t$ and on the state variable $p_t$. The sender's action $\lambda_t$ affects the stochastic evolution of $p_t$. In particular, $\lambda_t$ is the rate at which public information arrives, and $p_t$ is the common belief, which evolves according to Bayes's rule.

\subsection{State and payoffs}
\label{sec:model:payoffs}

There is a binary state $\theta \in \{0,1\}$.
The sender and the decision-maker have a common prior belief $p_0$ that the state is $\theta=1$.
(We will drop the common-prior assumption in §\ref{sec:disagree}.)
Time $t \in \R_+$ is continuous. The decision-maker takes an action $a \in \mathcal{A}$ at each moment, where $\mathcal{A}$ is a finite set.

When the decision-maker takes action $a$ and the common belief is $p$, the sender's and decision-maker's respective flow payoffs are $f_{\textsl{S}}(a,p)$ and $f_{\textsl{D}}(a,p)$, both continuous in $p$.
Expected utility ($f_{\textsl{S}}(a,\cdot)$ and $f_{\textsl{D}}(a,\cdot)$ affine) is a natural special case.
The sender and decision-maker discount flow payoffs at rates $r>0$ and $r_{\textsl{D}}>0$, respectively.

This abstract setting nests the examples in the \hyperref[sec:intro]{introduction} as follows.
In the first example,
an investor (decision-maker) decides in each period how much financing $a \in \mathcal{A} \subseteq \R_+$ to contribute to a firm.
The net return on the firm's project is $\alpha \theta - 1$, where $\alpha > 1$,
and thus the investor's flow payoff is $f_{\textsl{D}}(a,p) \coloneqq a (\alpha p - 1)$.%
	\footnote{Her expected discounted payoff
	is $\E\left( \int_0^\infty e^{-r_{\textsl{D}}t} a_t 
	(\alpha \theta - 1 ) \dd t \right)
	= \int_0^\infty e^{-r_{\textsl{D}}t} f_{\textsl{D}}(a_t,p_t) \dd t$.}
The firm's management (the sender) cares only about financing,
so $f_{\textsl{S}}(a,p) \coloneqq a$.
In the \hyperref[sec:intro]{introduction's} second example,
a policy-maker sets policy $a$ in each period,
and her and the sender's policy preferences depend on a persistent, unknown state $\theta$.

\subsection{The sender's information provision}
\label{sec:model:info}

At each instant, the sender can costlessly permit a small amount of public information to arrive. In particular, she chooses $\lambda_t \in \bigl[0,\widebar{\lambda}\bigr]$, and everyone observes the process
\begin{equation*}
	\dd X_t 
	= \theta \lambda_t \dd t 
	+ \sigma \sqrt{ \lambda_t } \, \dd \widetilde{B}_t ,
\end{equation*}
where $\widetilde{B}$ is a standard Brownian motion and $\sigma > 0$.
The parameter $\widebar{\lambda}>0$ quantifies the slackness of the graduality constraint $\lambda_t \leq \widebar{\lambda}$.
Our assumption that the noise is Brownian rules out information arriving in discrete lumps.

The signal process may be micro-founded as follows.
Write $\Lambda_t \coloneqq \int_0^t \lambda_s \dd s$ for cumulative information-production effort.
Total effort $\Lambda$ produces evidence, summarised by a `score':
write $Y_\Lambda$ for the cumulative score,
and assume that today's score $\dot{Y}_\Lambda$ has mean $\theta$, but is subject to iid noise.
Since white noise is the rate of change of a random walk, we may write
$\dd Y_\Lambda = \theta \dd \Lambda + \sigma \dd \widebar{B}_\Lambda$, where $\widebar{B}$ is a standard Brownian motion.
Then the evolution over time of the cumulative score $X_t \coloneqq Y_{\Lambda_t}$ follows
\begin{equation*}
	\dd X_t 
	= \theta \dd \Lambda_t
	+ \sigma \dd \widebar{B}_{\Lambda_t}
	= \theta \frac{ \dd \Lambda_t }{ \dd t } \dd t 
	+ \sigma \sqrt{ \frac{ \dd \Lambda_t }{ \dd t } } \, \dd \widetilde{B}_t 
	= \theta \lambda_t \dd t 
	+ \sigma \sqrt{ \lambda_t } \, \dd \widetilde{B}_t ,
\end{equation*}
where $\widetilde{B}$ is a(nother) standard Brownian motion.%
	\footnote{As is well-known, the `time-changed' process $t \mapsto \dd \widebar{B}_{\Lambda_t}$ has the same law as the \emph{scaled} process $\smash{ t \mapsto \sqrt{ \dd \Lambda_t / \dd t } \, \dd \widetilde{B}_t }$, where $\smash{\widetilde{B}}$ is a standard Brownian motion.}

As the players observe $( X_t )_{ t \in \R_+ }$, their common belief $p_t$ is updated according to Bayes's rule. By a well-known result from filtering theory, the belief evolves as
\begin{equation*}
	\dd p_t = \sqrt{ \lambda_t } \, \frac{ p_t (1-p_t) }{ \sigma } \dd B_t , 
\end{equation*}
where $B$ is a standard Brownian motion according to the common belief.%
	\footnote{See e.g. \textcite[§4.2.2]{Papanicolaou2016}.
	The result features assumptions about the process $( \lambda_t )_{ t \in \R_+ }$,
	which are satisfied by equilibrium processes in our model.
	The process $B$ is given by $\dd B_t = \dd X_t - p_t \dd t$ and $B_0=0$. It is \emph{not} a Brownian motion according to the `objective' law of $X$ under either $\theta=0$ or $\theta=1$, but it \emph{is} a Brownian motion from the point of view of an observer with belief $p_t$, as can seen from the Girsanov theorem (e.g. \textcite[§3.5]{KaratzasShreve1991}).}
See e.g. \textcite[Lemma 1]{BoltonHarris1999} for a heuristic derivation.
The belief process $( p_t )_{ t \in \R_+ }$ is a martingale with a.s. continuous sample paths.

\subsection{Strategies and equilibrium}
\label{sec:model:soln}

We focus on Markov perfect equilibria, in which players' behaviour depends on the past only through the current state $p_t$.
These equilibria are natural and standard, and avoid technical issues that can arise in continuous time.
It is worth noting, however,
that there may be other equilibria (suitably defined),
and that these may differ qualitatively from the ones that we study.

\begin{definition}
	\label{definition:pwc}
	A function $[0,1] \to \R^n$ is \emph{piecewise continuous} iff its discontinuities form a discrete subset of $(0,1)$.%
		\footnote{A subset of $(0,1)$ is \emph{discrete} iff each of its members is an isolated point: it lives in a neighbourhood that contains no other members.}
\end{definition}

Note that a piecewise continuous function is continuous at $0$ and $1$.
Recall that a discrete subset of $(0,1)$ is at most countable.

\begin{definition}
	\label{definition:strat}
	A \emph{strategy} of the sender (decision-maker) is a $\bigl[0,\widebar{\lambda}\bigr]$-valued ($\mathcal{A}$-valued) stochastic process $( \lambda_t )_{ t \in \R_+ }$ ($( a_t )_{ t \in \R_+ }$) adapted to the filtration generated by $( p_t )_{ t \in \R_+ }$ and actions.
	A \emph{(pure) Markov strategy of the sender} is a measurable map $\Lambda : [0,1] \to \bigl[0,\widebar{\lambda}\bigr]$.
	A \emph{Markov strategy of the decision-maker} is a piecewise continuous map $A : [0,1] \to \Delta(\mathcal{A})$,
	where $\Delta(\mathcal{A})$ denotes the set of all probability distributions over the (finite) set $\mathcal{A}$.
	We identify a Markov strategy $\Lambda$ ($A$) with the strategy (stochastic process) that it induces via $\lambda_t \coloneqq \Lambda(p_t)$ ($a_t$ distributed according to $A(p_t)$).
\end{definition}

The restriction to piecewise continuous strategies is a mild assumption on the decision-maker's tie-breaking that has no payoff consequences for her, provided her preferences are non-degenerate in a weak sense.
(See \cref{suppl:tie-breaking}. This is obvious for expected-utility preferences.)
The role of piecewise continuity is to ensure that the sender's best-reply problem satisfies a dynamic programming principle (\Cref{theorem:viscosity} in §\ref{sec:value:viscosity} below).

\begin{definition}
	\label{definition:BR}
	A strategy $( \lambda_t )_{ t \in \R_+ }$ of the sender is a \emph{best reply} at $\widebar{p} \in [0,1]$ to a Markov strategy $A : [0,1] \to \Delta(\mathcal{A})$ of the decision-maker iff it maximises
	\begin{gather*}
		\E\left( \int_0^\infty e^{-rt} 
		\left[ \int_{\mathcal{A}} f_{\textsl{S}}( a, p_t ) A( \dd a | p_t ) \right] 
		\dd t \right) ,
		\\
		\text{where} \quad
		\dd p_t = \sqrt{ \lambda_t } \, \frac{ p_t (1-p_t) }{ \sigma } \dd B_t 
		\quad \text{and} \quad p_0 = \widebar{p} ,
	\end{gather*}
	over all $\bigl[0,\widebar{\lambda}\bigr]$-valued processes $(\lambda_t)_{t \in \R_+}$ adapted to the filtration generated by $( p_t )_{ t \in \R_+ }$.%
		\footnote{In principle, the sender can use a process that is adapted to the filtration generated by $( p_t )_{ t \in \R_+ }$ \emph{and actions}. But when the decision-maker uses a Markov strategy, actions contain no additional information.}

	A strategy $( a_t )_{ t \in \R_+ }$ of the decision-maker is \emph{undominated} iff there is no other strategy $( a_t' )_{ t \in \R_+ }$ that yields the same expected payoff and has $f_{\textsl{D}}(a_t',p) > f_{\textsl{D}}(a_t,p)$ a.s. for some $p \in [0,1]$.
	A strategy $( a_t )_{ t \in \R_+ }$ of the decision-maker is a \emph{best reply} at $\widebar{p} \in [0,1]$ to a strategy $( \lambda_t )_{ t \in \R_+ }$ of the sender iff it is undominated and maximises
	\begin{gather*}
		\E\left( \int_0^\infty e^{-r_{\textsl{D}} t} 
		f_{\textsl{D}}( a_t, p_t ) 
		\dd t \right) ,
		\\
		\text{where} \quad
		\dd p_t = \sqrt{ \lambda_t } \, \frac{ p_t (1-p_t) }{ \sigma } \dd B_t 
		\quad \text{and} \quad p_0 = \widebar{p} ,
	\end{gather*}
	over all $\mathcal{A}$-valued processes adapted to the filtration generated by $( p_t )_{ t \in \R_+ }$.
\end{definition}

We rule out dominated strategies of the decision-maker as uninteresting.
Accommodating them merely complicates the statements of some results.

\begin{definition}
	\label{definition:MPE}
	A \emph{Markov perfect equilibrium (MPE)} is a pair of Markov strategies that are best replies to each other at each $\widebar{p} \in [0,1]$.
\end{definition}

\subsection{Unconstrained (static) benchmark}
\label{sec:model:static}

In unconstrained persuasion, the sender flexibly provides information once and for all, with no graduality constraint. If the sender induces belief $p$, then the decision-maker takes an action $A(p)$ that maximises $f_{\textsl{D}}(\cdot,p)$, giving the sender a payoff of $u(p) \coloneqq f_{\textsl{S}}( A(p), p )$.
Assume that the decision-maker breaks ties such that $u$ is upper semi-continuous.%
	\footnote{Without this assumption, an optimal policy may fail to exist.}

\textcite{KamenicaGentzkow2011} studied this model, and showed the following. The sender is able to induce all and only distributions of beliefs whose mean is $p_0$ (`splits' of the prior).
The sender's value at prior $p_0$ is $(\cav u)(p_0)$, where $\cav u$ is the concave envelope of $u$ (the smallest concave function that majorises $u$).
The sender has an optimal policy that induces either two beliefs (if $(\cav u)(p_0) > u(p_0)$) or one belief (if $(\cav u)(p_0) = u(p_0)$).

As we shall see, unconstrained persuasion typically provides a poor approximation to our sender's value and information provision,
but the approximation is good if the graduality constraint is slack ($\widebar{\lambda}$ large).

\section{Myopic behaviour by the decision-maker}
\label{sec:myopic}

To avoid uninteresting technicalities, we will focus on MPEs in which the decision-maker's tie-breaking is well-behaved.
Say that a Markov strategy $A : [0,1] \to \Delta(\mathcal{A})$ of the decision-maker is \emph{regular} iff it breaks ties such that the sender's induced payoff $u(p) \coloneqq \int_{\mathcal{A}} f_{\textsl{S}}(a,p) A(\dd a|p)$ is upper semi-continuous.
Regular Markov strategies exist---indeed, \emph{any} Markov strategy of the decision-maker need be modified only on a discrete (hence at most countable) set of beliefs $p$ to be made regular, and this modification leaves the decision-maker's flow payoff unchanged.%
	\footnote{Recall that a Markov strategy is piecewise continuous by definition, and that the decision-maker's flow payoff $f_{\textsl{D}}(a,\cdot)$ is continuous.}

Call a Markov strategy $A : [0,1] \to \Delta(\mathcal{A})$ of the decision-maker \emph{myopic} iff at each belief $p \in [0,1]$, every action in the support of $A(\cdot|p)$ maximises $f_{\textsl{D}}(\cdot,p)$.

\begin{observation}
	\label{observation:myopic}
	A regular strategy of the decision-maker is part of a MPE iff it is myopic.
\end{observation}

That is: all and only myopic behaviour can be supported in a MPE, modulo tie-breaking.
It follows that our analysis below of the sender's behaviour in MPEs carries over to a simpler model with a myopic decision-maker, or alternatively a sequence of short-lived decision-makers.

\begin{proof}
	If the sender uses a Markov strategy, then since the decision-maker cannot affect the evolution of the state, a regular strategy of hers is a best reply iff it is myopic. The `only if' part follows.

	For the `if' part, fix a regular and myopic Markov strategy $A : [0,1] \to \Delta(\mathcal{A})$ of the decision-maker.
	We show in §\ref{sec:beliefs:BR} that
	(given that $A$ is regular,)
	there is a Markov strategy $\Lambda$ of the sender
	which is a best reply to $A$ at every $\widebar{p} \in [0,1]$.
	Since $A$ is myopic,
	it is a best reply to $\Lambda$ at every $\widebar{p} \in [0,1]$.
	So $(\Lambda,A)$ is a MPE.
\end{proof}

\Cref{observation:myopic} implies that Markov perfect equilibria exist, and further that a sender-preferred MPE exists. It also follows that the decision-maker's behaviour can differ across MPEs only at beliefs at which she is exactly indifferent, which in turn implies that generically, under mild conditions, the MPE is partially unique.%
	\footnote{In particular, for Lebesgue-a.e. expected-utility $f_{\textsl{D}}$ (viewed as vectors in $\R^{2\abs*{A}}$), the decision-maker's strategy differs across MPEs only on a countable set of beliefs.
	We will see in §\ref{sec:beliefs:BR} that the sender's best reply is generically partially unique.}

In light of \Cref{observation:myopic}, it remains only to characterise the sender's best reply to a given regular and myopic strategy of the decision-maker. We will in fact characterise her best reply to an arbitrary regular Markov strategy. We proceed in two steps, studying the sender's value function §\ref{sec:value}, and then her best reply in §\ref{sec:beliefs}.

\section{The sender's value function}
\label{sec:value}

Fix a regular Markov strategy $A : [0,1] \to \Delta(\mathcal{A})$ of the decision-maker. The sender's induced preferences over beliefs are given by
\begin{equation*}
	u(p) \coloneqq \int_{\mathcal{A}} f_{\textsl{S}}(a,p) A( \dd a | p ) .
\end{equation*}
Note that $u$ is piecewise continuous and upper semi-continuous since $f_{\textsl{S}}(a,\cdot)$ is continuous and $A$ is piecewise continuous and regular.
We will study the sender's problem given an arbitrary piecewise continuous and upper semi-continuous flow payoff $u : [0,1] \to \R$.

The sender's best-reply problem, with (discounted) value function $v$, is
\begin{align}
	v(p_0) 
	&= \sup_{ ( \lambda_t )_{ t \in \R_+ } } 
	\E\left( r \int_0^\infty e^{-rt} u(p_t) \dd t \right)
	\nonumber
	\\
	&\text{s.t.} \quad
	\dd p_t = \sqrt{ \lambda_t } \, \frac{ p_t (1-p_t) }{ \sigma } \dd B_t ,
	\label{eq:v_defn}
	\tag{BRP}
\end{align}
where $( \lambda_t )_{ t \in \R_+ }$ is chosen among all $\bigl[0,\widebar{\lambda}\bigr]$-valued processes adapted to the filtration generated by $( p_t )_{ t \in \R_+ }$, and $p_0$ is given.

To understand the sender's incentives in \eqref{eq:v_defn}
and to motivate our solution technique,
we begin with an illustrative example.

\subsection{A two-action example}
\label{sec:value:example}

There is a risky action $a=1$ that yields a benefit of $3$ to both players (only) in state $\theta=1$.
Taking the risky action costs the sender and decision-maker $1$ and $2$, respectively,
so that expected utilities at belief $p \in [0,1]$ are
\begin{equation*}
	f_{\textsl{S}}(1,p) = 3 p - 1
	\quad \text{and} \quad
	f_{\textsl{D}}(1,p) = 3 p - 2 .
\end{equation*}
There is also a safe action $a=0$ giving both players a certain payoff of zero.

The decision-maker's (unique regular) myopic strategy is $A = \1_{[2/3,1]}$.
The sender's induced flow payoff $u$ is depicted in \Cref{fig:2act}.
\begin{figure}
	\begin{subfigure}{0.49\textwidth}
		\centering
		\input{tikz/2act_v_lo}
		\caption{Evidence accumulates rapidly.}
		\label{fig:2act_v_lo}
	\end{subfigure}
	\begin{subfigure}{0.49\textwidth}
		\centering
		\input{tikz/2act_v_hi}
		\caption{Evidence accumulates slowly.}
		\label{fig:2act_v_hi}
	\end{subfigure}
	\caption{The sender's value function in the two-action example.}
	\label{fig:2act}
\end{figure}%

Most of the sender's best-reply problem is easily solved.
When $p_t \in (0,2/3)$, she finds it strictly optimal to provide information (at full tilt, i.e. $\lambda_t=\widebar{\lambda}$), since her flow payoff can only improve.
When $p_t \in (2/3,1)$, it is weakly optimal to provide information, since $u$ is affine on this region.%
	\footnote{She could provide information only while $p_t > 2/3$.
	Then at each instant that she does, the belief changes by a mean-zero random increment $\dd p_t$, which since $u$ is affine on $[2/3,1]$ yields expected payoff $\E( u(p_t + \dd p_t) ) = u(p_t)$, the same as from not providing information.}

The sender faces a non-trivial trade-off at $2/3$, however.
By stopping information provision, she can lock in a moderate payoff of $1$ forever.
If she instead continues,
then she may increase her payoff toward $2$ (if $p_t$ rises),
but may equally suffer a flow payoff of zero in the near future (if $p_t$ declines).
The optimal resolution of this trade-off depends on how rapidly evidence can accumulate:
if quickly ($\widebar{\lambda}$ high), then she provides information, and if slowly ($\widebar{\lambda}$ low), then she does not.

The value function is easily computed in either case:
it has the shape depicted in \Cref{fig:2act_v_lo} if evidence accumulates quickly,
and the shape in \Cref{fig:2act_v_hi} if slowly.
The kink in the latter case
arises from the sender choosing to freeze the state variable $p_t$
once it hits the discontinuity point $2/3$ of the flow payoff $u$
(induced by the decision-maker switching actions).

For comparison, the value function in the unconstrained benchmark is the concave envelope $\cav u$, which is affine and strictly higher.

\subsection{The HJB equation and viscosity solutions}
\label{sec:value:viscosity}

The Hamilton--Jacobi--Bellman (HJB) equation for the sender's best-reply problem is the following differential equation in an unknown function $w : [0,1] \to \R$:
\begin{equation*}
	w(p) = \sup_{ \lambda \in [0,\widebar{\lambda}] } \left\{ 
	u(p) 
	+ \frac{1}{r} 
	\left( \sqrt{\lambda} \frac{ p (1-p) }{ \sigma } \right)^2 
	\frac{w''(p)}{2} 
	\right\} ,
\end{equation*}
or equivalently
\begin{equation}
	w(p) = u(p) + \widebar{\lambda} \frac{ p^2 (1-p)^2 }{ 2r\sigma^2 } \max\left\{ 0, w''(p) \right\} .
	\tag{HJB}
	\label{eq:hjb}
\end{equation}

In well-behaved problems, a \emph{dynamic programming principle} holds:
the value function $v$ is a \emph{classical solution} of \eqref{eq:hjb}, meaning that $v$ is twice continuously differentiable and that $v$ and $v''$ satisfy \eqref{eq:hjb} at every $p \in (0,1)$.
The familiar interpretation is that the value $v$ is today's flow payoff $u$ plus the expected rate of change of the value, discounted by $r$.
In the latter term, $\sqrt{\lambda} p (1-p) / \sigma$ is the rate of information arrival, and $v''(p)/2$ is the (local) value of information.

Our sender's problem is not so well-behaved.
Since the sender can freeze the state variable $p$ (by setting $\lambda=0$)
and her flow payoff $u$ may have discontinuities (arising from action switches by the decision-maker),
the value function may have kinks,
as seen in the two-action example of the previous section.
Since $v''$ does not exist at kinks, $v$ cannot be a classical solution of \eqref{eq:hjb}: the right-hand side is ill-defined.%
	\footnote{Even in the absence of a kink (as in \Cref{fig:2act_v_lo}), $v$ cannot be a classical solution of \eqref{eq:hjb} unless $u$ is continuous. For whenever $u$ jumps, $v''$ must also jump to balance \eqref{eq:hjb}, in which case $v$ fails to be twice \emph{continuously} differentiable.}

To be able to use \eqref{eq:hjb} to study the value function when the latter may have kinks, we require a broader notion of `solution' of a differential equation. Let $u^\star$ ($u_\star$) denote the upper (lower) semi-continuous envelope of $u$,
i.e. the pointwise smallest (largest) upper (lower) semi-continuous function that majorises (minorises) $u$.
The envelopes $u^\star$ and $u_\star$ differ only on a discrete set since $u$ is piecewise continuous, and we have $u_\star \leq u = u^\star$ since $u$ is upper semi-continuous.
\begin{definition}
	\label{definition:viscosity}
	$w : [0,1] \to \R$ is a \emph{viscosity sub-solution (super-solution)} of \eqref{eq:hjb} iff it is upper (lower) semi-continuous, and for any twice continuously differentiable $\phi : (0,1) \to \R$ and local minimum $p \in (0,1)$ of $\phi - w$ (of $w - \phi$),
	\begin{align*}
		w(p) &\leq u^\star(p) 
		+ \widebar{\lambda} \frac{ p^2 (1-p)^2 }{ 2r\sigma^2 } 
		\max\left\{ 0, \phi''(p) \right\}
		\\
		\biggl(
		w(p) &\geq u_\star(p)
		+ \widebar{\lambda} \frac{ p^2 (1-p)^2 }{ 2r\sigma^2 } 
		\max\left\{ 0, \phi''(p) \right\}
		\biggr) .
	\end{align*}
	$w$ is a \emph{viscosity solution} of \eqref{eq:hjb} iff it is both a sub- and a super-solution.
\end{definition}

\begin{remark}
	\label{remark:viscosity_contact}
	It is without loss of generality to restrict attention at each $p \in (0,1)$ to functions $\phi$ that satisfy $\phi(p) = w(p)$ and for which $\phi - w$ ($w - \phi$) has a \emph{strict global} minimum at $p$.
\end{remark}

A brief exposition of the theory of viscosity solutions is given in \cref{suppl:viscosity_intro}. Observe that if $w$ is a viscosity solution of \eqref{eq:hjb} and is twice continuously differentiable on a neighbourhood of $p \in (0,1)$, then it satisfies \eqref{eq:hjb} in the classical sense at $p$.%
	\footnote{Since we may choose a twice continuously differentiable $\phi$ that coincides with $w$ on a neighbourhood of $p$, so $\phi-w$ and $w-\phi$ are locally minimised at $p$ and $\phi''(p)=w''(p)$.}

Although the value function need not satisfy \eqref{eq:hjb} in the classical sense, it \emph{does} satisfy \eqref{eq:hjb} in the viscosity sense:

\begin{theorem}[dynamic programming principle]
	\label{theorem:viscosity}
	Assume that $u$ is piecewise continuous.
	Then $v$ is a viscosity solution of \eqref{eq:hjb}, with boundary condition $v=u$ on $\{0,1\}$.
\end{theorem}

The proof is in \cref{app:viscosity}.

We view \Cref{theorem:viscosity} as a technical contribution.
It extends a well-known theorem from the optimal control literature in which the flow payoff $u$ is assumed to be continuous.
That is an unacceptable hypothesis in economic applications such as ours, where $u$ depends on the endogenous strategic behaviour of other players.
\Cref{theorem:viscosity} may prove useful for studying other models of strategic interaction in continuous-time stochastic environments.

\subsection{Characterisation of the sender's value function}
\label{sec:value:charac}

We shall characterise the sender's value $v$ in terms of its local convexity, defined as follows.

\begin{definition}
	\label{definition:local_convexity}
	$w : [0,1] \to \R$ is \emph{locally (strictly) convex} at $p \in (0,1)$ iff
	\begin{equation*}
		w(p) \leq \mathrel{(<)} \gamma w(p') + (1-\gamma) w(p'')
	\end{equation*}
	for all $p' < p < p''$ sufficiently close to $p$, where $\gamma$ is such that $\gamma p' + (1-\gamma) p'' = p$.
	It is \emph{locally (strictly) concave} at $p$ iff the reverse (strict) inequality holds.
\end{definition}

By way of illustration, the function $v$ depicted in \Cref{fig:2act_v_hi} is locally strictly concave at $2/3$
(but is concave on no open neighbourhood of $2/3$).

Let $C \subseteq (0,1)$ be the beliefs at which $v$ is locally strictly convex,
and let $D \subseteq (0,1)$ be the (discrete) set of beliefs at which $u$ is discontinuous.

\begin{proposition}[value function]
	\label{proposition:value_fn}
	$v$ is continuous and satisfies $u \leq v \leq \cav u$.
	On $C$, we have $v < \cav u$, and $v$ is once continuously differentiable.
	On $C \setminus D$, we have further that $v$ is twice continuously differentiable and satisfies
	\begin{equation}
		v(p) = u(p) + \widebar{\lambda} \frac{ p^2 (1-p)^2 }{ 2r\sigma^2 } v''(p) 
		\quad \text{at each $p \in C \setminus D$.}
		\tag{$\partial$}
		\label{eq:diff_eqn_convex}
	\end{equation}
	On $(0,1) \setminus C$, we have $v=u$. On $\{0,1\}$, we have $u=v=\cav u$.
\end{proposition}

\Cref{proposition:value_fn} is summarised in \Cref{table:value_fn}, where $\mathcal{C}^k$ means `continuous and $k$ times continuously differentiable'.
\begin{table}
	\centering
	\begin{tabular}{l|llll}
		region
		& properties of $v$
		& 
		& 
		& 
		\\ \hline
		$C \setminus D$
		& locally strictly convex
		& $u \leq v < \cav u$
		& $\mathcal{C}^2$
		& equation \eqref{eq:diff_eqn_convex}
		\\
		$C \intersect D$
		& locally strictly convex
		& $u \leq v < \cav u$
		& $\mathcal{C}^1$
		& smooth pasting
		\\
		$(0,1) \setminus C$
		& 
		& $u = v \leq \cav u$
		& $\mathcal{C}^0$
		& 
		\\
		$\{0,1\}$
		& 
		& $u = v = \cav u$
		& $\mathcal{C}^0$
		& 
	\end{tabular}
	\caption{Summary of \Cref{proposition:value_fn}. $\mathcal{C}^k$ means `continuous and $k$ times continuously differentiable'.}
	\label{table:value_fn}
\end{table}
The `smooth pasting' property in the entry for the region $C \intersect D$ is the following consequence of continuous differentiability on $C$: for any sequence $( p_n )_{n \in \N}$ of beliefs in $C \setminus D$ converging to $p \in C \intersect D$,%
	\footnote{Every point in $C \intersect D$ can be reached by such a sequence, since every element of $D$ is isolated by piecewise continuity of $u$.}
we have $\lim_{n\to\infty} v'(p_n) = v'(p)$.
Remark that since $v$ is only $\mathcal{C}^0$ on $(0,1) \setminus C$, it may have (locally concave) kinks in this region---we saw an example of this in \Cref{fig:2act_v_hi}.

The characterisation of $v$ in \Cref{proposition:value_fn} is a generalisation of the concave envelope $\cav u$. Both are upper envelopes of $u$ that exceed $u$ when convex and coincide with $u$ when concave. But whereas $\cav u$ is affine whenever it exceeds $u$, $v$ is strictly convex when it exceeds $u$, due to impatience. The differential equation \eqref{eq:diff_eqn_convex} pins down the exact form of this strict convexity.

\Cref{proposition:value_fn} permits us to solve for the value function.
Given a candidate $C'$ for $C$, \eqref{eq:diff_eqn_convex} may be solved in closed form on each maximal interval of $C'$ up to constants.
There is at most one collection of constants that ensures the properties demanded by \Cref{proposition:value_fn}
(at least if $C'$ comprises \emph{finitely} many maximal intervals),
and if there is one then $C'=C$.
We give some details in \cref{suppl:hjb_details}.

\begin{namedthm}[Two-action example (§\ref{sec:value:example}), continued.]
	\label{two-action:value-shape}
	\Cref{proposition:value_fn} implies that the value function must have either the strictly convex shape in \Cref{fig:2act_v_lo} or the convex-affine shape in \Cref{fig:2act_v_hi}.
	It further rules out one of the two; for example, for $\widebar{\lambda}$ small, the convex candidate violates $u \leq v$ at $2/3$.
	(See \cref{suppl:hjb_details:2act} for details.)
\end{namedthm}

The proof relies heavily on \Cref{theorem:viscosity}.
In particular, all three lemmata below are derived using the fact that $v$ is a viscosity solution of \eqref{eq:hjb}.

\begin{proof}[Proof of \Cref{proposition:value_fn}]
	By \Cref{theorem:viscosity}, $v$ is a viscosity solution of \eqref{eq:hjb} satisfying $v=u$ on $\{0,1\}$.
	It follows that $v$ is continuous.
	The fact that $u = \cav u$ on $\{0,1\}$ follows from upper semi-continuity of $u$.
	We have $u \leq v$ because for any $p \in [0,1]$, the value $u(p)$ is attainable (by setting $\lambda=0$ forever), so must be lower than the optimal value $v(p)$.

	To show that $v \leq \cav u$, take any $p \in [0,1]$, and consider the auxiliary problem in which the sender may choose any $[0,1]$-valued process $(p_t)_{t \in \R_+}$ satisfying $\E(p_t) = p$ for every $t \in \R_+$.
	The value $V(p)$ of this problem must exceed $v(p)$ since any belief process the sender can induce in her best-reply problem is available in the auxiliary problem.
	And we have $V(p) = (\cav u)(p)$ since the auxiliary problem consists of a sequence of independent unconstrained persuasion problems (one for each instant $t$), in each of which the optimal value is $(\cav u)(p)$ since $u$ is upper semi-continuous.

	For $(0,1) \setminus C$, we show in \cref{app:lemma_v_u} that
	\begin{lemma}
		\label{lemma:v_u}
		On $(0,1) \setminus C$, we have $v=u$.
	\end{lemma}

	Now for $C$. In \cref{app:lemma_v_C1}, we prove that
	\begin{lemma}
		\label{lemma:v_C1}
		$v$ is continuously differentiable on $C$.
	\end{lemma}
	\noindent
	To show that $v < \cav u$ on $C$, take $p \in C$ and $p' < p < p''$ sufficiently close to $p$, and let $\gamma \in (0,1)$ satisfy $\gamma p' + (1-\gamma) p'' = p$. We have
	\begin{align*}
		v(p) 
		&< \gamma v(p') + (1-\gamma) v(p'')
		&& \quad\text{since $p \in C$}
		\\
		&\leq \gamma (\cav u)(p') + (1-\gamma) (\cav u)(p'')
		&& \quad\text{since $v \leq \cav u$}
		\\
		&\leq (\cav u)(p)
		&& \quad\text{since $\cav u$ is concave} .
	\end{align*}

	Finally, consider $C \setminus D$. We have
	\begin{lemma}
		\label{lemma:v_C2}
		On $C \setminus D$, $v$ is twice continuously differentiable and satisfies \eqref{eq:diff_eqn_convex}.
	\end{lemma}
	\noindent
	This lemma is proved in \cref{app:lemma_v_C2}.
\end{proof}

We prove in \cref{app:value_fn_limit}
that letting $\widebar{\lambda} \to \infty$ in \Cref{proposition:value_fn} yields

\begin{corollary}
	\label{corollary:value_fn_limit}
	As $\widebar{\lambda}$ increases, $v$ increases pointwise.
	As $\widebar{\lambda} \to \infty$, $v$ converges uniformly to $\cav u$.
\end{corollary}

Thus when the sender is able to provide information rapidly,
her equilibrium value is well-approximated by the unconstrained-persuasion model.
Beyond this case, the approximation is typically poor, as evidenced by the two-action example (\Cref{fig:2act_v_hi}, p. \pageref{fig:2act_v_hi}).

\section{Equilibrium information provision}
\label{sec:beliefs}

Having characterised the sender's value function (\Cref{proposition:value_fn}), we are ready to study her equilibrium behaviour.
We first show that she provides more information the less stringent the graduality constraint (\Cref{proposition:beliefs_comp_stat}, §\ref{sec:beliefs:comp_stats}).
We then establish (\Cref{proposition:beliefs_persuasion}, §\ref{sec:beliefs:persuasion})
that less information is provided than in the unconstrained benchmark,
but that the latter model provides a good approximation if the graduality constraint is sufficiently slack.
Finally, 
we highlight the role of discontinuities in the sender's flow payoff $u$:
any discontinuity, if paired with a sufficiently tight constraint, leads less-than-full information to be provided in equilibrium
(\Cref{proposition:disc_nonfull}, §\ref{sec:beliefs:disc}).

\subsection{Induced beliefs in the long run}
\label{sec:beliefs:BR}

As in §\ref{sec:value}, fix a regular Markov strategy $A : [0,1] \to \Delta(\mathcal{A})$.

\begin{corollary}
	\label{corollary:BR}
	The following Markov strategy is a best reply of the sender:
	\begin{equation*}
		\Lambda^\star(p)
		=
		\begin{cases}
			\widebar{\lambda} &\text{if $v(p) > u(p)$} \\
			0 &\text{otherwise} . 
		\end{cases}
	\end{equation*}
\end{corollary}

\begin{proof}
	When $v=u$, setting $\lambda=0$ clearly attains the value $v$ in the sender's problem \eqref{eq:v_defn}.
	When $v>u$, $\lambda>0$ must be optimal.
	By inspection of the sender's problem, $\lambda=\widebar{\lambda}$ is optimal whenever $\lambda>0$ is.
\end{proof}

The strategy $\Lambda^\star$ provides information at full tilt when it is strictly valuable, and provides none otherwise.
It is partially unique.%
	\footnote{\label{footnote:BR_unique}Precisely: any best reply must have $\Lambda>0$ on $\{v>u\}$, $\Lambda = \widebar{\lambda}$ a.e. on $\{v>u\}$, and $\Lambda = 0$ a.e. on $\{ v = u \} \setminus K$, where $K \subseteq (0,1)$ is the set on which $u$ is locally (weakly) convex.
	Anything is optimal on $\{ v = u \} \intersect K$.
	(In \Cref{fig:2act_v_hi} on p. \pageref{fig:2act_v_hi}, we have $\{ v = u \} \intersect K = (2/3,1)$.)}

Under this strategy, the belief $p_t$ evolves according to $\dd p_t = \sqrt{ \widebar{\lambda} / \sigma^2 } p_t (1-p_t) \dd B_t$ until it hits the (closed) set $\{ v = u \}$, then remains constant.
The belief process $( p_t )_{ t \in \R_+ }$ converges a.s. by the martingale convergence theorem
\parencite[e.g. Theorem 3.15 in][ch. 1]{KaratzasShreve1991}.
Write $F$ for the distribution of the limiting random variable.
The support of $F$ is the set of beliefs that the sender induces (with positive probability) in the long run.
Note that $F$ has mean $p_0$ since each $p_t$ does,
by the bounded convergence theorem
\parencite[e.g. Theorem 16.5 in][]{Billingsley1995}.

\begin{corollary}
	\label{corollary:beliefs}
	Fix a prior $p_0$.
	A best reply of the sender induces the beliefs $\{ p^-, p^+ \}$ in the long run, where
	\begin{align*}
		p^- 
		&\coloneqq \sup\{ p \in [0,p_0] : v(p) = u(p) \}
		\\
		\text{and} \quad
		p^+ 
		&\coloneqq \inf\{ p \in [p_0,1] : v(p) = u(p) \} .
	\end{align*}
\end{corollary}

\begin{proof}
	The best reply $\Lambda^\star$ in \Cref{corollary:BR} obviously induces $\{ p^-, p^+ \}$.
\end{proof}

Provided $v(p_0) > u(p_0)$ (the interesting case), the long-run beliefs $\{p^-,p^+\}$ that we study are generically the unique ones consistent with a best reply---see \cref{suppl:generic_lr} for a discussion.
In general, they are the least extreme beliefs induced by some best reply (by \cref{footnote:BR_unique}).

\subsection{Comparative statics}
\label{sec:beliefs:comp_stats}

A slacker graduality constraint
leads to more extreme beliefs:

\begin{proposition}[comparative statics]
	\label{proposition:beliefs_comp_stat}
	Fix any prior $p_0$.
	As $\widebar{\lambda}$ increases, $p^-$ decreases and $p^+$ increases.
\end{proposition}

\begin{proof}
	Write $\{ v > u \}$ for the set of beliefs at which the strategy $\Lambda^\star$ from \Cref{corollary:BR} provides information.
	As $\widebar{\lambda}$ increases, $v$ increases pointwise by \Cref{corollary:value_fn_limit}, so $\{ v > u \}$ increases in the sense of set inclusion, and thus
	\begin{equation*}
		p^-
		= \sup \left( [0,p_0] \setminus \{ v > u \} \right)
	\end{equation*}
	decreases and $p^+$ similarly increases.
\end{proof}

\begin{namedthm}[Two-action example (§\ref{sec:value:example}), continued.]
	\label{two-action:comp-stat}
	Let the prior be $p_0 \in (0,2/3)$.
	The sender induces the long-run beliefs $\{0,1\}$ if information can be produced quickly (\Cref{fig:2act_v_lo}, p. \pageref{fig:2act_v_lo}),
	and the less extreme $\{0,2/3\}$ if not (\Cref{fig:2act_v_hi}).
\end{namedthm}

\Cref{proposition:beliefs_comp_stat} states that more information is provided in the long run when evidence accumulates faster.
More extreme long-run beliefs are achieved
by providing information (at full tilt) for longer a.s.
Consequently, slackening the graduality constraint leads Blackwell-more information to be generated.
Thus the decision-maker's payoff improves,
provided she values information in the sense that $p \mapsto f_{\textsl{D}}(a,p)$ is convex for each $a \in \mathcal{A}$
(a property satisfied by expected utility).

\subsection{Slow vs. unconstrained persuasion}
\label{sec:beliefs:persuasion}

We next characterise how equilibrium information provision compares with that in the (static) unconstrained-persuasion benchmark.
It is well-known (see \textcite{KamenicaGentzkow2011}) that given $p_0$, an optimal policy in the unconstrained persuasion problem induces the beliefs $\{ P^-, P^+ \}$, where
\begin{align*}
	P^- 
	&\coloneqq \sup\{ p \in [0,p_0] : (\cav u)(p) = u(p) \}
	\\
	P^+ 
	&\coloneqq \inf\{ p \in [p_0,1] : (\cav u)(p) = u(p) \} .
\end{align*}
It follows that either one or two beliefs are induced, depending on $p_0$.

Given \Cref{proposition:beliefs_comp_stat}, it is intuitive that less information will be provided in the dynamic model than in the benchmark, since the latter corresponds (informally) to the case $\widebar{\lambda} = \infty$ of instantaneous information arrival.
The following result shows that information provision is close to that in the unconstrained benchmark when evidence accumulates fast (high $\widebar{\lambda}$).

\begin{proposition}[slow vs. unconstrained]
	\label{proposition:beliefs_persuasion}
	Fix a prior $p_0$.
	For any $\widebar{\lambda} > 0$, we have $P^- \leq p^- \leq p^+ \leq P^+$.
	As $\widebar{\lambda} \to \infty$, $p^- \to P^-$ and $p^+ \to P^+$.
\end{proposition}

\begin{namedthm}[Two-action example (§\ref{sec:value:example}), continued.]
	\label{two-action:beliefs-persuasion}
	Let the prior be $p_0 \in (0,2/3)$.
	If the graduality constraint is tight (\Cref{fig:2act_v_hi}, p. \pageref{fig:2act_v_hi}),
	then less information is provided than with no constraint:
	$p^- = 0 = P^-$ and $p^+ = 2/3 < 1 = P^+$.
	For a sufficiently slack constraint (\Cref{fig:2act_v_lo}),
	there is no gap.%
		\footnote{Examples also exist
		in which there is a gap for all $\widebar{\lambda}>0$,
		which closes only in the limit.}
\end{namedthm}

Although it is intuitive that long-run induced beliefs should converge to the unconstrained ones as $\widebar{\lambda} \to \infty$, the result is not obvious.
To see why, observe that the analogous result for the `slow limit' $\widebar{\lambda} \to 0$ is false!
The natural static benchmark here is the trivial model in which no information is available, so that the belief stays put at the prior $p_0$.
It is \emph{not} true that $p^-$ and $p^+$ converge to $p_0$ as $\widebar{\lambda} \to 0$: indeed, in the two-action example, the (uniquely optimal) long-run induced beliefs are $\{p^-,p^+\} = \{0,2/3\}$ for \emph{every} low value of $\widebar{\lambda} > 0$.%
	\footnote{The key formal difference between the two limits is that the convergence of $v$ to $\cav u$ as $\widebar{\lambda} \to \infty$ is uniform by \Cref{corollary:value_fn_limit} (p. \pageref{corollary:value_fn_limit}), whereas the convergence of $v$ to $u$ as $\widebar{\lambda}\to 0$ is merely pointwise (unless $u$ is continuous).}

\begin{proof}[Proof of \Cref{proposition:beliefs_persuasion}]
	To emphasise dependence on parameters, write $v_{\widebar{\lambda}}$, $p^-_{\widebar{\lambda}}$ and $p^+_{\widebar{\lambda}}$ for the value and long-run beliefs.
	Let $\{ v_{\widebar{\lambda}} > u \}$ be the set of beliefs at which the strategy $\Lambda^\star$ in \Cref{corollary:BR} provides information.

	For the first part, fix a $\widebar{\lambda}>0$.
	Clearly $P^- \leq P^+$.
	By \Cref{proposition:value_fn}, $u(p) < v_{\widebar{\lambda}}(p)$ implies $u(p) < (\cav u)(p)$, so that $\{ v_{\widebar{\lambda}} > u \} \subseteq \{ \cav u > u \}$. Therefore
	\begin{equation*}
		P^-
		= \sup\left( [0,p_0] \setminus \{ \cav u > u \} \right)
		\leq \sup\left( [0,p_0] \setminus \{ v_{\widebar{\lambda}} > u \} \right)
		= p^-_{\widebar{\lambda}} ,
	\end{equation*}
	and similarly $p^+_{\widebar{\lambda}} \leq P^+$.

	Now for the second part.
	Since $p^-_{\widebar{\lambda}}$ decreases monotonically as $\widebar{\lambda}$ increases by \Cref{proposition:beliefs_comp_stat}, and lives in the compact set $[P^-,p_0]$, it converges to some limit $p^-_\infty \in [P^-,p_0]$.
	We wish to show that $p^-_\infty = P^-$, so suppose toward a contradiction that $p^-_\infty > P^-$.
	On the one hand,
	\begin{equation}
		(\cav u)\left(p^-_{\widebar{\lambda}}\right) 
		\to (\cav u)\left(p^-_\infty\right) > u\left(p^-_\infty\right)
		\label{eq:beliefs_persuasion}
	\end{equation}
	by continuity of $\cav u$ and $p^-_\infty > P^-$ (recall the definition of $P^-$).
	On the other hand, (recalling the definition of $p^-_{\widebar{\lambda}}$,)
	\begin{equation*}
		\abs*{ (\cav u)\left(p^-_{\widebar{\lambda}}\right) 
		- u\left(p^-_{\widebar{\lambda}}\right) } 
		= \abs*{ (\cav u)\left(p^-_{\widebar{\lambda}}\right) 
		- v_{\widebar{\lambda}}\left(p^-_{\widebar{\lambda}}\right) } 
		\to 0
	\end{equation*}
	since $v_{\widebar{\lambda}}$ converges \emph{uniformly} to $\cav u$ by \Cref{corollary:value_fn_limit} (p. \pageref{corollary:value_fn_limit}),
	by a standard property of uniform convergence
	\parencite[e.g. Theorem 7.11 in][]{Rudin1976}.
	It follows by upper semi-continuity of $u$ that
	\begin{equation*}
		(\cav u)\left(p^-_{\widebar{\lambda}}\right) 
		\to \lim_{ \widebar{\lambda} \to \infty } u\left(p^-_{\widebar{\lambda}}\right) 
		\leq u\left(p^-_\infty\right)
	\end{equation*}
	a contradiction with \eqref{eq:beliefs_persuasion}.
	A similar argument shows that $p^+_{\widebar{\lambda}} \to P^+$.
\end{proof}

\subsection{The role of discontinuities}
\label{sec:beliefs:disc}

The \hyperref[two-action:beliefs-persuasion]{two-action example}
suggests that
discontinuities in the sender's flow payoff $u$
generate a particularly stark dampening effect of the graduality constraint on information provision.
The following proposition expresses this idea:
any discontinuity, when paired with a sufficiently tight constraint,
will preclude full information from being provided in equilibrium.

\begin{proposition}
	\label{proposition:disc_nonfull}
	Fix a prior $p_0$.
	If the sender's flow payoff $u$ has an interior discontinuity,
	then $\left\{ p^-, p^+ \right\} \neq \{0,1\}$
	provided $\widebar{\lambda} > 0$ is small enough.
\end{proposition}

\begin{proof}
	Let $u$ be discontinuous at $p \in (0,1)$.
	Since $u$ is piecewise continuous,
	it is continuous on a left-neighbourhood $(p',p)$ of $p$
	and on a right-neighbourhood $(p,p'')$.

	Fix a value of $\widebar{\lambda}>0$
	such that $\left\{ p^-, p^+ \right\} = \{0,1\}$.
	(Recall that $p^-$, $p^+$ and $v$ all depend on $\widebar{\lambda}$, albeit our notation leaves this implicit.)
	Then $v>u$ on $(0,1)$ by definition of $p^-$ and $p^+$.
	It follows by \Cref{proposition:value_fn} that on the intervals $(p',p)$ and $(p,p'')$, the value $v$
	(is everywhere locally strictly convex, and thus)
	satisfies equation \eqref{eq:diff_eqn_convex} on p. \pageref{eq:diff_eqn_convex}.
	By inspection, this implies that $v$ is strictly convex on $(p',p)$ and $(p,p'')$.
	And since $v$ pastes smoothly at $p$, it follows $v$ is convex on all of $(p',p'')$.
	To summarise: if $\left\{ p^-, p^+ \right\} = \{0,1\}$, then $v$ must be a convex majorant of $u$ on $(p',p'')$.

	It thus suffices to show that $v$ fails to be a convex majorant of $u$ on $(p',p'')$ if $\widebar{\lambda} > 0$ is small enough.
	And that follows from the pointwise (monotone) convergence of $v$ to $u$ as $\widebar{\lambda} \downarrow 0$
	and the fact that $u$ is discontinuous at $p$.
\end{proof}

\section{Belief-based conflict of interest}
\label{sec:disagree}

In this section, we drop the common-prior assumption.
The resulting differences in beliefs may engender a conflict of interest even if the players' preferences are aligned conditional on the state of the world.

Such \emph{belief-based} conflicts are pervasive.
They arise, for instance, where pre-existing contracts have largely aligned all parties' interests conditional on the state,
as in many organisational settings.
Disagreements about the best course of action remain ubiquitous in such environments, but originate in agents' differing assessments of the evidence.

Some policy problems also feature well-aligned preferences.
For example, disagreements over actions to mitigate climate change
usually play out as debates about the likely extent of (anthropogenic) global warming,
rather than about what policies are desirable in any given physical scenario.

In this section, we show that a purely belief-based conflict may preclude full information being provided in slow-persuasion equilibrium,
contrary to the unconstrained-persuasion model.
We further argue that belief disagreement harms the sender:
the greater the prior gap, the lower her value.

We extend our preceding results to heterogeneous priors (maintaining arbitrary preferences) in the next section,
then specialise in §\ref{sec:disagree:belief-based} to the case of ex-post aligned preferences.

\subsection{Equilibrium characterisation}
\label{sec:disagree:charac}

The model is as in §\ref{sec:model}, except that the priors $p_0,p_{\textsl{D},0} \in (0,1)$ of the sender and decision-maker may differ.
The priors are commonly known (that is, the players agree to disagree).
Write $p_t$ and $p_{\textsl{D},t}$ for the sender's and decision-maker's beliefs at time $t$.

The model remains tractable because we need not keep track of the decision-maker's belief,
as it may be backed out from the sender's belief and the priors via Bayes's rule:

\begin{observation}
	\label{observation:Pa_beliefs}
	The decision-maker's time-$t$ belief is $p_{\textsl{D},t} = \phi(p_t,p_0,p_{\textsl{D},0})$, where
	\begin{equation*}
		\phi(p,p_0,p_{\textsl{D},0})
		\coloneqq \frac{ p }{ p + (1-p)
		\left. \frac{ p_0 }{ 1 - p_0 } 
		\middle/
		\frac{ p_{\textsl{D},0} }{ 1 - p_{\textsl{D},0} } \right. } .
	\end{equation*}
\end{observation}

In light of \Cref{observation:Pa_beliefs}, MPEs have all of the same properties as in the common-prior case.
For the same reason as in \Cref{observation:myopic} (§\ref{sec:myopic}), a regular Markov strategy $A : [0,1] \to \Delta(\mathcal{A})$ is part of a MPE iff it is myopic.
Given a regular Markov strategy $A$, the sender's induced flow payoff is now
\begin{equation*}
	u(p) \coloneqq \int_{\mathcal{A}} f_{\textsl{S}}( a, p ) 
	A( \dd a | \phi(p,p_0,p_{\textsl{D},0}) ) 
\end{equation*}
since the decision-maker's belief is $\phi(p,p_0,p_{\textsl{D},0})$ when the sender's is $p$. (Note that $u$ depends on the priors.)
It remains true that $u$ is piecewise continuous and upper semi-continuous.
Given $u$, the sender's best-reply problem is unchanged, noting again that $p_t$ is the \emph{sender's} belief.

All of our preceding results therefore remain valid: the sender's value function is a generalised concave envelope (\Cref{proposition:value_fn}), she provides more information the faster evidence accumulates and the more patient she is (\Cref{proposition:beliefs_comp_stat}), and her information provision is well-approximated by unconstrained persuasion when information can be generated quickly (\Cref{proposition:beliefs_persuasion}).

\subsection{Belief-based conflict of interest}
\label{sec:disagree:belief-based}

We now specialise the model of the previous section by assuming that interests are aligned ex-post: $f_{\textsl{S}} = f_{\textsl{D}} = f$.
(In the special case of expected utility, this is equivalent to players having the same preferences conditional on the state $\theta$.)
Whatever conflict remains arises from the prior disagreement alone.

In the unconstrained-persuasion benchmark,
the sender provides full information when the conflict is purely belief-based \parencite[e.g.][]{AlonsoCamara2016}.
This follows from two observations.
First, the sender would be better-off if she were in charge of choosing the action,
and were this the case, then her payoff would be highest under full information.%
	\footnote{This is true for expected-utility preferences $f$,
	and more generally for preferences $f$ that value information
	in the sense that $p \mapsto f(a,p)$ is convex for each $a \in \mathcal{A}$.}
Secondly, the sender attains this upper bound on her payoff by providing full information:
the decision-maker then chooses as the sender would,
since the players' posteriors always agree (they both assign probability 1 to the true state).

When persuasion is constrained to be slow, this argument remains approximately valid if the graduality constraint is loose. In particular, near-full information is provided in equilibrium by \Cref{proposition:beliefs_persuasion}.

Otherwise, the argument breaks down.
The long-run benefit of providing full information
must then be weighed against its potential cost over the short run.
In fact, outside of trivial cases, the cost is sure to dominate whenever the constraint is tight enough.
To formalise this, call an action $a \in \mathcal{A}$ \emph{redundant} iff it is never strictly optimal, i.e.
\begin{equation*}
	f(a,p) \leq \max_{a' \in \mathcal{A} \setminus \{a\}} f(a',p)
	\quad \text{for every $p \in [0,1]$.}
\end{equation*}

\begin{proposition}
	\label{proposition:disagree_nonfull}
	If the actions $\mathcal{A}$ can be totally ordered so that $f = f_{\textsl{S}} = f_{\textsl{D}}$ is strictly single-crossing,%
		\footnote{That is: there is a total order $\succeq$ on $\mathcal{A}$ such that for $a' \succ a$ and $p' > p$, $f(a',p) \geq f(a,p)$ implies $f(a',p') > f(a,p')$.
		Totality can be weakened: it is enough that $\succeq$ be a partial order
		such that $(\mathcal{A},\mathord{\succeq})$ is a lattice and $f(\cdot,p)$ is quasi-supermodular for each $p \in [0,1]$.}
	and at least two actions are non-redundant,
	then for any fixed priors $p_0 \neq p_{\textsl{D},0}$,
	we have $\left\{ p^-, p^+ \right\} \neq \{0,1\}$
	provided $\widebar{\lambda} > 0$ is small enough.
\end{proposition}

Any expected-utility preference $f$ satisfies strict single-crossing.%
	\footnote{An expected-utility $f$ has $f(a,p) = (1-p) u_0(a) + p u_1(a)$ for some $u_0,u_1 : \mathcal{A} \to \R$.
	Define $\succeq$ by $a' \succeq a$ iff $u_1(a') - u_0(a') \geq u_1(a) - u_0(a)$.
	Then $f(a',p) \geq f(a,p)$ implies $f(a',p') > f(a,p')$
	for any $a' \succ a$ and $p' > p$.
	This $\succeq$ is a total order (in particular, anti-symmetric)
	once $\mathcal{A}$ is pruned of strictly dominated actions and duplicates.}
We prove \Cref{proposition:disagree_nonfull} in \cref{app:disagree_nonfull_proof}.

\begin{namedthm}[Two-action example (§\ref{sec:value:example}), continued.]
	\label{example:two-actions_het-priors}
	Modify the example so that the decision-maker shares the sender's preference $f_{\textsl{S}}$,
	but has a prior $p_{\textsl{D},0} = 1/5$ 
	different from the sender's $p_0 = 1/2$.
	The decision-maker's (unique regular) Markov strategy remains $A = \1_{[2/3,1]}$,%
		\footnote{In terms of the \emph{sender's} belief $p$,
		the decision-maker's payoff from action $a=1$ is 
		$3 \phi(p,1/2,1/5) - 1
		= 3 \tfrac{p}{4-3p} - 1$,
		which exceeds zero (the payoff of $a=0$)
		iff $p \geq 2/3$.}
	so our preceding analysis remains valid.
	Thus when evidence accumulates slowly (\Cref{fig:2act_v_hi}, p. \pageref{fig:2act_v_hi}),
	the sender induces the imperfectly-informative
	long-run beliefs $\{p^-,p^+\} = \{0,2/3\}$.
\end{namedthm}

In unconstrained persuasion, the sender's value is invariant to the decision-maker's prior.
This is because she provides full information,
leading both players' posterior beliefs always to agree (assigning probability 1 to the true state),
so that the decision-maker chooses actions just like the sender would.
By contrast, prior disagreement harms the sender in slow persuasion,
because it induces a conflict of interest.
Under natural conditions, the sender is better-off the smaller the disagreement:

\begin{proposition}
	\label{proposition:disagree_value}
	If the actions $\mathcal{A}$ can be totally ordered so that $f = f_{\textsl{S}} = f_{\textsl{D}}$ is strictly single-crossing,
	then for any fixed prior $p_0$,
	$v(p_0)$ increases pointwise
	as $p_{\textsl{D},0} < p_0$ increases toward $p_0$
	(as $p_{\textsl{D},0} > p_0$ decreases toward $p_0$).
\end{proposition}

The proof is in \cref{app:disagree_value_proof}.



\begin{appendices}
\crefalias{section}{appsec}
\crefalias{subsection}{appsec}
\crefalias{subsubsection}{appsec}

\renewcommand*{\thesubsection}{\Alph{subsection}}

\section*{Appendices}
\label{app}
\addcontentsline{toc}{section}{Appendices}

\subsection{Proof of \texorpdfstring{\Cref{theorem:viscosity}}{Theorem \ref{theorem:viscosity}} (p. \pageref{theorem:viscosity})}
\label{app:viscosity}

The boundary condition $v=u$ on $\{0,1\}$ holds by inspection of the sender's best-reply problem \eqref{eq:v_defn}, since these are absorbing states.
For the remainder, let $v_\star$ and $v^\star$ be the lower and upper semi-continuous envelopes of $v$.
In \Cref{theorem:viscosity}(a), we show that piecewise continuity of $u$ suffices for $v_\star$ to be a viscosity super-solution.
In \Cref{theorem:viscosity}(b), we prove that $v^\star$ is a viscosity sub-solution, without requiring piecewise continuity.
In both cases, we adapt a standard argument.
Finally, we establish in \Cref{theorem:viscosity}(c) that $v$ is continuous, so that $v_\star = v^\star = v$.%
	\footnote{As we explain in \cref{suppl:viscosity_intro:prop}, it is typical to replace the last step with an appeal to a comparison principle.
	Since we are not aware of a comparison principle that requires only piecewise continuity of $u$, we prove continuity directly instead.}
We will make occasional use of the fact that $v \geq u$, which holds since the value $u$ is attainable (by never providing any information).

First, the super-solution property of $v_\star$, which relies on piecewise continuity of $u$:

\begin{namedthm}[\Cref{theorem:viscosity}(a).]
	\label{namedthm:lemma1sup}
	If $u$ is piecewise continuous, then $v_\star$ is a viscosity super-solution of \eqref{eq:hjb}.
\end{namedthm}

\begin{proof}
	We follow the standard argument \parencite[e.g.][Proposition 4.3.1]{Pham2009}, which assumes that $u$ is continuous.
	We sketch the steps that are unchanged, and emphasise the juncture at which a new argument is needed to accommodate merely piecewise continuity of $u$.

	Take any $p \in (0,1)$ and any twice continuously differentiable $\phi : (0,1) \to \R$ such that $v_\star-\phi$ has a local minimum at $p$.
	In light of \Cref{remark:viscosity_contact}, we may assume without loss of generality that $v_\star(p) - \phi(p) = 0$ and that $v_\star - \phi \geq 0$ (i.e. $p$ is a \emph{global} minimum of $v_\star-\phi$.)
	We wish to show that
	\begin{equation*}
		v_\star(p) 
		\geq u_\star(p)
		+ \widebar{\lambda} \frac{ p^2 (1-p)^2 }{ 2r\sigma^2 } 
		\max\left\{ 0, \phi''(p) \right\} .
	\end{equation*}
	We have $v_\star \geq u_\star$ since $v \geq u$, so what must be shown is that
	\begin{equation}
		v_\star(p) 
		\geq u_\star(p)
		+ \widebar{\lambda} \frac{ p^2 (1-p)^2 }{ 2r\sigma^2 } 
		\phi''(p) .
		\label{eq:super_fulltilt}
	\end{equation}

	By definition of $v_\star$ and since $\phi$ is continuous with $v_\star(p) - \phi(p) = 0$, we may find a sequence $(p_n)_{n \in \N}$ in $(0,1)$ converging to $p$ along which
	\begin{equation*}
		\gamma_n \coloneqq v(p_n) - \phi(p_n)
	\end{equation*}
	vanishes.
	Choose any strictly positive sequence $(h_n)_{n \in \N}$ in $\R$ such that $h_n \to 0$ and $\gamma_n/h_n \to 0$ as $n \to \infty$.

	Consider the `full tilt forever' strategy which sets $\lambda_t=\widebar{\lambda}$ a.s. no matter what happens.
	Write $P_s^n$ for the induced (belief) process when the initial condition is $P_0 = p_n$.
	Since $v(p_0)$ is the optimal value, it must exceed the expected discounted payoff obtained by using the `full tilt forever' control process until time $h_n$, then reverting to optimal behaviour:
	\begin{equation*}
		v(p_n)
		\geq \E\left(
		r \int_0^{h_n} e^{-rt} u\left( P_t^n \right) \dd t
		+ e^{-rh_n} v\left( P_{h_n}^n \right)
		\right)
		\quad \forall n \in \N .
	\end{equation*}
	(The integrand on the right-hand side is in fact measurable,
	so that the expectation is well-defined; see e.g. \textcite[Theorem 3.3.1]{Pham2009}.)
	Using $v-\phi \geq v_\star-\phi \geq 0$ and the definition of $\gamma_n$ yields
	\begin{equation}
		\phi(p_n) + \gamma_n 
		\geq \E\left(
		r \int_0^{h_n} e^{-rt} u\left( P_t^n \right) \dd t
		+ e^{-rh_n} \phi\left( P_{h_n}^n \right)
		\right)
		\quad \forall n \in \N .
		\label{eq:DPPsup1}
	\end{equation}

	Since $\phi$ is twice continuously differentiable and $P_t^n$ evolves as
	\begin{equation*}
		\dd P_t^n 
		= \sqrt{\widebar{\lambda}} \frac{ P_t^n \left( 1 - P_t^n \right) }{ \sigma } \dd B_t
	\end{equation*}
	where $(B_t)_{t \in \R_+}$ is a standard Brownian motion,
	we may apply Itô's lemma to the process $\left( e^{-rt} \phi\left( P_t^n \right)\right)_{t \in \R_+}$ to obtain, for each $n \in \N$,
	\begin{multline*}
		e^{-rh_n} \phi\left( P_{h_n}^n \right)
		= \phi\left( p_n \right)
		- r \int_0^{h_n}
		e^{-rt} \phi\left( P_t^n \right)
		\dd t
		\\
		+ \frac{1}{2} \int_0^{h_n}
		\left( \sqrt{\widebar{\lambda}} \frac{ P_t^n \left( 1 - P_t^n \right) }{ \sigma } \right)^2
		e^{-rt}
		\phi''\left( P_t^n \right)
		\dd t .
	\end{multline*}
	Substituting in \eqref{eq:DPPsup1} and rearranging slightly yields
	\begin{multline}
		\frac{ \gamma_n }{ h_n }
		\geq \E\left(
		r \frac{1}{h_n} \int_0^{h_n} e^{-rt} u\left( P_t^n \right) \dd t
		- r \frac{1}{h_n} \int_0^{h_n}
		e^{-rt} \phi\left( P_t^n \right)
		\dd t
		{}\right.
		\\
		\left.{}
		+ r \frac{1}{h_n} \int_0^{h_n}
		e^{-rt}
		\widebar{\lambda}
		\frac{ \left( P_t^n \right)^2 \left( 1 - P_t^n \right)^2 }
		{ 2 r \sigma^2 }
		\phi''\left( P_t^n \right)
		\dd t 
		\right)
		\quad \forall n \in \N .
		\label{eq:DPPsup2}
	\end{multline}
	We will obtain \eqref{eq:super_fulltilt} as the limit of this inequality as $n \to \infty$.

	Since $\phi$ is twice continuously differentiable and the sample paths of $\left( P_t^n \right)_{t \in \R_+}$ are continuous a.s., the mean-value theorem may be applied path-by-path to the second and third terms inside the expectation in \eqref{eq:DPPsup2} to conclude that they converge a.s. to, respectively, $-\phi(p)$ and
	\begin{equation*}
		\widebar{\lambda}
		\frac{ p^2 (1-p)^2 }
		{ 2 r \sigma^2 }
		\phi''(p) .
	\end{equation*}

	It remains to show that the first term converges a.s. to a limit that exceeds $u_\star(p)$.
	If $p$ is a continuity point of $u$, then $u$ is continuous on a neighbourhood of $p$ by piecewise continuity, so the same mean-value-theorem argument implies that the first term converges a.s. to $u(p) \geq u_\star(p)$, as desired.

	Suppose instead that $p$ is a discontinuity point of $u$; this requires an additional argument relative to the standard proof.
	By piecewise continuity, $u$ is continuous on a left- and a right-neighbourhood of $p$.
	Thus for any sufficiently small $\eps>0$, we may apply the mean-value theorem on either side of $p$ to obtain the existence of a $p_-^\eps \in (p-\eps,p)$ and a $p_+^\eps \in (p,p+\eps)$ such that
	\begin{equation*}
		\frac{1}{\eps} \int_{p-\eps}^p u
		= u\left( p_-^\eps \right)
		\quad \text{and}\quad
		\frac{1}{\eps} \int_p^{p+\eps} u
		= u\left( p_+^\eps \right) ,
	\end{equation*}
	so that
	\begin{equation*}
		\frac{1}{2\eps} \int_{p-\eps}^{p+\eps} u
		\geq \min\left\{ u\left( p_-^\eps \right), u\left( p_+^\eps \right) \right\} .
	\end{equation*}
	The left-hand side converges as $\eps \downarrow 0$, and the right-hand side converges to $\min\{ u(p-), u(p+) \}$. Thus
	\begin{equation*}
		\lim_{ \eps \to 0 } \frac{1}{2\eps} \int_{p-\eps}^{p+\eps} u
		\geq \min\left\{ u\left( p- \right), u\left( p+ \right) \right\} 
		\geq u_\star(p) .
	\end{equation*}
	As with the second and third terms in \eqref{eq:DPPsup2}, we may apply this argument to a.e. path of the first term since a.e. sample path of $\left( P_t^n \right)_{t \in \R_+}$ is continuous.
	Thus the first term in \eqref{eq:DPPsup2} converges a.s. to a limit that exceeds $u_\star(p)$.

	Next, observe that all three terms inside the expectation in \eqref{eq:DPPsup2} are bounded off $D$ by a constant independent of $n$ because $\phi$, $\phi''$ and $u$ are continuous off $D$.
	Furthermore, the set $D$ is null under the occupancy measure of $\left( P_t^n \right)_{ t \in \R_+}$, for every $n \in \N$.%
		\footnote{This is because the occupancy measure is absolutely continuous with respect to Lebesgue measure, and $D$ is Lebesgue-null since it is discrete.}
	It follows by the bounded convergence theorem that the right-hand side of \eqref{eq:DPPsup2} converges to a limit exceeding
	\begin{equation*}
		u_\star(p)
		- \phi(p)
		+ \widebar{\lambda}
		\frac{ p^2 (1-p)^2 }
		{ 2 r \sigma^2 }
		\phi''(p) .
	\end{equation*}
	The left-hand side of \eqref{eq:DPPsup2} vanishes by construction of $(h_n)_{n \in \N}$.
	Thus
	\begin{equation*}
		0 \geq
		u_\star(p)
		- \phi(p)
		+ \widebar{\lambda}
		\frac{ p^2 (1-p)^2 }
		{ 2 r \sigma^2 }
		\phi''(p) .
	\end{equation*}
	Using $\phi(p) = v_\star(p)$ and rearranging yields the desired inequality \eqref{eq:super_fulltilt}.
\end{proof}

Unlike the super-solution property of $v_\star$, the sub-solution property of $v^\star$ holds for general $u$:

\begin{namedthm}[\Cref{theorem:viscosity}(b).]
	\label{namedthm:lemma1sub}
	$v^\star$ is a viscosity sub-solution of \eqref{eq:hjb}.
\end{namedthm}

\begin{proof}
	Again, we follow the standard line of reasoning (e.g. \textcite[Proposition 4.3.2]{Pham2009}, noting the errata \parencite{Pham2012}) for the case in which $u$ is continuous.
	Where continuity of $u$ is usually invoked, we shall make do with the (definitional) upper semi-continuity of $u^\star$.

	Take any $p \in (0,1)$ and any twice continuously differentiable $\phi : (0,1) \to \R$ such that $\phi-v^\star$ has a local minimum at $p$.
	By \Cref{remark:viscosity_contact}, we may assume without loss that $\phi(p) - v^\star(p) = 0$.
	Suppose that the viscosity sub-solution property fails at $p$:
	\begin{equation*}
		\phi(p)
		= v^\star(p) 
		> u^\star(p)
		+ \widebar{\lambda}
		\frac{ p^2 (1-p)^2 }{ 2r\sigma^2 } 
		\max\left\{ 0, \phi''(p) \right\} .
	\end{equation*}
	We shall derive a contradiction.

	By \Cref{remark:viscosity_contact} again, we may assume that $\phi - v^\star$ has a strict global minimum at $p$.
	For $\eta>0$, write
	\begin{equation*}
		B_\eta \coloneqq \{ q \in (0,1) : \abs*{ q - p } < \eta \}
	\end{equation*}
	for the open ball of radius $\eta$ around $p$, and $\partial B_\eta$ for its boundary.
	Define
	\begin{equation*}
		k_\eta 
		\coloneqq \min_{ q \in \partial B_\eta } \abs*{ \phi(q) - v^\star(q) } ,
	\end{equation*}
	noting that it is strictly positive for $\eta>0$ because the minimum of $\phi-v^\star$ at $p$ is strict.
	Since $\phi$ and $\phi''$ are continuous and $u^\star$ is upper semi-continuous,
	we may find an $\eta > 0$ and an $\eps \in (0,k_\eta]$ small enough that
	\begin{equation}
		\phi(q) 
		\geq u^\star(q)
		+ \widebar{\lambda}
		\frac{ q^2 (1-q)^2 }{ 2r\sigma^2 } 
		\max\left\{ 0, \phi''(q) \right\} 
		+ \eps
		\quad \text{for all $q \in B_\eta$.}
		\label{eq:DPPsub1}
	\end{equation}

	By definition of $v^\star$ and since $\phi$ is continuous with $\phi(p) - v^\star(p) = 0$, we may find a sequence $(p_n)_{n \in \N}$ in $B_\eta$ converging to $p$ along which
	\begin{equation*}
		\gamma_n \coloneqq \phi(p_n) - v(p_n)
	\end{equation*}
	vanishes.
	Let $(\lambda_t^n)_{t \in \R_+}$ be an $\eps/2$-best reply in the sender's best-reply problem with prior $p_0 = p_n$, and write $\left( P_t^n \right)_{ t \in \R_+ }$ for the belief process induced by this strategy.
	Let $\tau_n$ be the first exit time of $\left( P_t^n \right)_{ t \in \R_+ }$ from $B_\eta$.
	Using $(\lambda_t^n)_{t \in \R_+}$ only until time $\tau_n$ and then reverting to optimal behaviour is even better, so certainly attains value at least $v(p_n) - \eps/2$:
	\begin{equation*}
		v(p_n) - \frac{\eps}{2}
		\leq \E\left(
		r \int_0^{\tau_n} e^{-rt} u\left( P_t^n \right) \dd t
		+ e^{-r\tau_n} v\left( P_{\tau_n}^n \right)
		\right) .
	\end{equation*}
	(It is non-trivial but true that the integrand is measurable, so that the expectation-well defined;
	see e.g. \textcite[Theorem 3.3.1]{Pham2009}.)
	Subtracting $\phi(p_n)$ from both sides and using the fact that
	\begin{equation*}
		\phi - v \geq \phi - v^\star \geq k_\eta \geq \eps
		\quad \text{on $\partial B_\eta$}
	\end{equation*}
	yields
	\begin{equation}
		- \gamma_n - \frac{\eps}{2}
		\leq \E\left(
		- e^{-r\tau_n} \eps
		+ r \int_0^{\tau_n} e^{-rt} u\left( P_t^n \right) \dd t
		+ e^{-r\tau_n} \phi\left( P_{\tau_n}^n \right)
		- \phi(p_n)
		\right) .
		\label{eq:DPPsub2}
	\end{equation}

	Since $\phi$ is twice continuously differentiable and $P_t^n$ evolves as
	\begin{equation*}
		\dd P_t^n 
		= \lambda_t^n \frac{ P_t^n \left( 1 - P_t^n \right) }{ \sigma } \dd B_t
	\end{equation*}
	where $(B_t)_{t \in \R_+}$ is a standard Brownian motion,
	we may apply Itô's lemma to the process $\left( e^{-rt} \phi\left( P_t^n \right)\right)_{t \in \R_+}$ to obtain, for each $n \in \N$,
	\begin{multline*}
		e^{-r\tau_n} \phi\left( P_{\tau_n}^n \right)
		= \phi\left( p_n \right)
		- r \int_0^{\tau_n}
		e^{-rt} \phi\left( P_t^n \right)
		\dd t
		\\
		+ \frac{1}{2} \int_0^{\tau_n}
		\left( \sqrt{ \lambda_t^n } \frac{ P_t^n \left( 1 - P_t^n \right) }{ \sigma } \right)^2
		e^{-rt}
		\phi''\left( P_t^n \right)
		\dd t .
	\end{multline*}
	Substituting in \eqref{eq:DPPsub2} and using \eqref{eq:DPPsub1} yields
	\begin{multline*}
		- \gamma_n - \frac{\eps}{2}
		\leq \E\Biggl(
			- e^{-r\tau_n} \eps
		\\
		\begin{aligned}
			&\quad
			+ r \int_0^{\tau_n} e^{-rt} \left[ 
			- \phi\left( P_t^n \right)
			+ u\left( P_t^n \right) 
			+ \lambda_t^n
			\frac{ \left( P_t^n \right)^2 \left( 1 - P_t^n \right)^2 }
			{ 2r\sigma^2 }
			\phi''\left( P_t^n \right)
			\right] \dd t
			\Biggr) 
			\\
			&\leq \E\Biggl(
			- e^{-r\tau_n} \eps
			\\
			&\quad
			+ r \int_0^{\tau_n} e^{-rt} \left[ 
			- \phi\left( P_t^n \right)
			+ u\left( P_t^n \right) 
			+ \widebar{\lambda}
			\frac{ \left( P_t^n \right)^2 \left( 1 - P_t^n \right)^2 }
			{ 2r\sigma^2 }
			\max\left\{ 0, \phi''\left( P_t^n \right) \right\}
			\right] \dd t
			\Biggr) 
			\\
			&\leq \E\left(
			- e^{-r\tau_n} \eps
			+ r \int_0^{\tau_n} e^{-rt} (-\eps) \dd t
			\right) 
			= -\eps .
		\end{aligned}
	\end{multline*}
	Since $\gamma_n$ vanishes as $n \to \infty$, we have the contradiction $-\eps/2 \leq -\eps$.
\end{proof}

It remains only to show that $v = v_\star = v^\star$, i.e. that $v$ is continuous.

\begin{namedthm}[\Cref{theorem:viscosity}(c).]
	\label{namedthm:v_conts}
	If $u$ is piecewise continuous, then $v$ is continuous.
\end{namedthm}

\begin{proof}
	We deal separately with $\{0,1\}$ and $(0,1)$.
	Begin with continuity at $0$; the argument at $1$ is analogous.
	Take a sequence $(p_n)_{n \in \N}$ in $(0,1)$ converging to $0$; we will show that $v(p_n) \to u(0) = v(0)$.
	
	At each $n \in \N$, consider the auxiliary problem in which the sender may choose any process $(p_t)_{t \in \R_+}$ satisfying $\E(p_t) = p_n$ for every $t \in \R_+$.
	The value $V(p_n)$ of this problem must exceed $v(p_n)$ since any belief process the sender can induce in her best-reply problem is available in the auxiliary problem.
	And we have $V(p_n) \leq (\cav u)(p_n)$ since the auxiliary problem may broken down into a sequence of independent static persuasion problems, in each of which the optimal value is at most $(\cav u)(p_n)$.
	Thus we have
	\begin{equation*}
		u(p_n) \leq v(p_n) \leq V(p_n) \leq (\cav u)(p_n) 
		\quad \text{for every $n \in \N$.}
	\end{equation*}
	As $n \to \infty$, $u(p_n) \to u(0)$ since $u$ is continuous at $0$ by piecewise continuity, and $(\cav u)(p_n) \to u(0)$ since $\cav u$ is continuous and $(\cav u)(0) = u(0)$ because $u$ is continuous at $0$.
	It follows that $u(0) \leq \lim_{n \to \infty} v(p_n) \leq u(0)$.

	To establish that $v$ is continuous on $(0,1)$, fix a $p \in (0,1)$.
	It suffices to show that $\underline{v} \geq \widebar{v}$, where
	\begin{equation*}
		\underline{v} \coloneqq \liminf_{q \to p} v(q)
		\quad \text{and} \quad
		\widebar{v} \coloneqq \limsup_{q \to p} v(q) .
	\end{equation*}
	By construction, there exist sequences $\bigl( \underline{p}_n \bigr)_{n \in \N}$ and $\left( \widebar{p}_n \right)_{n \in \N}$ converging to $p$ along which
	\begin{equation*}
		v\bigl( \underline{p}_n \bigr) \to \underline{v}
		\quad \text{and} \quad
		v\left( \widebar{p}_n \right) \to \widebar{v} 
		\quad \text{as $n \to \infty$.}
	\end{equation*}
	Note that $v$ is bounded since $u$ is (being piecewise continuous).

	Suppose first that these sequences may both be chosen to lie in $(0,p)$; the case in which they may be chosen to lie in $(p,1)$ is analogous.
	Then we may choose them so that $\widebar{p}_{n-1} \leq \underline{p}_n \leq \widebar{p}_n$ for every $n \in \N$, where $\widebar{p}_0 \coloneqq 0$ by convention.
	For the sender's best-reply problem with prior $p_0 = \underline{p}_n$, consider a strategy that sets $\lambda = \widebar{\lambda}$ while $p_t \in \left( \widebar{p}_{n-1}, \widebar{p}_n \right)$ and $\lambda = 0$ otherwise, and write $\left( P_t^n \right)_{t \in \R_+}$ for the induced belief process. Write $\tau_n$ for the first time that $\left( P_t^n \right)_{t \in \R_+}$ hits $\left\{ \widebar{p}_{n-1}, \widebar{p}_n \right\}$.
	Since this strategy cannot be better than optimal, we have
	\begin{equation*}
		v\bigl( \underline{p}_n \bigr)
		\geq \E\left(
		r \int_0^{\tau_n} e^{-rt} u\left( P_t^n \right) \dd t
		+ e^{-r\tau_n} v\left( P_{\tau_n}^n \right)
		\right)
		\quad \text{for each $n \in \N$.}
	\end{equation*}
	(A standard result ensures that the right-hand-side integrand is measurable, making the expectation well-defined; e.g. \textcite[Theorem 3.3.1]{Pham2009}.)
	The left-hand side converges to $\underline{v}$ as $n \to \infty$.
	The hitting time $\tau_n$ vanishes a.s., and $v\left( P_{\tau_n}^n \right)$ converges a.s. to $\widebar{v}$.
	Furthermore, $u$ and $v$ are bounded by piecewise continuity.%
		\footnote{$v$ is bounded below by $u$, and is bounded above by $V \leq \cav u$, where $V$ is the value of the auxiliary problem in the first part of the proof.}
	Hence the right-hand side converges to $\widebar{v}$ by the bounded convergence theorem, so that $\underline{v} \geq \widebar{v}$.

	Suppose instead that the sequences cannot be chosen to lie on the same side of $p$---without loss of generality, $\underline{p}_n < p < \widebar{p}_n$ for every $n \in \N$.
	For the sender's problem with $p_0 = \underline{p}_n$, consider a strategy that sets $\lambda = \widebar{\lambda}$ while $p_t \in \bigl( \underline{p}_{n-1}, \widebar{p}_n \bigr)$ and $\lambda = 0$ otherwise, and write $\left( P_t^n \right)_{t \in \R_+}$ for the induced belief process. Let $\tau_n$ be the first time that $\left( P_t^n \right)_{t \in \R_+}$ hits $\bigl\{ \underline{p}_{n-1}, \widebar{p}_n \bigr\}$.
	The optimal value must exceed the value from using this strategy:
	\begin{equation}
		v\bigl( \underline{p}_n \bigr)
		\geq \E\left(
		r \int_0^{\tau_n} e^{-rt} u\left( P_t^n \right) \dd t
		+ e^{-r\tau_n} v\left( P_{\tau_n}^n \right)
		\right)
		\quad \text{for each $n \in \N$.}
		\label{eq:v_cont_ineq}
	\end{equation}
	(Again, the right-hand side is well-defined.)
	The left-hand side converges to $\underline{v}$ as $n \to \infty$.
	The hitting time $\tau_n$ vanishes a.s. since $\bigl| \widebar{p}_n - \underline{p}_n \bigr| \to 0$.
	For each $n \in \N$, we have
	\begin{equation*}
		\E\left( v\left( P_{\tau_n}^n \right) \right)
		= \gamma_n v\bigl( \underline{p}_{n-1} \bigr) 
		+ (1-\gamma_n) v\left( \widebar{p}_n \right)
	\end{equation*}
	for some $\gamma_n \in (0,1)$, and the sequences $\bigl( \underline{p}_n \bigr)_{n \in \N}$ and $\left( \widebar{p}_n \right)_{n \in \N}$ may be chosen so that $(\gamma_n)_{n \in \N}$ converges to some $\gamma < 1$.
	Thus, applying the bounded convergence theorem (using the boundedness of $u$ and $v$) to \eqref{eq:v_cont_ineq} yields $\underline{v} \geq \gamma \underline{v} + (1-\gamma) \widebar{v}$, which is equivalent to $\underline{v} \geq \widebar{v}$ since $\gamma<1$.
\end{proof}

\subsection{Proof of \texorpdfstring{\Cref{lemma:v_u}}{Lemma \ref{lemma:v_u}} (p. \pageref{lemma:v_u})}
\label{app:lemma_v_u}

Take $p \in (0,1) \setminus C$, and suppose toward a contradiction that $v(p) > u(p)$.
Since $v$ is continuous and $u$ is upper semi-continuous, we have $v > u$ on an open neighbourhood $N$ of $p$.
We will derive a contradiction assuming that $p \notin D$.
The result for $p \in D$ then follows from the observation that if $v(p) > u(p)$ for $p \in D$, then since $D$ is discrete, the neighbourhood $N$ also contains a $p' \notin D$ at which $v(p') > u(p')$.

We may choose $N$ to not intersect $D$ since the latter is discrete.
By \Cref{theorem:viscosity} (p. \pageref{theorem:viscosity}) and the fact that $v>u$ on $N$, $v$ is a viscosity solution of
\begin{equation}
	w(p) = u(p) + \widebar{\lambda} \frac{ p^2 (1-p)^2 }{ 2r\sigma^2 } w''(p) 
	\label{eq:diff_eqn_convex2}
\end{equation}
on $N$. Observe that $u$ is continuous on $N$.

We show (constructively) in \cref{suppl:hjb_details} that \eqref{eq:diff_eqn_convex2} has a classical (hence viscosity) solution $w^\dag$ on $N$ which satisfies the (Dirichlet) boundary condition $w^\dag=v$ on $\partial N$.
By the comparison principle (e.g. Theorem 3.3. in \textcite{CrandallIshiiLions1992}), $w^\dag$ is the unique viscosity solution of \eqref{eq:diff_eqn_convex2} on $N$ satisfying this boundary condition.
It follows that $v=w^\dag$.

Since $v>u$ on $N$, \eqref{eq:diff_eqn_convex2} requires that $v''>0$ on $N$.
But then $N \subseteq C$, contradicting the supposition that $p$ lies in $(0,1) \setminus C$.
\qed

\subsection{Proof of \texorpdfstring{\Cref{lemma:v_C1}}{Lemma \ref{lemma:v_C1}} (p. \pageref{lemma:v_C1})}
\label{app:lemma_v_C1}

Since a differentiable locally convex function is continuously differentiable (see e.g. Theorem 24.1 in \textcite{Rockafellar1970}), it suffices to show that $v$ is differentiable on $C$.
By local convexity, the left- and right-hand derivatives $v'_-$ and $v'_+$ of $v$ exist on $C$ and satisfy $v'_- \leq v'_+$ (again, see Theorem 24.1 in \textcite{Rockafellar1970}).
We must show that $v'_- = v'_+$.

To that end, take a $p \in C$, and suppose toward a contradiction that $v'_-(p) < v'_+(p)$. (That is, there is a convex kink at $p$.)
Then for any $k > 0$, we may find a twice continuously differentiable $\phi : (0,1) \to \R$ with $\phi''(p) = k$ such that $v - \phi$ is locally minimised at $p$.%
	\footnote{For example, $\phi(q) \coloneqq v(p) + \frac{1}{2} [ v'_-(p) + v'_+(p) ] (q-p) + \frac{1}{2} k (q-p)^2$.}
Since $v$ is a viscosity super-solution of \eqref{eq:hjb} by \Cref{theorem:viscosity}, it follows that
\begin{equation*}
	v(p) 
	\geq u_\star(p) + \widebar{\lambda} \frac{ p^2 (1-p)^2 }{ 2r\sigma^2 } k
\end{equation*}
for any $k>0$.
For large enough $k$, this contradicts the previously-established fact that $v(p) \leq (\cav u)(p)$.
\qed

\subsection{Proof of \texorpdfstring{\Cref{lemma:v_C2}}{Lemma \ref{lemma:v_C2}} (p. \pageref{lemma:v_C2})}
\label{app:lemma_v_C2}

Since $v$ is locally convex on $C \setminus D$, $v''$ is non-negative whenever it exists.
Thus by \Cref{theorem:viscosity} (p. \pageref{theorem:viscosity}), $v$ is a viscosity solution of the differential equation
\begin{equation}
	w(p) = u(p) + \widebar{\lambda} \frac{ p^2 (1-p)^2 }{ 2r\sigma^2 } w''(p) .
	\label{eq:diff_eqn_convex3}
\end{equation}
on $C \setminus D$. Note that $u$ is continuous on $C \setminus D$.

In \cref{suppl:hjb_details}, we show constructively that \eqref{eq:diff_eqn_convex3} has a classical solution on $C \setminus D$ that can be extended to a continuous function $w^\dag : C \to \R$ satisfying the (Dirichlet) boundary condition $w^\dag=u$ on $\partial C$.
By the comparison principle (e.g. Theorem 3.3. in \textcite{CrandallIshiiLions1992}), $w^\dag$ is the unique viscosity solution of \eqref{eq:diff_eqn_convex3} that satisfies this boundary condition.
Since $w^\dag$ is twice continuously differentiable and satisfies \eqref{eq:diff_eqn_convex3} on $C \setminus D$, it suffices to show that $v=w^\dag$ on $C \setminus D$.

Since $v$ is locally convex on $C \setminus D$, it is twice differentiable a.e. on $C \setminus D$ by the Aleksandrov theorem (e.g. Theorem A.2 in \textcite{CrandallIshiiLions1992}), so $v''$ exists on a dense subset $B$ of $C \setminus D$.
Being a derivative, $v''$ is continuous on a dense subset $A$ of $B$.%
	\footnote{This is a consequence of the Baire category theorem---see \textcite[p. 27]{BrucknerLeonard1966}.}
It follows that $v$ satisfies \eqref{eq:diff_eqn_convex3} on $A$.
We have already shown that it satisfies the boundary condition $v=u$ on $\partial C \subseteq [0,1] \setminus C$.

Since the solution $w^\dag$ is unique, $v$ coincides with $w^\dag$ on $A$.
Because $A$ is dense in $C \setminus D$, $v|_A$ admits at most one continuous extension to $C \setminus D$.
Since $w^\dag$ is continuous, it follows that $v=w^\dag$ on $C \setminus D$.
\qed

\subsection{Proof of \texorpdfstring{\Cref{corollary:value_fn_limit}}{Corollary \ref{corollary:value_fn_limit}} (p. \pageref{corollary:value_fn_limit})}
\label{app:value_fn_limit}

To emphasise dependence on $\widebar{\lambda}$, write the value as $v_{\widebar{\lambda}}$.

For the first part, fix an arbitrary $p_0 \in [0,1]$.
Since increasing $\widebar{\lambda}$ slackens the constraint $\lambda \leq \widebar{\lambda}$ in the best-reply problem \eqref{eq:v_defn} in every period, it raises the value $v_{\widebar{\lambda}}(p_0)$.

Now for the second part.
We have established for every $p \in [0,1]$ that the sequence $( v_{\widebar{\lambda}}(p) )_{ \widebar{\lambda} > 0 }$ is increasing.
Since it lives in the compact set $[u(p),(\cav u)(p)]$ by \Cref{proposition:value_fn}, it must converge to some $v_\infty(p) \in [u(p),(\cav u)(p)]$.
In other words, $(v_{\widebar{\lambda}} )_{\widebar{\lambda}>0}$ converges pointwise to some function $v_\infty : [0,1] \to \R$ satisfying $u \leq v_\infty \leq \cav u$.
We claim that $v_\infty = \cav u$.
Since $\cav u$ is by definition the pointwise smallest concave majorant of $u$, it suffices to show that $v_\infty$ is concave.

To that end, take $0 < p' < p < p'' < 1$ in $[0,1]$, and let $\gamma \in (0,1)$ be such that $\gamma p' + (1-\gamma) p'' = p$; we will establish that $\gamma v_\infty(p') + (1-\gamma) v_\infty(p'') \leq v_\infty(p)$.
By changing the units in which time is measured
and adjusting discounting accordingly,
the sender's best-reply problem \eqref{eq:v_defn} may be reformulated as
\begin{equation*}
	v_{\widebar{\lambda}}(p) 
	= \sup_{ ( \lambda_t' )_{ t \in \R_+ } } 
	\E\left( r' \int_0^\infty e^{-r' t} u(p_t) \dd t \right)
	\quad \text{s.t.} \quad
	\dd p_t 
	= \sqrt{ \lambda_t' } \, p_t (1-p_t) \dd \widehat{B}_t ,
\end{equation*}
where $r' \coloneqq r \sigma^2 / \widebar{\lambda}$,
$\widehat{B}_t \coloneqq \sqrt{ r/r' } B_{t r'/r}$ is a standard Brownian motion, and
\begin{equation*}
	( \lambda_t' )_{ t \in \R_+ } 
	\coloneqq \left( \lambda_t \middle/ \widebar{\lambda} \right)_{ t \in \R_+ }
\end{equation*}
is chosen among all $[0,1]$-valued processes adapted to the filtration generated by $(p_t)_{t \in \R_+}$.

Consider the strategy that always sets $\lambda'=1$, and let $( p_t )_{t \in \R_+}$ be the induced belief process.
Write $\tau$ for the first time that $( p_t )_{t \in \R_+}$ hits $\{p',p''\}$.
Following the proposed strategy until time $\tau$ and then behaving optimally cannot be better than optimal, so for every $\widebar{\lambda} > 0$ we have
\begin{equation*}
	\E\left( r' \int_0^\tau e^{-r' t} u(p_t) \dd t 
	+ e^{-r' \tau} v_{\widebar{\lambda}}(p_\tau) \right)
	\leq v_{\widebar{\lambda}}(p)
	\leq v_\infty(p) ,
\end{equation*}
where the second inequality holds since $v_{\widebar{\lambda}}$ increases pointwise as $\widebar{\lambda}$ increases.
As $\widebar{\lambda} \to \infty$, the first term inside the expectation on the left-hand side vanishes a.s., and the second term converges a.s. to $v_\infty(p_\tau)$.
Since both terms are bounded, the left-hand side converges to $\E( v_\infty(p_\tau) )$ by the bounded convergence theorem.
And we have
\begin{equation*}
	\E( v_\infty(p_\tau) ) = \gamma v_\infty(p') + (1-\gamma) v_\infty(p'') 
\end{equation*}
by the optional sampling theorem.%
	\footnote{When $0 < p' < p'' < 1$, we have $\E(\tau) < \infty$, so the optional sampling theorem \parencite[][ch. 1]{KaratzasShreve1991} yields $\E(p_\tau) = p$, whence $\PP(p_\tau=p') = \gamma$ and $\PP(p_\tau=p'') = 1-\gamma$ by definition of $\gamma$.
	For the case in which $0 < p'< p''=1$ (the other cases are analogous), let $\tau_n$ be the first time that $(p_t)_{t \in \R_+}$ hits $\{p',1-1/n\}$, for each $n \in \N$.
	Then $\E(\tau_n) < \infty$, so $\E\left( p_{\tau_n} \right) = p$ by the optional sampling theorem.
	Since $\left( p_{\tau_n} \right)_{n \in \N}$ is bounded and converges a.s. to $p_\tau$ as $n \to \infty$, the bounded convergence theorem yields $\E\left( p_\tau \right) = p$.}

We have proved that $v_{\widebar{\lambda}}$ converges monotonically, pointwise, to $\cav u$. Since $v_{\widebar{\lambda}}$ and $\cav u$ are continuous and defined on a compact domain, it follows by Dini's theorem
\parencite[e.g. Theorem 7.13 in][]{Rudin1976}
that the convergence of $v_{\widebar{\lambda}}$ to $\cav u$ is uniform.
\qed

\subsection{Proof of \texorpdfstring{\Cref{proposition:disagree_nonfull}}{Proposition \ref{proposition:disagree_nonfull}} (p. \pageref{proposition:disagree_nonfull})}
\label{app:disagree_nonfull_proof}

By \Cref{proposition:disc_nonfull} (p. \pageref{proposition:disc_nonfull}), it suffices to find an interior belief at which $u$ is discontinuous.
Since
$\mathcal{A}$ is finite with at least two non-redundant elements,
$f(a,\cdot)$ is continuous for each $a \in \mathcal{A}$,
and $f$ is strictly single-crossing,
the monotone selection theorem of \textcite{MilgromShannon1994}
yields a finite collection of two or more open intervals of $[0,1]$,
on each of which some action is strictly optimal,
with no two intervals associated with the same action;
furthermore, the closure of the union of these intervals is $[0,1]$ itself.
It follows that there is at least one interior belief $p \in (0,1)$ that belongs to the boundary of two such open intervals.
At this belief, we have $f(a,p) = f(a',p)$ for two actions $a \neq a'$,
with $a$ strictly optimal on a left-neighbourhood of $p$
and $a'$ strictly optimal on a right-neighbourhood.
Note that $a' \succ a$, where $\succeq$ is the total order on $\mathcal{A}$ with respect to which $f$ is strictly single-crossing.

Assume that $p_{\textsl{D},0} > p_0$; the argument if $p_{\textsl{D},0} < p_0$ is symmetric.
Write $p_{\textsl{D}} \in (0,1)$ for the decision-maker's belief when the sender has belief $p$,
i.e.
$\phi\left(p_{\textsl{D}},p_0,p_{\textsl{D},0}\right) = p$.
Since $f(a',p) = f(a,p)$, $a' \succ a$
and $p_{\textsl{D}} > p$,
the strict single-crossing property of $f$ yields that
$f\left(a',p_{\textsl{D}}\right)
> f\left(a,p_{\textsl{D}}\right)$.
Since $u=f(a,\cdot)$ on a left-neighbourhood of $p_{\textsl{D}}$
and $u=f(a',\cdot)$ on a right-neighbourhood,
it follows that $u$ is discontinuous at $p_{\textsl{D}}$.
\qed

\subsection{Proof of \texorpdfstring{\Cref{proposition:disagree_value}}{Proposition \ref{proposition:disagree_value}} (p. \pageref{proposition:disagree_value})}
\label{app:disagree_value_proof}

We shall use the following comparative statics lemma.

\begin{lemma}
	\label{lemma:abst_comp_stat}
	Let $(\mathcal{T},\mathord{\gtrsim})$ and $(\mathcal{X},\succeq)$ be totally ordered sets,
	let $f : \mathcal{T} \times \mathcal{X} \to \R$ be strictly single-crossing,
	and let $X: \mathcal{T} \to \Delta(\mathcal{X})$
	satisfy $\supp X(\cdot|t) \subseteq \argmax_{\mathcal{X}} f(\cdot,t)$ for every $t \in \mathcal{T}$.
	Then for any $t_1 \gtrsim t_2 \gtrsim t_3$ in $\mathcal{T}$, we have
	\begin{align*}
		\int_{\mathcal{X}} f(x,t_1) X( \dd x | t_2 )
		&\geq \int_{\mathcal{X}} f(x,t_1) X( \dd x | t_3 )
		\\
		\text{and} \quad
		\int_{\mathcal{X}} f(x,t_3) X( \dd x | t_2 )
		&\geq \int_{\mathcal{X}} f(x,t_3) X( \dd x | t_1 ).
	\end{align*}
\end{lemma}

\begin{proof}%
	\renewcommand{\qedsymbol}{$\square$}
	Fix arbitrary $x_2 \in \supp X(\cdot|t_2)$ and $x_3 \in \supp X(\cdot|t_3)$.
	Since $f$ is strictly single-crossing,
	we have $x_2 \succeq x_3$
	by the monotone selection theorem of \textcite{MilgromShannon1994}.
	As $f(x_2,t_2) \geq f(x_3,t_2)$ and $t_1 \gtrsim t_2$,
	single-crossing then yields $f(x_2,t_1) \geq f(x_3,t_1)$.
	Since $x_2 \in \supp X(\cdot|t_2)$ and $x_3 \in \supp X(\cdot|t_3)$ were arbitrary, it follows that
	$\int_{\mathcal{X}} f(x,t_1) X( \dd x | t_2 )
	\geq \int_{\mathcal{X}} f(x,t_1) X( \dd x | t_3 )$.
	The argument for the second inequality is analogous.
\end{proof}%
\renewcommand{\qedsymbol}{$\blacksquare$}

To prove the proposition, fix $p_0$.
Further fix $p_{\textsl{D},0} < p_{\textsl{D},0}' < p_0$; the other case is analogous.
Write $A : [0,1] \to \Delta(\mathcal{A})$ for the decision-maker's myopic and regular Markov strategy,
let $u,u'$ be the sender's induced flow payoffs under the two decision-maker priors $p_{\textsl{D},0},p_{\textsl{D},0}'$,
and write $v,v'$ for the corresponding value functions of the sender.

Fix an arbitrary belief $p \in [0,1]$ of the sender.
The decision-maker's beliefs $p_{\textsl{D}} \coloneqq \phi(p,p_0,p_{\textsl{D},0})$ and $p_{\textsl{D}}' \coloneqq \phi(p,p_0,p_{\textsl{D},0}')$
satisfy $p_{\textsl{D}} \leq p_{\textsl{D}}' \leq p$.
We furthermore have $\supp A(\cdot|p') \subseteq \argmax_{\mathcal{A}} f(\cdot,p')$ at every $p' \in [0,1]$ because $A$ is myopic.
Hence since $f$ is strictly single-crossing, \Cref{lemma:abst_comp_stat} yields
\begin{equation*}
	u'(p)
	= \int_{\mathcal{A}} f(a,p) A( \dd a | p_{\textsl{D}}' )
	\geq \int_{\mathcal{A}} f(a,p) A( \dd a | p_{\textsl{D}} )
	= u(p) .
\end{equation*}
Thus $u' \geq u$ since $p \in [0,1]$ was arbitrary,
whence $v'(p_0) \geq v(p_0)$.
\qed

\crefalias{section}{supplsec}
\crefalias{subsection}{supplsec}
\crefalias{subsubsection}{supplsec}

\section*{Supplemental appendices}
\label{suppl}
\addcontentsline{toc}{section}{Supplemental appendices}

\subsection{Solving for the value function}
\label{suppl:hjb_details}

In this appendix, we explain how to solve for the sender's value function using \Cref{proposition:value_fn}.
We detail in particular how the value is computed in the two-action example, allowing us to draw \Cref{fig:2act}.

Partition $C \setminus D$ into maximal intervals $(R_k)_{k=1}^K$.
(In the two-action example, the maximal intervals are $[0,2/3)$ and $(2/3,1]$.)
Fix a continuity interval $R_k$.
The homogeneous part (without the $u$) of the differential equation \eqref{eq:diff_eqn_convex} has general solution $A H_1 - B H_2$ for constants $A,B \in \R$, where
\begin{equation*}
	H_1(p) \coloneqq p^\xi (1-p)^{1-\xi} ,
	\xi \coloneqq 1/2 + \sqrt{ 1/4 + 2r\sigma^2/\widebar{\lambda} } 
	\quad \text{and} \quad
	H_2(p) \coloneqq H_1(1-p) .
\end{equation*}

A particular solution may be obtained from formula (6.2) in \textcite[ch. 3]{Coddington1961}.
Things are easier when the sender has expected-utility preferences, so that $f(a,\cdot)$ is affine, as $u$ itself is then a particular solution.
This is the case in the two-action example,
and in the three-action example below.
In the expected-utility case, the value function is given on each maximal interval $R_k$ of $C \setminus D$ as
\begin{equation*}
	v(p) = u(p) + A_{R_k} H_1(p) - B_{R_k} H_2(p) 
	\quad \text{for all $p \in R_k$,}
\end{equation*}
where the constants $\left( A_{R_k}, B_{R_k} \right)_{k=1}^K$ are the unique ones that ensure that the properties in \Cref{proposition:value_fn} are satisfied: the boundary condition $v=u$ on $\{0,1\}$, the continuity of $v$ on $D$, and smooth pasting on $C \union D$.

\subsubsection{The two-action example (§\ref{sec:value:example}, p. \pageref{sec:value:example})}
\label{suppl:hjb_details:2act}

Here $D = \{2/3\}$, and $C$ contains $[0,2/3)$ and may or may not contain $[2/3,1]$.
In either case,
\begin{equation*}
	v(p)
	= 
	\begin{cases}
		A_{[0,2/3)} H_1(p) - B_{[0,2/3)} H_2(p)
		& \text{for $p \in [0,1/2)$}
		\\
		\alpha p - \beta
		+ A_{(2/3,1]} H_1(p) - B_{(2/3,1]} H_2(p)
		& \text{for $p \in (2/3,1]$,}
	\end{cases}
\end{equation*}
where $\alpha = 3/2$ and $\beta = 1/2$.%
	\footnote{If $C$ contains $[2/3,1]$ then the expression for $p \in (2/3,1]$ holds since \eqref{eq:hjb} must be satisfied in the classical sense by \Cref{proposition:value_fn}. If not, then \Cref{proposition:value_fn} requires that $v=u$, which amounts to setting $A_{(2/3,1]} = B_{(2/3,1]} = 0$.}
The boundary conditions require that $B_{[0,2/3)} = 0$ and $A_{(2/3,1]} = 0$.
Continuity of $v$ at $2/3$ requires that
\begin{equation*}
	A_{[0,2/3)} H_1(2/3)
	= \alpha (2/3) - \beta - B_{(2/3,1]} H_2(2/3) .
\end{equation*}

If $\widebar{\lambda}$ is sufficiently high, then $2/3 \in C$, in which case smooth pasting must hold at $2/3$:
\begin{equation*}
	A_{[0,2/3)} H_1'(2/3)
	= \alpha - B_{(2/3,1]} H_2'(2/3) .
\end{equation*}
Thus the constants are uniquely pinned down.

If $\widebar{\lambda}$ is low, then $2/3 \notin C$, in which case $v = u$ on $[2/3,1]$.
Thus $B_{(2/3,1]} = 0$, whence $A_{[0,2/3)}$ is pinned down by the continuity condition.

To determine which case applies for a given value of $\widebar{\lambda}$, calculate $A_{[0,2/3)}$ assuming that the first case applies.
If
\begin{equation*}
	A_{[0,2/3)} H_1(2/3) \geq u(1/2) = 1/2 ,
\end{equation*}
then the first case does indeed apply; if not, then not.

\subsubsection{A three-action example}
\label{suppl:hjb_details:3act}

Consider the flow payoff $u$ depicted in \Cref{fig:3act}.
The underlying model has actions $\mathcal{A} = \{0,1,3\}$, flow payoff $f_{\textsl{S}}(a,p) = a$ for the sender, and payoffs $f_{\textsl{D}}(0,p) = 0$, $f_{\textsl{D}}(1,p) = 2 p - 1$ and $f_{\textsl{D}}(3,p) = \frac{14}{3} p - 3$ for the decision-maker.
\Cref{fig:3act} depicts the concave envelope,
as well as the sender's value function for high and low values of $\widebar{\lambda}$.%
\begin{figure}
	\begin{subfigure}{0.49\textwidth}
		\centering
		\input{tikz/3act_v_lo}
		\caption{Value $v$ for high $\widebar{\lambda}$.}
		\label{fig:3act_v_lo}
	\end{subfigure}
	\begin{subfigure}{0.49\textwidth}
		\centering
		\input{tikz/3act_v_hi}
		\caption{Value $v$ for low $\widebar{\lambda}$.}
		\label{fig:3act_v_hi}
	\end{subfigure}
	\caption{Three-action example.}
	\label{fig:3act}
\end{figure}%

Clearly $C$ contains $[0,1/2)$ and $(1/2,3/4)$, and does not contain $[3/4,1]$.
Thus the value function off $D$ is
\begin{equation*}
	v(p)
	= 
	\begin{cases}
		A_{[0,1/2)} H_1(p) - B_{[0,1/2)} H_2(p)
		& \text{for $p \in [0,1/2)$}
		\\
		\ell 
		+ A_{(1/2,3/4)} H_1(p) - B_{(1/2,3/4)} H_2(p)
		& \text{for $p \in (1/2,3/4)$}
		\\
		h ,
	\end{cases}
\end{equation*}
where $\ell = 1$ and $h = 3$.
The boundary condition at $p=0$ again requires that $B_{[0,1/2)} = 0$.
Continuity of $v$ at $1/2$ and at $3/4$ requires that
\begin{align*}
	&A_{[0,1/2)} H_1(1/2)
	= \ell 
	+ A_{(1/2,3/4)} H_1(1/2) - B_{(1/2,3/4)} H_2(1/2)	
	\\
	\text{and}\quad
	&\ell 
	+ A_{(1/2,3/4)} H_1(3/4) - B_{(1/2,3/4)} H_2(3/4)
	= h .
\end{align*}
These are two equations in three unknowns.

If $\widebar{\lambda}$ is sufficiently high that $1/2 \in C$, then smooth pasting must hold at $1/2$, giving us the third equation
\begin{equation*}
	A_{[0,1/2)} H_1'(1/2)
	= A_{(1/2,3/4)} H_1'(1/2) - B_{(1/2,3/4)} H_2'(1/2) .
\end{equation*}
If $\widebar{\lambda}$ is sufficiently low that $1/2 \notin C$, then $v(1/2) = u(1/2) = \ell$.
We thus obtain a third equation from the requirement that $v$ be continuous at $1/2$:
\begin{equation*}
	A_{[0,1/2)} H_1(1/2)
	= \ell .
\end{equation*}

To discern which case applies, compute $A_{[0,1/2)}$ assuming that the first (information arrives fast) case applies. If
$A_{[0,1/2)} H_1(1/2) \geq \ell$,
then the fast-arrival case does indeed apply; if not, then not.

\subsection{Generic uniqueness of long-run beliefs}
\label{suppl:generic_lr}

We claimed in §\ref{sec:beliefs:BR} that provided $v(p_0) > u(p_0)$, generically, all best replies of the sender induce the same long-run beliefs (viz. the beliefs $\{p^-,p^+\}$ defined in \Cref{corollary:beliefs} (p. \pageref{corollary:beliefs})).

\begin{figure}
	\begin{subfigure}{0.49\textwidth}
		\centering
		\input{tikz/3act_jeff}
		\caption{Three-action example from \Cref{fig:3act}.}
		\label{fig:3act_jeff}
	\end{subfigure}
	\begin{subfigure}{0.49\textwidth}
		\centering
		\input{tikz/2act_jeff}
		\caption{Two-action example from §\ref{sec:value:example}.}
		\label{fig:2act_jeff}
	\end{subfigure}
	\caption{Knife-edge cases in which long-run beliefs are not unique.}
	\label{fig:jeff}
\end{figure}%
To see how uniqueness can fail, consider the three-action example in \cref{suppl:hjb_details:3act} (p. \pageref{fig:3act}).
\Cref{fig:3act_jeff} depicts the knife-edge case in which $\widebar{\lambda}$ is such that the fast-information value function with the convex-flat shape in \Cref{fig:3act_v_lo} touches $u$ at $1/2$.%
	\footnote{We thank Jeff Ely for pointing out this scenario.}
In this case, the sender is indifferent between providing and not providing information at $1/2$, and strictly prefers to do so on $(0,1/2)$ and $(0,3/4)$.
The best reply $\Lambda^\star$ from \Cref{corollary:BR} (p. \pageref{corollary:BR}) stops at $1/2$, inducing the long-run beliefs $\{p^-,p^+\} = \{1/2,3/4\}$ from \Cref{corollary:beliefs}.
But since the sender is indifferent at $1/2$, she also has a best reply that provides information at $1/2$, which induces long-run beliefs $\{0,3/4\}$.

This scenario is non-generic in the sense that slightly increasing $\widebar{\lambda}$ puts us back in \Cref{fig:3act_v_lo} (p. \pageref{fig:3act_v_lo}), where the sender strictly prefers to provide information at full tilt at $1/2$, whereas slightly decreasing $\widebar{\lambda}$ puts us in \Cref{fig:3act_v_hi}, where she strictly prefers to stop at $1/2$.

Similarly, \Cref{fig:2act_jeff} depicts the case in §\ref{sec:value:example} in which $\widebar{\lambda}$ has exactly the value needed for the fast-information value function with the convex shape in \Cref{fig:2act_v_lo} to just touch $u$ at $2/3$.
In this example, there is more multiplicity: the sender is indifferent on $[1/2,1]$, so has best replies that induce any mean-$p_0$ distribution of long-run beliefs supported on $\{0\} \union [2/3,1]$. (The best reply $\Lambda^\star$ induces the beliefs $\{0,2/3\}$.)
Again, perturbing $\widebar{\lambda}$ makes the sender's preference strict at $2/3$, so that long-run induced beliefs are unique (either $\{0,2/3\}$ or $\{0,1\}$).

The non-genericity of multiplicity in these examples is a general phenomenon.
Multiplicity occurs for some prior $p_0$ with $v(p_0) > u(p_0)$ precisely if the sender is indifferent between stopping and continuing at some $p \in (0,1)$ and weakly prefers to continue on a neighbourhood of $p$.
In such cases, her preference becomes strict when $\widebar{\lambda}$ is perturbed slightly.

\subsection{Piecewise continuity is merely tie-breaking}
\label{suppl:tie-breaking}

We asserted in §\ref{sec:model:soln} that provided the decision-maker's flow payoff $f_{\textsl{D}}$ is non-degenerate in a mild sense, it is without loss of optimality for her to restrict attention to piecewise continuous Markov strategies $A : [0,1] \to \Delta(\mathcal{A})$.

To justify this claim, begin by recalling from §\ref{sec:myopic} that the decision-maker best-replies to a Markov strategy of the sender by myopically maximising $f_{\textsl{D}}(a,p)$ at each $p$.
Fix two actions $a,a' \in \mathcal{A}$, and write
\begin{equation*}
	\psi(p) \coloneqq f_{\textsl{D}}(a,p) - f_{\textsl{D}}(a',p)
\end{equation*}
for their payoff difference.
Say that $\psi$ \emph{strictly up-crosses} at $p \in (0,1)$ iff $\psi(p)=0$ and for any $\eps>0$, there are $p' \in (p-\eps,p)$ and $p'' \in (p,p+\eps)$ such that $\psi(p') < 0 < \psi(p'')$, \emph{strictly down-crosses} if the reverse inequalities hold, and simply \emph{strictly crosses} if either is the case.
Write $K \subseteq (0,1)$ for the set on which $\psi$ strictly crosses.
We claim that given some weak non-degeneracy condition on $f_{\textsl{D}}$, the crossing set $K$ is discrete, so that the decision-maker strictly prefers to switch actions only on a discrete set.
(It suffices to consider only two arbitrary actions $a,a' \in \mathcal{A}$ because $\mathcal{A}$ is finite.)

To see what can go wrong, suppose that $f_{\textsl{D}}(a,p) = 0$ and that $p \mapsto f_{\textsl{D}}(a',p)$ is a typical path of a standard Brownian motion.
Then $\psi$ is continuous, but the strict crossing set $K$ is non-empty with no isolated points (see e.g. Theorem 9.6 in \textcite[ch. 2]{KaratzasShreve1991}).
This preference dithers maniacally, wishing to switch actions back and forth extremely frequently.

As a first pass, observe that if $\psi$ is monotone, or more generally if $\psi$ or $-\psi$ has the single-crossing property ($\psi(p) \geq \mathrel{(>)} 0$ implies $\psi(p') \geq \mathrel{(>)} 0$ for $p < p'$), then $K$ is empty or a singleton, so certainly discrete. These assumptions are satisfied by expected-utility preferences.

A weak non-degeneracy condition that suffices is \emph{local single-crossing:} for each $p \in K$, we have either $\psi \geq 0$ or $\psi \leq 0$ on a left-neighbourhood of $p$, and similarly on a right-neighbourhood.
Then each $p \in K$ is manifestly the unique strict crossing of $\psi$ on a neighbourhood, hence isolated.
A sufficient condition for this is \emph{local monotonicity:} for each $p \in K$, we have $\psi(p-\eps) \leq 0 \leq \psi(p+\eps)$ for all sufficiently small $\eps>0$, or the reverse inequality.

\subsection{A very brief introduction to viscosity solutions}
\label{suppl:viscosity_intro}

\textcite{Crandall1997}, \textcite{Katzourakis2015} and \textcite{CrandallIshiiLions1992} provide overviews of the theory of viscosity solutions of second-order differential equations. \textcite{Moll2017}, \textcite[ch. 10]{Evans2010}, \textcite{Calder2018} and \textcite{Bressan2011} give easier treatments that deal mostly with first-order equations.

The general idea of viscosity solutions is as follows. If $w$ is a viscosity solution of \eqref{eq:hjb}, then it must satisfy \eqref{eq:hjb} in the classical sense on any neighbourhood on which $w''$ exists and is continuous.
If $w''$ does not exist at $p \in [0,1]$, we require instead that \eqref{eq:hjb} hold with the appropriate inequality when $w''(p)$ is replaced by $\phi''(p)$ for some twice continuously differentiable local approximation $\phi$ to $w$ at $p$. (The formal definition was given on p. \pageref{definition:viscosity}.)

\subsubsection{Illustration of the definition}
\label{suppl:viscosity_intro:ill}

Consider the three-action example from \cref{suppl:hjb_details:3act} (\Cref{fig:3act}, p. \pageref{fig:3act}).
Write $\mathcal{C}^2$ for the set of twice continuously differentiable functions $(0,1) \to \R$. Begin by observing that $v$ is continuous, hence upper and lower semi-continuous.

Consider a $p$ in whose vicinity $v$ is twice continuously differentiable, e.g. $p=2/5$. We may easily find $\phi_1,\phi_2 \in \mathcal{C}^2$ such that $\phi_1-v$ and $v-\phi_2$ are locally minimised at $p$, as in \Cref{fig:3act2_25_subsupersoln}.
\begin{figure}
	\begin{subfigure}{0.49\textwidth}
		\centering
		\input{tikz/3act2_25_subsupersoln}
		\caption{$\phi_1,\phi_2 \in \mathcal{C}^2$ such that $\phi_1-v$ and $v-\phi_2$ have local minima at $2/5$.}
		\label{fig:3act2_25_subsupersoln}
	\end{subfigure}
	\begin{subfigure}{0.49\textwidth}
		\centering
		\input{tikz/3act2_12_subsoln1}
		\caption{$\phi \in \mathcal{C}^2$ such that $\phi-v$ has a local minimum at $1/2$.}
		\label{fig:3act2_12_subsoln1}
	\end{subfigure}
	\\
	\begin{subfigure}{0.49\textwidth}
		\centering
		\input{tikz/3act2_12_subsoln2}
		\caption{$\phi \in \mathcal{C}^2$ such that $\phi-v$ has a local minimum at $1/2$ and $\phi''(1/2)=0$.}
		\label{fig:3act2_12_subsoln2}
	\end{subfigure}
	\begin{subfigure}{0.49\textwidth}
		\centering
		\input{tikz/3act2_12_supersoln}
		\caption{$\phi \in \mathcal{C}^2$ for which $v-\phi$ does not have a local minimum at $1/2$.}
		\label{fig:3act2_12_supersoln}
	\end{subfigure}
	\caption{Functions $\phi \in \mathcal{C}^2$ that approximate $v$ locally.}
	\label{fig:3act2}
\end{figure}%
But in particular, we may choose $\phi \in \mathcal{C}^2$ to coincide with $v$ on a neighbourhood of $p$. Then $\phi-v$ and $v-\phi$ are \emph{both} locally minimised at $p$, and $\phi''(p) = v''(p)$. Since $v$ is a viscosity sub-solution (super-solution) by \Cref{theorem:viscosity} (p. \pageref{theorem:viscosity}), and $u(p) = u^\star(p) = u_\star(p)$, it follows that
\begin{equation*}
	v(p) \leq \mathrel{(\geq)}
	u(p) + \widebar{\lambda} \frac{ p^2 (1-p)^2 }{ 2r\sigma^2 } 
	\max\left\{ 0, v''(p) \right\} .
\end{equation*}
So \eqref{eq:hjb} must be satisfied in the classical sense at $p$.

Next consider a point at which $v''$ is undefined, e.g. $p=1/2$. There are many $\phi \in \mathcal{C}^2$ such that $\phi-v$ has a local minimum at $p$; an example is depicted in \Cref{fig:3act2_12_subsoln1}. Since $v$ is a viscosity sub-solution of \eqref{eq:hjb} and $u^\star(p) = u(p)$, we must have
\begin{equation*}
	v(p) \leq 
	u(p) + \widebar{\lambda} \frac{ p^2 (1-p)^2 }{ 2r\sigma^2 } 
	\max\left\{ 0, \phi''(p) \right\} 
\end{equation*}
for any such $\phi$. In fact, $\phi$ can be chosen so that $\phi''(p) \leq 0$: the $\phi$ depicted in \Cref{fig:3act2_12_subsoln2} is affine, so has $\phi''(p)=0$. The sub-solution condition therefore requires precisely that
\begin{equation*}
	v(p) \leq \inf_{ \substack{ \phi \in \mathcal{C}^2 : \\ \text{$\phi-v$ loc. min. at $p$} } }
	\left\{ 
	u(p) + \widebar{\lambda} \frac{ p^2 (1-p)^2 }{ 2r\sigma^2 } 
	\max\left\{ 0, \phi''(p) \right\}
	\right\}
	= u(p) ,
\end{equation*}
which holds (with equality, in fact).

By contrast, there are no $\phi \in \mathcal{C}^2$ such that $v-\phi$ has a local minimum at $p$; a (failed) attempt to find such a $\phi$ is drawn in \Cref{fig:3act2_12_supersoln}. The fact that $v$ is a viscosity super-solution of \eqref{eq:hjb} therefore has no bite at $p=1/2$.

\subsubsection{Some properties of viscosity solutions}
\label{suppl:viscosity_intro:prop}

There are other non-classical notions of `solution' of a differential equation, most importantly distributional solutions (e.g. \textcite[chs. 5--9]{Evans2010}). But for many differential equations, including HJB equations, viscosity solutions are the appropriate notion. The chief reasons are twofold: viscosity solutions exist, and they satisfy a comparison principle.

Begin with existence. Many HJB equations, including ours, fail to have a classical solution. Many also fail to have non-classical solutions of e.g. the distributional variety. By contrast, HJB equations always have a viscosity solution.

The other virtue of viscosity solutions is that they satisfy a comparison principle (also called a `maximum principle') of the following kind: if $\underline{w}$ is a sub-solution on $(a,b)$, $\widebar{w}$ is a super-solution on $(a,b)$, and $\underline{w} \leq \widebar{w}$ on $\{a,b\}$, then $\underline{w} \leq \widebar{w}$ on $(a,b)$. (See \textcite[Theorem 3.3]{CrandallIshiiLions1992}.)
Classical sub- and super-solutions also satisfy a comparison principle, but other non-classical notions of `solution' do not.

The comparison principle may be used to obtain uniqueness results; a standard one is that the HJB equation has at most one viscosity solution with the right boundary conditions satisfying a linear-growth condition. It follows that the value function is the unique solution with the right boundary conditions and linear growth. (See \textcite[ch. V]{FlemingSoner2006}.)
We use the comparison principle in this manner in the proofs of \Cref{lemma:v_u,lemma:v_C2} (\cref{app:lemma_v_u,app:lemma_v_C2}).

The comparison principle may also be used to establish the continuity of solutions, and thus of the value function.
In particular, suppose that we have shown that the upper (lower) semi-continuous envelope $v^\star$ ($v_\star$) of the value $v$ is a sub-solution (super-solution) of the HJB equation, and that $v_\star = v^\star$ on $\{0,1\}$. (We do precisely this in the proof of \Cref{theorem:viscosity} in \cref{app:viscosity}.)
A comparison principle then yields $v^\star \leq v_\star$, which since $v_\star \leq v \leq v^\star$ implies that $v$ is itself a viscosity solution, hence continuous.

In our proof of \Cref{theorem:viscosity} (\cref{app:viscosity}), we eschew this approach in favour of a direct proof that $v$ is continuous.
We do this because we are not aware of a comparison principle that applies assuming only piecewise continuity of $u$.
The closest result that we know of is Theorem 3.3 in \textcite{Soravia2006}, which would be applicable under the additional hypotheses that $u$ has only \emph{finitely} many discontinuities and satisfies $u(p) \in [ u(p-) \meet u(p+), u(p-) \join u(p+) ]$ at every $p \in (0,1)$.

\end{appendices}



\printbibliography[heading=bibintoc]


\end{document}

%% file: preamble.tex
\DeclareTextFontCommand{\emph}{\slshape}

\usepackage[T1]{fontenc}
\usepackage{lmodern}
\usepackage[utf8]{inputenc}

\usepackage[american,german,french,british]{babel}

\usepackage[activate={true,nocompatibility}]{microtype}

\usepackage{csquotes}

\usepackage[backend=biber,autolang=other,style=apa,maxcitenames=5,sortcites=false]{biblatex}
\DeclareLanguageMapping{british}{british-apa}
\DeclareLanguageMapping{american}{american-apa}
\DeclareLanguageMapping{german}{german-apa}
\DeclareLanguageMapping{french}{french-apa}
\DefineBibliographyExtras{french}{\restorecommand\mkbibnamefamily}

\usepackage{amsmath}
\usepackage{amssymb}
\usepackage{amsfonts}
\usepackage{amsthm}
\usepackage[retainorgcmds]{IEEEtrantools}
\usepackage{mathtools}
\usepackage{stmaryrd}
\usepackage{mleftright}

\usepackage{graphicx}
\usepackage{tikz,pgfplots}
\usetikzlibrary{arrows}
\pgfplotsset{compat=1.10}
\usepackage{etoolbox}
\AtBeginEnvironment{tikzpicture}{\shorthandoff{;}}

\let\originalleft\left
\let\originalright\right
\renewcommand{\left}{\mathopen{}\mathclose\bgroup\originalleft}
\renewcommand{\right}{\aftergroup\egroup\originalright}

\usepackage[title]{appendix}

\usepackage{datetime}
\newdateformat{datestyle}{ {\THEDAY} \monthname[\THEMONTH] \THEYEAR }

\usepackage{enumerate}
\usepackage{enumitem}
\setlist[enumerate,1]{label=(\arabic*)}
\setlist[itemize,1]{label=--}
\setlist[itemize,2]{label=--}
\setlist[itemize,3]{label=--}
\setlist[itemize,4]{label=--}

\usepackage{xcolor}
\usepackage{verbatim}
\usepackage{gensymb}
\usepackage{rotating}
\usepackage{subcaption}
\usepackage{floatpag}
\floatpagestyle{plain}
\usepackage{marvosym}

\usepackage{hyphenat}
\theoremstyle{definition}

\newtheorem{theorem}{Theorem}
\newtheorem{proposition}{Proposition}
\newtheorem{lemma}{Lemma}
\newtheorem{corollary}{Corollary}
\newtheorem{remark}{Remark}
\newtheorem{observation}{Observation}
\newtheorem{definition}{Definition}

\newtheoremstyle{named}
	{\topsep}					
	{\topsep}					
	{}							
	{0pt}						
	{\bfseries}					
	{}							
	{5pt plus 1pt minus 1pt}	
	{\thmnote{#3}}				
\theoremstyle{named}
\newtheorem{namedthm}{}

\renewcommand{\qedsymbol}{$\blacksquare$}

\usepackage{xpatch}
\xpatchcmd{\proof}{\itshape}{\proofheadfont}{}{}
\newcommand{\proofheadfont}{\slshape}

\usepackage{hyperref}
\hypersetup{
	colorlinks=true,
	linkcolor=black,
	citecolor=black,
	filecolor=black,
	urlcolor=black,
}

\usepackage[nameinlink]{cleveref}
\crefname{page}{p.}{pp.}
\crefname{equation}{equation}{equations}
\crefname{section}{section}{sections}
\crefname{subsection}{section}{sections}
\crefname{subsubsection}{section}{sections}
\crefname{appsec}{appendix}{appendices}
\crefname{supplsec}{supplemental appendix}{supplemental appendices}
\crefname{footnote}{footnote}{footnotes}
\crefname{figure}{figure}{figures}
\crefname{table}{table}{tables}
\crefname{theorem}{theorem}{theorems}
\crefname{proposition}{proposition}{propositions}
\crefname{lemma}{lemma}{lemmata}
\crefname{corollary}{corollary}{corollaries}
\crefname{remark}{remark}{remarks}
\crefname{observation}{observation}{observations}
\crefname{example}{example}{examples}
\crefname{fact}{fact}{facts}
\crefname{definition}{definition}{definitions}
\crefname{assumption}{assumption}{assumptions}
\crefname{exercise}{exercise}{exercises}
\crefname{notation}{notation}{notation}
\crefname{claim}{claim}{claims}
\crefname{conjecture}{conjecture}{conjectures}

\newcommand{\eps}{\varepsilon}

\newcommand{\dd}{\mathrm{d}}

\DeclareMathOperator*{\argmax}{arg\,max}

\DeclareMathOperator*{\supp}{supp}

\DeclareMathOperator*{\cav}{cav}

\newcommand{\E}{\mathbf{E}}

\newcommand{\PP}{\mathbf{P}}

\newcommand{\R}{\mathbf{R}}

\newcommand{\N}{\mathbf{N}}

\newcommand{\1}{\boldsymbol{1}}

\newcommand{\join}{\vee}

\newcommand{\meet}{\wedge}

\newcommand{\union}{\cup}

\newcommand{\intersect}{\cap}

\DeclarePairedDelimiter\abs{\lvert}{\rvert}

\makeatletter
\let\save@mathaccent\mathaccent
\newcommand*\if@single[3]{%
	\setbox0\hbox{${\mathaccent"0362{#1}}^H$}%
	\setbox2\hbox{${\mathaccent"0362{\kern0pt#1}}^H$}%
	\ifdim\ht0=\ht2 #3\else #2\fi
	}
\newcommand*\rel@kern[1]{\kern#1\dimexpr\macc@kerna}
\newcommand*\widebar[1]{\@ifnextchar^{{\wide@bar{#1}{0}}}{\wide@bar{#1}{1}}}
\newcommand*\wide@bar[2]{\if@single{#1}{\wide@bar@{#1}{#2}{1}}{\wide@bar@{#1}{#2}{2}}}
\newcommand*\wide@bar@[3]{%
	\begingroup
	\def\mathaccent##1##2{%
	  \let\mathaccent\save@mathaccent
	  \if#32 \let\macc@nucleus\first@char \fi
	  \setbox\z@\hbox{$\macc@style{\macc@nucleus}_{}$}%
	  \setbox\tw@\hbox{$\macc@style{\macc@nucleus}{}_{}$}%
	  \dimen@\wd\tw@
	  \advance\dimen@-\wd\z@
	  \divide\dimen@ 3
	  \@tempdima\wd\tw@
	  \advance\@tempdima-\scriptspace
	  \divide\@tempdima 10
	  \advance\dimen@-\@tempdima
	  \ifdim\dimen@>\z@ \dimen@0pt\fi
	  \rel@kern{0.6}\kern-\dimen@
	  \if#31
	    \overline{\rel@kern{-0.6}\kern\dimen@\macc@nucleus\rel@kern{0.4}\kern\dimen@}%
	    \advance\dimen@0.4\dimexpr\macc@kerna
	    \let\final@kern#2%
	    \ifdim\dimen@<\z@ \let\final@kern1\fi
	    \if\final@kern1 \kern-\dimen@\fi
	  \else
	    \overline{\rel@kern{-0.6}\kern\dimen@#1}%
	  \fi
	}%
	\macc@depth\@ne
	\let\math@bgroup\@empty \let\math@egroup\macc@set@skewchar
	\mathsurround\z@ \frozen@everymath{\mathgroup\macc@group\relax}%
	\macc@set@skewchar\relax
	\let\mathaccentV\macc@nested@a
	\if#31
	  \macc@nested@a\relax111{#1}%
	\else
	  \def\gobble@till@marker##1\endmarker{}%
	  \futurelet\first@char\gobble@till@marker#1\endmarker
	  \ifcat\noexpand\first@char A\else
	    \def\first@char{}%
	  \fi
	  \macc@nested@a\relax111{\first@char}%
	\fi
	\endgroup
}
\makeatother
\newcommand\test[1]{%
$#1{M}$ $#1{A}$ $#1{g}$ $#1{\beta}$ $#1{\mathcal A}^q$
$#1{AB}^\sigma$ $#1{H}^C$ $#1{\sin z}$ $#1{W}_n$}


%% file: tikz/2act_v_lo.tex


\begin{tikzpicture}[scale=5.2, line cap=round,
	declare function={
		u(\p,\slope,\intercept)
			= \slope * \p - \intercept;
		D(\prior,\priora)
			= ( \prior / (1-\prior) ) / ( \priora / (1-\priora) );
		phiinv(\p,\prior,\priora)
			= \p / ( \p + (1-\p) / D(\prior,\priora) );
		H1(\p,\exponent) 
			= \p^\exponent * (1-\p)^(1-\exponent);
		H2(\p,\exponent)
			= H1( 1-\p, \exponent );
		H1deriv(\p,\exponent)
			= H1(\p,\exponent) * ( (\exponent/\p) + ((\exponent-1)/(1-\p)) );
		H2deriv(\p,\exponent)
			= - H1deriv(1-\p,\exponent);
		valuefn_lo(\p,\exponent,\A,\B)
			= \A * H1(\p,\exponent) - \B * H2(\p,\exponent);
		valuefn_hi(\p,\exponent,\slope,\intercept,\A,\B)
			= u(\p,\slope,\intercept)
			+ \A * H1(\p,\exponent) - \B * H2(\p,\exponent);
		}]

	\pgfmathsetmacro{\yscale}{1}; 
	\pgfmathsetmacro{\ticklength}{\yscale/60}; 
	\pgfmathsetmacro{\samples}{200}; 
	\pgfmathsetmacro{\radius}{\pgflinewidth/2}; 

	\pgfmathsetmacro{\prior}{1/2}; 
	\pgfmathsetmacro{\priora}{1/5}; 
	\pgfmathsetmacro{\intercept}{1/2}; 
	\pgfmathsetmacro{\slope}{3/2}; 
	\pgfmathsetmacro{\vol}{1/4}; 
	\pgfmathsetmacro{\r}{1}; 

	\pgfmathsetmacro{\pstar}{ \intercept / \slope };
	\pgfmathsetmacro{\pstara}{ phiinv(\pstar,\prior,\priora) };
	\pgfmathsetmacro{\exponent}
		{ (1/2) + sqrt( (1/4) + 2*\vol*\vol*\r ) }; 
	\pgfmathsetmacro{\Ahi}{ 0 };
	\pgfmathsetmacro{\Blo}{ 0 };
	\pgfmathsetmacro{\Aloimp}
		{ ( u(\pstara,\slope,\intercept) ) / H1(\pstara,\exponent) ) };
	\pgfmathsetmacro{\Bhipat}
		{ ( u(\pstara,\slope,\intercept) * H1deriv(\pstara,\exponent)
		- \slope * H1(\pstara,\exponent) )
		/ ( 2 * \exponent - 1 ) };
	\pgfmathsetmacro{\Alopat}
		{ ( \slope - H2deriv(\pstara,\exponent ) * \Bhipat )
		/ H1deriv(\pstara,\exponent) };
	\pgfmathsetmacro{\test}{ 
		valuefn_lo(\pstara,\exponent,\Alopat,\Blo)
		- u(\pstara,\slope,\intercept) };

	\ifdim \test pt < 0.0 pt \relax

		\draw [domain=0:\pstara, variable=\p, smooth] 
			plot ( {\p}, 
			{ \yscale * valuefn_lo(\p,\exponent,\Aloimp,\Blo) } );
		\draw [domain=\pstara:1, variable=\p, smooth] 
			plot ( {\p}, 
			{ \yscale * u(\p,\slope,\intercept) } );
		\draw ( { \pstara*9/10 }, 
			{ \yscale * valuefn_lo(\pstara*9/10,\exponent,\Aloimp,\Blo) } )
			node[anchor=north west] {$v$};

	\else

		\draw [domain=0:\pstara, variable=\p, smooth] 
			plot ( {\p}, 
			{ \yscale * valuefn_lo(\p,\exponent,\Alopat,\Blo) } );
		\draw [domain=\pstara:1, variable=\p, smooth] 
			plot ( {\p}, 
			{ \yscale * valuefn_hi(\p,\exponent,\slope,\intercept,\Ahi,\Bhipat) } );
		\draw ( { \pstara*2/3 }, 
			{ \yscale * valuefn_lo(\pstara*2/3,\exponent,\Alopat,\Blo) } )
			node[anchor=north west] {$v$};

	\fi

	\draw [domain=0:1, variable=\p, samples=2, dotted]
		plot ( {\p}, 
		{ (u(1,\slope,\intercept)) * \p } );
	\draw ( { 0.5 }, { \yscale * (u(1,\slope,\intercept)) * 0.5 } )
		node[anchor=south east]
		{$\text{cav} u$};

	\draw [domain=0:\pstara, variable=\p, samples=2, very thick]
		plot ( {\p}, { \yscale * 0 } );
	\draw [domain=\pstara:1, variable=\p, samples=2, very thick]
		plot ( {\p}, { \yscale * u(\p,\slope,\intercept) ) } );
	\draw ( { (2*\pstara+1)/3 }, 
		{ \yscale * u((2*\pstara+1)/3,\slope,\intercept) } )
		node[anchor=north west] {$\boldsymbol{u}$};

	\draw[-]  (0,0) -- ( 1, \yscale * 0 ); 
	\draw[->] (0,0) -- ( 0, { \yscale * (u(1,\slope,\intercept))*1.05 } ); 
	
	\draw ( \pstara, \yscale * 0 ) node[anchor=north] {$2/3\vphantom{\tfrac{2}{3}}$};
	\draw ( 1,       \yscale * 0 ) node[anchor=north] {$1\vphantom{\tfrac{2}{3}}$};

	\draw ( 1, \yscale * 0 ) node[anchor=west] {$p$};

	\draw[-] ( \pstara, - \ticklength ) -- ( \pstara, \ticklength );
	\draw[-] ( 1,       - \ticklength ) -- ( 1,       \ticklength );

	\draw ( 0, { \yscale * (u(\pstara,\slope,\intercept)) } ) 
		node[anchor=east] {$1$};
	\draw ( 0, { \yscale * (u(1,\slope,\intercept)) } ) 
		node[anchor=east] {$2$};

	\draw[-] ( - \ticklength, { \yscale * (u(\pstara,\slope,\intercept)) } ) 
		-- ( \ticklength, { \yscale * (u(\pstara,\slope,\intercept)) } );
	\draw[-] ( - \ticklength, { \yscale * (u(1,\slope,\intercept)) } ) 
		-- ( \ticklength, { \yscale * (u(1,\slope,\intercept)) } );

	\fill[white] ( \pstara, \yscale * 0 ) circle[radius=\radius pt];
	\draw[thick] ( \pstara, \yscale * 0 ) circle[radius=\radius pt];

\end{tikzpicture}


%% file: tikz/2act_v_hi.tex


\begin{tikzpicture}[scale=5.2, line cap=round,
	declare function={
		u(\p,\slope,\intercept)
			= \slope * \p - \intercept;
		D(\prior,\priora)
			= ( \prior / (1-\prior) ) / ( \priora / (1-\priora) );
		phiinv(\p,\prior,\priora)
			= \p / ( \p + (1-\p) / D(\prior,\priora) );
		H1(\p,\exponent) 
			= \p^\exponent * (1-\p)^(1-\exponent);
		H2(\p,\exponent)
			= H1( 1-\p, \exponent );
		H1deriv(\p,\exponent)
			= H1(\p,\exponent) * ( (\exponent/\p) + ((\exponent-1)/(1-\p)) );
		H2deriv(\p,\exponent)
			= - H1deriv(1-\p,\exponent);
		valuefn_lo(\p,\exponent,\A,\B)
			= \A * H1(\p,\exponent) - \B * H2(\p,\exponent);
		valuefn_hi(\p,\exponent,\slope,\intercept,\A,\B)
			= u(\p,\slope,\intercept)
			+ \A * H1(\p,\exponent) - \B * H2(\p,\exponent);
		}]

	\pgfmathsetmacro{\yscale}{1}; 
	\pgfmathsetmacro{\ticklength}{\yscale/60}; 
	\pgfmathsetmacro{\samples}{200}; 
	\pgfmathsetmacro{\radius}{\pgflinewidth/2}; 

	\pgfmathsetmacro{\prior}{1/2}; 
	\pgfmathsetmacro{\priora}{1/5}; 
	\pgfmathsetmacro{\intercept}{1/2}; 
	\pgfmathsetmacro{\slope}{3/2}; 
	\pgfmathsetmacro{\vol}{2}; 
	\pgfmathsetmacro{\r}{1}; 

	\pgfmathsetmacro{\pstar}{ \intercept / \slope };
	\pgfmathsetmacro{\pstara}{ phiinv(\pstar,\prior,\priora) };
	\pgfmathsetmacro{\exponent}
		{ (1/2) + sqrt( (1/4) + 2*\vol*\vol*\r ) }; 
	\pgfmathsetmacro{\Ahi}{ 0 };
	\pgfmathsetmacro{\Blo}{ 0 };
	\pgfmathsetmacro{\Aloimp}
		{ ( u(\pstara,\slope,\intercept) ) / H1(\pstara,\exponent) ) };
	\pgfmathsetmacro{\Bhipat}
		{ ( u(\pstara,\slope,\intercept) * H1deriv(\pstara,\exponent)
		- \slope * H1(\pstara,\exponent) )
		/ ( 2 * \exponent - 1 ) };
	\pgfmathsetmacro{\Alopat}
		{ ( \slope - H2deriv(\pstara,\exponent ) * \Bhipat )
		/ H1deriv(\pstara,\exponent) };
	\pgfmathsetmacro{\test}{ 
		valuefn_lo(\pstara,\exponent,\Alopat,\Blo)
		- u(\pstara,\slope,\intercept) };

	\ifdim \test pt < 0.0 pt \relax

		\draw [domain=0:\pstara, variable=\p, smooth] 
			plot ( {\p}, 
			{ \yscale * valuefn_lo(\p,\exponent,\Aloimp,\Blo) } );
		\draw [domain=\pstara:1, variable=\p, smooth] 
			plot ( {\p}, 
			{ \yscale * u(\p,\slope,\intercept) } );
		\draw ( { \pstara*9/10 }, 
			{ \yscale * valuefn_lo(\pstara*9/10,\exponent,\Aloimp,\Blo) } )
			node[anchor=north west] {$v$};

	\else

		\draw [domain=0:\pstara, variable=\p, smooth] 
			plot ( {\p}, 
			{ \yscale * valuefn_lo(\p,\exponent,\Alopat,\Blo) } );
		\draw [domain=\pstara:1, variable=\p, smooth] 
			plot ( {\p}, 
			{ \yscale * valuefn_hi(\p,\exponent,\slope,\intercept,\Ahi,\Bhipat) } );
		\draw ( { \pstara*2/3 }, 
			{ \yscale * valuefn_lo(\pstara*2/3,\exponent,\Alopat,\Blo) } )
			node[anchor=north west] {$v$};

	\fi

	\draw [domain=0:1, variable=\p, samples=2, dotted]
		plot ( {\p}, 
		{ (u(1,\slope,\intercept)) * \p } );
	\draw ( { 0.5 }, { \yscale * (u(1,\slope,\intercept)) * 0.5 } )
		node[anchor=south east]
		{$\text{cav} u$};

	\draw [domain=0:\pstara, variable=\p, samples=2, very thick]
		plot ( {\p}, { \yscale * 0 } );
	\draw [domain=\pstara:1, variable=\p, samples=2, very thick]
		plot ( {\p}, { \yscale * u(\p,\slope,\intercept) ) } );
	\draw ( { (2*\pstara+1)/3 }, 
		{ \yscale * u((2*\pstara+1)/3,\slope,\intercept) } )
		node[anchor=north west] {$\boldsymbol{u}$};

	\draw[-]  (0,0) -- ( 1, \yscale * 0 ); 
	\draw[->] (0,0) -- ( 0, { \yscale * (u(1,\slope,\intercept))*1.05 } ); 
	
	\draw ( \pstara, \yscale * 0 ) node[anchor=north] {$2/3\vphantom{\tfrac{2}{3}}$};
	\draw ( 1,       \yscale * 0 ) node[anchor=north] {$1\vphantom{\tfrac{2}{3}}$};

	\draw ( 1, \yscale * 0 ) node[anchor=west] {$p$};

	\draw[-] ( \pstara, - \ticklength ) -- ( \pstara, \ticklength );
	\draw[-] ( 1,       - \ticklength ) -- ( 1,       \ticklength );

	\draw ( 0, { \yscale * (u(\pstara,\slope,\intercept)) } ) 
		node[anchor=east] {$1$};
	\draw ( 0, { \yscale * (u(1,\slope,\intercept)) } ) 
		node[anchor=east] {$2$};

	\draw[-] ( - \ticklength, { \yscale * (u(\pstara,\slope,\intercept)) } ) 
		-- ( \ticklength, { \yscale * (u(\pstara,\slope,\intercept)) } );
	\draw[-] ( - \ticklength, { \yscale * (u(1,\slope,\intercept)) } ) 
		-- ( \ticklength, { \yscale * (u(1,\slope,\intercept)) } );

	\fill[white] ( \pstara, \yscale * 0 ) circle[radius=\radius pt];
	\draw[thick] ( \pstara, \yscale * 0 ) circle[radius=\radius pt];

\end{tikzpicture}


%% file: tikz/3act_v_lo.tex


\begin{tikzpicture}[scale=5.2, line cap=round,
	declare function={
		H1(\p,\exponent) 
			= \p^\exponent * (1-\p)^(1-\exponent);
		H2(\p,\exponent)
			= H1( 1-\p, \exponent );
		valuefn(\p,\exponent,\k,\A,\B)
			= \k + \A * H1(\p,\exponent) - \B * H2(\p,\exponent);
	}]

	\pgfmathsetmacro{\yscale}{1/3}; 
	\pgfmathsetmacro{\ticklength}{\yscale/20}; 
	\pgfmathsetmacro{\samples}{200}; 
	\pgfmathsetmacro{\radius}{\pgflinewidth/2}; 

	\pgfmathsetmacro{\pstar}{1/2}; 
	\pgfmathsetmacro{\pstarstar}{3/4}; 
	\pgfmathsetmacro{\l}{1}; 
	\pgfmathsetmacro{\h}{3}; 
	\pgfmathsetmacro{\prior}{5/8}; 
	\pgfmathsetmacro{\vol}{1/2}; 
	\pgfmathsetmacro{\r}{1}; 

	\pgfmathsetmacro{\exponent}{ (1/2) + sqrt( (1/4) + 2*\vol*\vol*\r ) }; 
	\pgfmathsetmacro{\prod}{ ( \exponent - 1/2 )^2 / 2 - 1/8 }; 
	\pgfmathsetmacro{\BElllogamma}{ 
		\l * ( \exponent - \pstar ) / ( H2(\pstar,\exponent) * ( 2*\exponent - 1 ) ) };
	\pgfmathsetmacro{\AElllogamma}{ 
		( \h-\l + \BElllogamma * H2(\pstarstar,\exponent) ) / H1(\pstarstar,\exponent) };
	\pgfmathsetmacro{\higammatest}{ 
		\AElllogamma * H1(\pstar,\exponent) - \BElllogamma * H2(\pstar,\exponent) };

	\ifdim \higammatest pt < 0.0 pt \relax

		\pgfmathsetmacro{\AZero}{ 
			\l / H1(\pstar,\exponent) };
		\pgfmathsetmacro{\BEll}{ 
			( (\h-\l) * H1(\pstar,\exponent) ) /
			( H1(\pstarstar,\exponent) * H2(\pstar,\exponent) 
			- H1(\pstar,\exponent) * H2(\pstarstar,\exponent) ) }
		\pgfmathsetmacro{\AEll}{ 
			\BEll * H2(\pstar,\exponent) / H1(\pstar,\exponent) };

	\else

		\pgfmathsetmacro{\AEll}{ \AElllogamma };
		\pgfmathsetmacro{\BEll}{ \BElllogamma };
		\pgfmathsetmacro{\AZero}{
			\AEll + ( \l - \BEll * H2(\pstar,\exponent) ) / H1(\pstar,\exponent) };

	\fi

	\draw [domain=0:\pstarstar, variable=\p, samples=2, dotted]
		plot ( {\p}, 
		{ \yscale * \h * \p / \pstarstar } );
	\draw ( { \pstar }, { \yscale * \h * \pstar / \pstarstar } )
		node[anchor=south east ]
		{$\text{cav} u$};

	\draw [domain=0:\pstar, variable=\p,          samples=2, very thick]
		plot ( {\p}, { \yscale * 0 } );
	\draw [domain=\pstar:\pstarstar, variable=\p, samples=2, very thick]
		plot ( {\p}, { \yscale * \l } );
	\draw [domain=\pstarstar:1, variable=\p,      samples=2, very thick]
		plot ( {\p}, { \yscale * \h } );
	\draw ( { (\pstarstar+1)/2 }, { \yscale * \h } )
		node[anchor=south] {$\boldsymbol{u}$};

	\draw [domain=0:\pstar, variable=\p, smooth] 
		plot ( {\p}, 
		{ \yscale * valuefn(\p,\exponent,0,\AZero,0) } );
	\draw [domain=\pstar:\pstarstar*1.001, variable=\p, smooth]
		plot ( {\p}, 
		{ \yscale * min( valuefn(\p,\exponent,\l,\AEll,\BEll), \h ) } );
	\draw ( { 0.2*\pstar + 0.8*\pstarstar }, 
		{ \yscale * valuefn(0.2*\pstar + 0.8*\pstarstar,\exponent,\l,\AEll,\BEll) } )
		node[anchor=north west] {$v$};



	\draw[-]  (0,0) -- ( 1, \yscale * 0       ); 
	\draw[->] (0,0) -- ( 0, \yscale * \h*1.05 ); 
	
	\draw ( \pstar,     \yscale * 0 ) node[anchor=north] {$1/2$};
	\draw ( \pstarstar, \yscale * 0 ) node[anchor=north] {$3/4$};
	\draw ( 1,          \yscale * 0 ) node[anchor=north] {$1$};

	\draw ( 1, \yscale * 0 ) node[anchor=west] {$p$};

	\draw[-] ( \pstar,     - \ticklength ) -- ( \pstar,     \ticklength );
	\draw[-] ( \pstarstar, - \ticklength ) -- ( \pstarstar, \ticklength );
	\draw[-] ( 1,          - \ticklength ) -- ( 1,          \ticklength );

	\draw ( 0, \yscale * \l ) node[anchor=east] {$\l$};
	\draw ( 0, \yscale * \h ) node[anchor=east] {$\h$};

	\draw[-] ( - \ticklength, \yscale * \l ) -- ( \ticklength, \yscale * \l );
	\draw[-] ( - \ticklength, \yscale * \h ) -- ( \ticklength, \yscale * \h );

	\fill[white] ( \pstar, \yscale * 0 ) circle[radius=\radius pt];
	\draw[thick] ( \pstar, \yscale * 0 ) circle[radius=\radius pt];
	\fill[white] ( \pstarstar, \yscale * \l ) circle[radius=\radius pt];
	\draw[thick] ( \pstarstar, \yscale * \l ) circle[radius=\radius pt];

\end{tikzpicture}


%% file: tikz/3act_v_hi.tex


\begin{tikzpicture}[scale=5.2, line cap=round,
	declare function={
		H1(\p,\exponent) 
			= \p^\exponent * (1-\p)^(1-\exponent);
		H2(\p,\exponent)
			= H1( 1-\p, \exponent );
		valuefn(\p,\exponent,\k,\A,\B)
			= \k + \A * H1(\p,\exponent) - \B * H2(\p,\exponent);
	}]

	\pgfmathsetmacro{\yscale}{1/3}; 
	\pgfmathsetmacro{\ticklength}{\yscale/20}; 
	\pgfmathsetmacro{\samples}{200}; 
	\pgfmathsetmacro{\radius}{\pgflinewidth/2}; 

	\pgfmathsetmacro{\pstar}{1/2}; 
	\pgfmathsetmacro{\pstarstar}{3/4}; 
	\pgfmathsetmacro{\l}{1}; 
	\pgfmathsetmacro{\h}{3}; 
	\pgfmathsetmacro{\prior}{5/8}; 
	\pgfmathsetmacro{\vol}{2}; 
	\pgfmathsetmacro{\r}{1}; 

	\pgfmathsetmacro{\exponent}{ (1/2) + sqrt( (1/4) + 2*\vol*\vol*\r ) }; 
	\pgfmathsetmacro{\prod}{ 0.5 * ( \exponent - 0.5 )^2 - 0.125 }; 
	\pgfmathsetmacro{\BElllogamma}{ 
		\l * ( \exponent - \pstar ) / ( H2(\pstar,\exponent) * ( 2*\exponent - 1 ) ) };
	\pgfmathsetmacro{\AElllogamma}{ 
		( \h-\l + \BElllogamma * H2(\pstarstar,\exponent) ) / H1(\pstarstar,\exponent) };
	\pgfmathsetmacro{\higammatest}{ 
		\AElllogamma * H1(\pstar,\exponent) - \BElllogamma * H2(\pstar,\exponent) };

	\ifdim \higammatest pt < 0.0 pt \relax

		\pgfmathsetmacro{\AZero}{ 
			\l / H1(\pstar,\exponent) };
		\pgfmathsetmacro{\BEll}{ 
			( (\h-\l) * H1(\pstar,\exponent) ) /
			( H1(\pstarstar,\exponent) * H2(\pstar,\exponent) 
			- H1(\pstar,\exponent) * H2(\pstarstar,\exponent) ) }
		\pgfmathsetmacro{\AEll}{ 
			\BEll * H2(\pstar,\exponent) / H1(\pstar,\exponent) };

	\else

		\pgfmathsetmacro{\AEll}{ \AElllogamma };
		\pgfmathsetmacro{\BEll}{ \BElllogamma };
		\pgfmathsetmacro{\AZero}{
			\AEll + ( \l - \BEll * H2(\pstar,\exponent) ) / H1(\pstar,\exponent) };

	\fi

	\draw [domain=0:\pstarstar, variable=\p, samples=2, dotted]
		plot ( {\p}, 
		{ \yscale * \h * \p / \pstarstar } );
	\draw ( { \pstar }, { \yscale * \h * \pstar / \pstarstar } )
		node[anchor=south east ]
		{$\text{cav} u$};

	\draw [domain=0:\pstar, variable=\p,          samples=2, very thick]
		plot ( {\p}, { \yscale * 0 } );
	\draw [domain=\pstar:\pstarstar, variable=\p, samples=2, very thick]
		plot ( {\p}, { \yscale * \l } );
	\draw [domain=\pstarstar:1, variable=\p,      samples=2, very thick]
		plot ( {\p}, { \yscale * \h } );
	\draw ( { (\pstarstar+1)/2 }, { \yscale * \h } )
		node[anchor=south] {$\boldsymbol{u}$};

	\draw [domain=0:\pstar, variable=\p, smooth] 
		plot ( {\p}, 
		{ \yscale * valuefn(\p,\exponent,0,\AZero,0) } );
	\draw [domain=\pstar:\pstarstar*1.001, variable=\p, smooth]
		plot ( {\p}, 
		{ \yscale * min( valuefn(\p,\exponent,\l,\AEll,\BEll), \h ) } );
	\draw ( { 0.2*\pstar + 0.8*\pstarstar }, 
		{ \yscale * valuefn(0.2*\pstar + 0.8*\pstarstar,\exponent,\l,\AEll,\BEll) } )
		node[anchor=north west] {$v$};



	\draw[-]  (0,0) -- ( 1, \yscale * 0       ); 
	\draw[->] (0,0) -- ( 0, \yscale * \h*1.05 ); 
	
	\draw ( \pstar,     \yscale * 0 ) node[anchor=north] {$1/2$};
	\draw ( \pstarstar, \yscale * 0 ) node[anchor=north] {$3/4$};
	\draw ( 1,          \yscale * 0 ) node[anchor=north] {$1$};

	\draw ( 1, \yscale * 0 ) node[anchor=west] {$p$};

	\draw[-] ( \pstar,     - \ticklength ) -- ( \pstar,     \ticklength );
	\draw[-] ( \pstarstar, - \ticklength ) -- ( \pstarstar, \ticklength );
	\draw[-] ( 1,          - \ticklength ) -- ( 1,          \ticklength );

	\draw ( 0, \yscale * \l ) node[anchor=east] {$\l$};
	\draw ( 0, \yscale * \h ) node[anchor=east] {$\h$};

	\draw[-] ( - \ticklength, \yscale * \l ) -- ( \ticklength, \yscale * \l );
	\draw[-] ( - \ticklength, \yscale * \h ) -- ( \ticklength, \yscale * \h );

	\fill[white] ( \pstar, \yscale * 0 ) circle[radius=\radius pt];
	\draw[thick] ( \pstar, \yscale * 0 ) circle[radius=\radius pt];
	\fill[white] ( \pstarstar, \yscale * \l ) circle[radius=\radius pt];
	\draw[thick] ( \pstarstar, \yscale * \l ) circle[radius=\radius pt];

\end{tikzpicture}


%% file: tikz/3act_jeff.tex


\begin{tikzpicture}[scale=5.2, line cap=round,
	declare function={
		H1(\p,\exponent) 
			= \p^\exponent * (1-\p)^(1-\exponent);
		H2(\p,\exponent)
			= H1( 1-\p, \exponent );
		valuefn(\p,\exponent,\k,\A,\B)
			= \k + \A * H1(\p,\exponent) - \B * H2(\p,\exponent);
	}]

	\pgfmathsetmacro{\yscale}{1/3}; 
	\pgfmathsetmacro{\ticklength}{\yscale/20}; 
	\pgfmathsetmacro{\samples}{200}; 
	\pgfmathsetmacro{\radius}{\pgflinewidth/2}; 

	\pgfmathsetmacro{\pstar}{1/2}; 
	\pgfmathsetmacro{\pstarstar}{3/4}; 
	\pgfmathsetmacro{\l}{1}; 
	\pgfmathsetmacro{\h}{3}; 
	\pgfmathsetmacro{\prior}{5/8}; 
	\pgfmathsetmacro{\vol}{0.9495}; 
	\pgfmathsetmacro{\r}{1}; 

	\pgfmathsetmacro{\exponent}{ (1/2) + sqrt( (1/4) + 2*\vol*\vol*\r ) }; 
	\pgfmathsetmacro{\prod}{ 0.5 * ( \exponent - 0.5 )^2 - 0.125 }; 
	\pgfmathsetmacro{\BElllogamma}{ 
		\l * ( \exponent - \pstar ) / ( H2(\pstar,\exponent) * ( 2*\exponent - 1 ) ) };
	\pgfmathsetmacro{\AElllogamma}{ 
		( \h-\l + \BElllogamma * H2(\pstarstar,\exponent) ) / H1(\pstarstar,\exponent) };
	\pgfmathsetmacro{\higammatest}{ 
		\AElllogamma * H1(\pstar,\exponent) - \BElllogamma * H2(\pstar,\exponent) };

	\ifdim \higammatest pt < 0.0 pt \relax

		\pgfmathsetmacro{\AZero}{ 
			\l / H1(\pstar,\exponent) };
		\pgfmathsetmacro{\BEll}{ 
			( (\h-\l) * H1(\pstar,\exponent) ) /
			( H1(\pstarstar,\exponent) * H2(\pstar,\exponent) 
			- H1(\pstar,\exponent) * H2(\pstarstar,\exponent) ) }
		\pgfmathsetmacro{\AEll}{ 
			\BEll * H2(\pstar,\exponent) / H1(\pstar,\exponent) };

	\else

		\pgfmathsetmacro{\AEll}{ \AElllogamma };
		\pgfmathsetmacro{\BEll}{ \BElllogamma };
		\pgfmathsetmacro{\AZero}{
			\AEll + ( \l - \BEll * H2(\pstar,\exponent) ) / H1(\pstar,\exponent) };

	\fi

	\draw [domain=0:\pstarstar, variable=\p, samples=2, dotted]
		plot ( {\p}, 
		{ \yscale * \h * \p / \pstarstar } );
	\draw ( { \pstar }, { \yscale * \h * \pstar / \pstarstar } )
		node[anchor=south east ]
		{$\text{cav} u$};

	\draw [domain=0:\pstar, variable=\p,          samples=2, very thick]
		plot ( {\p}, { \yscale * 0 } );
	\draw [domain=\pstar:\pstarstar, variable=\p, samples=2, very thick]
		plot ( {\p}, { \yscale * \l } );
	\draw [domain=\pstarstar:1, variable=\p,      samples=2, very thick]
		plot ( {\p}, { \yscale * \h } );
	\draw ( { (\pstarstar+1)/2 }, { \yscale * \h } )
		node[anchor=south] {$\boldsymbol{u}$};

	\draw [domain=0:\pstar, variable=\p, smooth] 
		plot ( {\p}, 
		{ \yscale * valuefn(\p,\exponent,0,\AZero,0) } );
	\draw [domain=\pstar:\pstarstar*1.001, variable=\p, smooth]
		plot ( {\p}, 
		{ \yscale * min( valuefn(\p,\exponent,\l,\AEll,\BEll), \h ) } );
	\draw ( { 0.2*\pstar + 0.8*\pstarstar }, 
		{ \yscale * valuefn(0.2*\pstar + 0.8*\pstarstar,\exponent,\l,\AEll,\BEll) } )
		node[anchor=north west] {$v$};



	\draw[-]  (0,0) -- ( 1, \yscale * 0       ); 
	\draw[->] (0,0) -- ( 0, \yscale * \h*1.05 ); 
	
	\draw ( \pstar,     \yscale * 0 ) node[anchor=north] {$\frac{1}{2}$};
	\draw ( \pstarstar, \yscale * 0 ) node[anchor=north] {$\frac{3}{4}$};
	\draw ( \prior,     \yscale * 0 ) node[anchor=north] {$\phantom{\frac{1}{2}}p_0\phantom{\frac{1}{2}}$};
	\draw ( 1,          \yscale * 0 ) node[anchor=north] {$\phantom{\frac{1}{2}}1\phantom{\frac{1}{2}}$};

	\draw ( 1, \yscale * 0 ) node[anchor=west] {$p$};

	\draw[-] ( \pstar,     - \ticklength ) -- ( \pstar,     \ticklength );
	\draw[-] ( \pstarstar, - \ticklength ) -- ( \pstarstar, \ticklength );
	\draw[-] ( \prior,     - \ticklength ) -- ( \prior,     \ticklength );
	\draw[-] ( 1,          - \ticklength ) -- ( 1,          \ticklength );

	\draw ( 0, \yscale * \l ) node[anchor=east] {$\l$};
	\draw ( 0, \yscale * \h ) node[anchor=east] {$\h$};

	\draw[-] ( - \ticklength, \yscale * \l ) -- ( \ticklength, \yscale * \l );
	\draw[-] ( - \ticklength, \yscale * \h ) -- ( \ticklength, \yscale * \h );

	\fill[white] ( \pstar, \yscale * 0 ) circle[radius=\radius pt];
	\draw[thick] ( \pstar, \yscale * 0 ) circle[radius=\radius pt];
	\fill[white] ( \pstarstar, \yscale * \l ) circle[radius=\radius pt];
	\draw[thick] ( \pstarstar, \yscale * \l ) circle[radius=\radius pt];

\end{tikzpicture}


%% file: tikz/2act_jeff.tex


\begin{tikzpicture}[scale=5.2, line cap=round,
	declare function={
		u(\p,\slope,\intercept)
			= \slope * \p - \intercept;
		D(\prior,\priora)
			= ( \prior / (1-\prior) ) / ( \priora / (1-\priora) );
		phiinv(\p,\prior,\priora)
			= \p / ( \p + (1-\p) / D(\prior,\priora) );
		H1(\p,\exponent) 
			= \p^\exponent * (1-\p)^(1-\exponent);
		H2(\p,\exponent)
			= H1( 1-\p, \exponent );
		H1deriv(\p,\exponent)
			= H1(\p,\exponent) * ( (\exponent/\p) + ((\exponent-1)/(1-\p)) );
		H2deriv(\p,\exponent)
			= - H1deriv(1-\p,\exponent);
		valuefn_lo(\p,\exponent,\A,\B)
			= \A * H1(\p,\exponent) - \B * H2(\p,\exponent);
		valuefn_hi(\p,\exponent,\slope,\intercept,\A,\B)
			= u(\p,\slope,\intercept)
			+ \A * H1(\p,\exponent) - \B * H2(\p,\exponent);
		}]

	\pgfmathsetmacro{\yscale}{1}; 
	\pgfmathsetmacro{\ticklength}{\yscale/60}; 
	\pgfmathsetmacro{\samples}{200}; 
	\pgfmathsetmacro{\radius}{\pgflinewidth/2}; 

	\pgfmathsetmacro{\prior}{1/2}; 
	\pgfmathsetmacro{\priora}{1/5}; 
	\pgfmathsetmacro{\intercept}{1/2}; 
	\pgfmathsetmacro{\slope}{3/2}; 
	\pgfmathsetmacro{\vol}{1/2.12088}; 
	\pgfmathsetmacro{\r}{1}; 

	\pgfmathsetmacro{\pstar}{ \intercept / \slope };
	\pgfmathsetmacro{\pstara}{ phiinv(\pstar,\prior,\priora) };
	\pgfmathsetmacro{\exponent}
		{ (1/2) + sqrt( (1/4) + 2*\vol*\vol*\r ) }; 
	\pgfmathsetmacro{\Ahi}{ 0 };
	\pgfmathsetmacro{\Blo}{ 0 };
	\pgfmathsetmacro{\Aloimp}
		{ ( u(\pstara,\slope,\intercept) ) / H1(\pstara,\exponent) ) };
	\pgfmathsetmacro{\Bhipat}
		{ ( u(\pstara,\slope,\intercept) * H1deriv(\pstara,\exponent)
		- \slope * H1(\pstara,\exponent) )
		/ ( 2 * \exponent - 1 ) };
	\pgfmathsetmacro{\Alopat}
		{ ( \slope - H2deriv(\pstara,\exponent ) * \Bhipat )
		/ H1deriv(\pstara,\exponent) };
	\pgfmathsetmacro{\test}{ 
		valuefn_lo(\pstara,\exponent,\Alopat,\Blo)
		- u(\pstara,\slope,\intercept) };

	\ifdim \test pt < 0.0 pt \relax

		\draw [domain=0:\pstara, variable=\p, smooth] 
			plot ( {\p}, 
			{ \yscale * valuefn_lo(\p,\exponent,\Aloimp,\Blo) } );
		\draw [domain=\pstara:1, variable=\p, smooth] 
			plot ( {\p}, 
			{ \yscale * u(\p,\slope,\intercept) } );
		\draw ( { \pstara*9/10 }, 
			{ \yscale * valuefn_lo(\pstara*9/10,\exponent,\Aloimp,\Blo) } )
			node[anchor=north west] {$v$};

	\else

		\draw [domain=0:\pstara, variable=\p, smooth] 
			plot ( {\p}, 
			{ \yscale * valuefn_lo(\p,\exponent,\Alopat,\Blo) } );
		\draw [domain=\pstara:1, variable=\p, smooth] 
			plot ( {\p}, 
			{ \yscale * valuefn_hi(\p,\exponent,\slope,\intercept,\Ahi,\Bhipat) } );
		\draw ( { \pstara*2/3 }, 
			{ \yscale * valuefn_lo(\pstara*2/3,\exponent,\Alopat,\Blo) } )
			node[anchor=north west] {$v$};

	\fi

	\draw [domain=0:1, variable=\p, samples=2, dotted]
		plot ( {\p}, 
		{ (u(1,\slope,\intercept)) * \p } );
	\draw ( { 0.5 }, { \yscale * (u(1,\slope,\intercept)) * 0.5 } )
		node[anchor=south east]
		{$\text{cav} u$};

	\draw [domain=0:\pstara, variable=\p, samples=2, very thick]
		plot ( {\p}, { \yscale * 0 } );
	\draw [domain=\pstara:1, variable=\p, samples=2, very thick]
		plot ( {\p}, { \yscale * u(\p,\slope,\intercept) ) } );
	\draw ( { (2*\pstara+1)/3 }, 
		{ \yscale * u((2*\pstara+1)/3,\slope,\intercept) } )
		node[anchor=north west] {$\boldsymbol{u}$};

	\draw[-]  (0,0) -- ( 1, \yscale * 0 ); 
	\draw[->] (0,0) -- ( 0, { \yscale * (u(1,\slope,\intercept))*1.05 } ); 
	
	\draw ( \prior,  \yscale * 0 ) node[anchor=north] {$\phantom{\frac{2}{3}}p_0\phantom{\frac{2}{3}}$};
	\draw ( \pstara, \yscale * 0 ) node[anchor=north] {$2/3\vphantom{\frac{2}{3}}$};
	\draw ( 1,       \yscale * 0 ) node[anchor=north] {$1\vphantom{\frac{2}{3}}$};

	\draw ( 1, \yscale * 0 ) node[anchor=west] {$p$};

	\draw[-] ( \prior,  - \ticklength ) -- ( \prior,  \ticklength );
	\draw[-] ( \pstara, - \ticklength ) -- ( \pstara, \ticklength );
	\draw[-] ( 1,       - \ticklength ) -- ( 1,       \ticklength );

	\draw ( 0, { \yscale * (u(\pstara,\slope,\intercept)) } ) 
		node[anchor=east] {$1$};
	\draw ( 0, { \yscale * (u(1,\slope,\intercept)) } ) 
		node[anchor=east] {$2$};

	\draw[-] ( - \ticklength, { \yscale * (u(\pstara,\slope,\intercept)) } ) 
		-- ( \ticklength, { \yscale * (u(\pstara,\slope,\intercept)) } );
	\draw[-] ( - \ticklength, { \yscale * (u(1,\slope,\intercept)) } ) 
		-- ( \ticklength, { \yscale * (u(1,\slope,\intercept)) } );

	\fill[white] ( \pstara, \yscale * 0 ) circle[radius=\radius pt];
	\draw[thick] ( \pstara, \yscale * 0 ) circle[radius=\radius pt];

\end{tikzpicture}


%% file: tikz/3act2_25_subsupersoln.tex


\begin{tikzpicture}[scale=5.2, line cap=round,
	declare function={
		H1(\p,\exponent) 
			= \p^\exponent * (1-\p)^(1-\exponent);
		H2(\p,\exponent)
			= H1( 1-\p, \exponent );
		valuefn(\p,\exponent,\k,\A,\B)
			= \k + \A * H1(\p,\exponent) - \B * H2(\p,\exponent);
	}]

	\pgfmathsetmacro{\yscale}{1/3}; 
	\pgfmathsetmacro{\ticklength}{\yscale/20}; 
	\pgfmathsetmacro{\samples}{200}; 
	\pgfmathsetmacro{\radius}{\pgflinewidth/2}; 

	\pgfmathsetmacro{\pstar}{1/2}; 
	\pgfmathsetmacro{\pstarstar}{3/4}; 
	\pgfmathsetmacro{\l}{1}; 
	\pgfmathsetmacro{\h}{3}; 
	\pgfmathsetmacro{\prior}{1/4}; 
	\pgfmathsetmacro{\vol}{2}; 
	\pgfmathsetmacro{\r}{1}; 

	\pgfmathsetmacro{\ptouch}{ 2/5 };
	\pgfmathsetmacro{\phiaupper}{ 100 };
	\pgfmathsetmacro{\phialower}{ -79 };
	\pgfmathsetmacro{\pupperlo}{ 0.25 };
	\pgfmathsetmacro{\pupperhi}{ 0.51 };
	\pgfmathsetmacro{\plowerlo}{ 0.36 };
	\pgfmathsetmacro{\plowerhi}{ 0.55 };

	\pgfmathsetmacro{\exponent}{ (1/2) + sqrt( (1/4) + 2*\vol*\vol*\r ) }; 
	\pgfmathsetmacro{\prod}{ 0.5 * ( \exponent - 0.5 )^2 - 0.125 }; 
	\pgfmathsetmacro{\BElllogamma}{ 
		\l * ( \exponent - \pstar ) / ( H2(\pstar,\exponent) * ( 2*\exponent - 1 ) ) };
	\pgfmathsetmacro{\AElllogamma}{ 
		( \h-\l + \BElllogamma * H2(\pstarstar,\exponent) ) / H1(\pstarstar,\exponent) };
	\pgfmathsetmacro{\higammatest}{ 
		\AElllogamma * H1(\pstar,\exponent) - \BElllogamma * H2(\pstar,\exponent) };

	\ifdim \higammatest pt < 0.0 pt \relax

		\pgfmathsetmacro{\AZero}{ 
			\l / H1(\pstar,\exponent) };
		\pgfmathsetmacro{\BEll}{ 
			( (\h-\l) * H1(\pstar,\exponent) ) /
			( H1(\pstarstar,\exponent) * H2(\pstar,\exponent) 
			- H1(\pstar,\exponent) * H2(\pstarstar,\exponent) ) }
		\pgfmathsetmacro{\AEll}{ 
			\BEll * H2(\pstar,\exponent) / H1(\pstar,\exponent) };

	\else

		\pgfmathsetmacro{\AEll}{ \AElllogamma };
		\pgfmathsetmacro{\BEll}{ \BElllogamma };
		\pgfmathsetmacro{\AZero}{
			\AEll + ( \l - \BEll * H2(\pstar,\exponent) ) / H1(\pstar,\exponent) };

	\fi


	\draw [domain=0:\pstar, variable=\p,          samples=2, very thick]
		plot ( {\p}, { \yscale * 0 } );
	\draw [domain=\pstar:\pstarstar, variable=\p, samples=2, very thick]
		plot ( {\p}, { \yscale * \l } );
	\draw [domain=\pstarstar:1, variable=\p,      samples=2, very thick]
		plot ( {\p}, { \yscale * \h } );
	\draw ( { (\pstarstar+1)/2 }, { \yscale * \h } )
		node[anchor=south] {$\boldsymbol{u}$};

	\draw [domain=0:\pstar, variable=\p, smooth] 
		plot ( {\p}, 
		{ \yscale * valuefn(\p,\exponent,0,\AZero,0) } );
	\draw [domain=\pstar:\pstarstar*1.001, variable=\p, smooth]
		plot ( {\p}, 
		{ \yscale * min( valuefn(\p,\exponent,\l,\AEll,\BEll), \h ) } );
	\draw ( { 0.2*\pstar + 0.8*\pstarstar }, 
		{ \yscale * valuefn(0.2*\pstar + 0.8*\pstarstar,\exponent,\l,\AEll,\BEll) } )
		node[anchor=north west] {$v$};

	\draw [domain=\pupperlo:\pupperhi, variable=\p, smooth, dotted] 
		plot ( {\p}, 
		{ \yscale * ( valuefn(\p,\exponent,0,\AZero,0)
		+ \phiaupper * ( \p - \ptouch )^2 ) } );
	\draw ( { \pupperlo }, 
		{ \yscale * ( valuefn(\pupperlo,\exponent,0,\AZero,0)
		+ \phiaupper * ( \pupperlo - \ptouch )^2 ) } )
		node[anchor=north east] {$\phi_1$};

	\draw [domain=\plowerlo:\plowerhi, variable=\p, smooth, dotted] 
		plot ( {\p}, 
		{ \yscale * ( valuefn(\p,\exponent,0,\AZero,0)
		+ \phialower * ( \p - \ptouch )^2 ) } );
	\draw ( { 1/2 }, 
		{ \yscale * ( valuefn(1/2,\exponent,0,\AZero,0)
		+ \phialower * ( 1/2 - \ptouch )^2 ) } )
		node[anchor=south west] {$\phi_2$};



	\draw[-]  (0,0) -- ( 1, \yscale * 0       ); 
	\draw[->] (0,0) -- ( 0, \yscale * \h*1.05 ); 
	
	\draw ( \pstar,     \yscale * 0 ) node[anchor=north] {$1/2$};
	\draw ( \pstarstar, \yscale * 0 ) node[anchor=north] {$3/4$};
	\draw ( 1,          \yscale * 0 ) node[anchor=north] {$1$};

	\draw ( 1, \yscale * 0 ) node[anchor=west] {$p$};

	\draw[-] ( \pstar,     - \ticklength ) -- ( \pstar,     \ticklength );
	\draw[-] ( \pstarstar, - \ticklength ) -- ( \pstarstar, \ticklength );
	\draw[-] ( 1,          - \ticklength ) -- ( 1,          \ticklength );

	\draw ( 0, \yscale * \l ) node[anchor=east] {$\l$};
	\draw ( 0, \yscale * \h ) node[anchor=east] {$\h$};

	\draw[-] ( - \ticklength, \yscale * \l ) -- ( \ticklength, \yscale * \l );
	\draw[-] ( - \ticklength, \yscale * \h ) -- ( \ticklength, \yscale * \h );

	\fill[white] ( \pstar, \yscale * 0 ) circle[radius=\radius pt];
	\draw[thick] ( \pstar, \yscale * 0 ) circle[radius=\radius pt];
	\fill[white] ( \pstarstar, \yscale * \l ) circle[radius=\radius pt];
	\draw[thick] ( \pstarstar, \yscale * \l ) circle[radius=\radius pt];

\end{tikzpicture}


%% file: tikz/3act2_12_subsoln1.tex


\begin{tikzpicture}[scale=5.2, line cap=round,
	declare function={
		H1(\p,\exponent) 
			= \p^\exponent * (1-\p)^(1-\exponent);
		H2(\p,\exponent)
			= H1( 1-\p, \exponent );
		valuefn(\p,\exponent,\k,\A,\B)
			= \k + \A * H1(\p,\exponent) - \B * H2(\p,\exponent);
	}]

	\pgfmathsetmacro{\yscale}{1/3}; 
	\pgfmathsetmacro{\ticklength}{\yscale/20}; 
	\pgfmathsetmacro{\samples}{200}; 
	\pgfmathsetmacro{\radius}{\pgflinewidth/2}; 

	\pgfmathsetmacro{\pstar}{1/2}; 
	\pgfmathsetmacro{\pstarstar}{3/4}; 
	\pgfmathsetmacro{\l}{1}; 
	\pgfmathsetmacro{\h}{3}; 
	\pgfmathsetmacro{\prior}{1/4}; 
	\pgfmathsetmacro{\vol}{2}; 
	\pgfmathsetmacro{\r}{1}; 

	\pgfmathsetmacro{\phislope}{4}; 
	\pgfmathsetmacro{\phiintercept}{15}; 

	\pgfmathsetmacro{\exponent}{ (1/2) + sqrt( (1/4) + 2*\vol*\vol*\r ) }; 
	\pgfmathsetmacro{\prod}{ 0.5 * ( \exponent - 0.5 )^2 - 0.125 }; 
	\pgfmathsetmacro{\BElllogamma}{ 
		\l * ( \exponent - \pstar ) / ( H2(\pstar,\exponent) * ( 2*\exponent - 1 ) ) };
	\pgfmathsetmacro{\AElllogamma}{ 
		( \h-\l + \BElllogamma * H2(\pstarstar,\exponent) ) / H1(\pstarstar,\exponent) };
	\pgfmathsetmacro{\higammatest}{ 
		\AElllogamma * H1(\pstar,\exponent) - \BElllogamma * H2(\pstar,\exponent) };

	\pgfmathsetmacro{\phia}{ 4*(\phiintercept-1) + 2*\phislope };
	\pgfmathsetmacro{\phib}{ \phislope - \phia };
	\pgfmathsetmacro{\philop}{ 
		( -\phib - sqrt( (\phib)^2 - 4*\phia*(\phiintercept-\h) ) ) 
		/ (2*\phia) };
	\pgfmathsetmacro{\phihip}{ 
		( -\phib + sqrt( (\phib)^2 - 4*\phia*(\phiintercept-\h) ) ) 
		/ (2*\phia) };

	\ifdim \higammatest pt < 0.0 pt \relax

		\pgfmathsetmacro{\AZero}{ 
			\l / H1(\pstar,\exponent) };
		\pgfmathsetmacro{\BEll}{ 
			( (\h-\l) * H1(\pstar,\exponent) ) /
			( H1(\pstarstar,\exponent) * H2(\pstar,\exponent) 
			- H1(\pstar,\exponent) * H2(\pstarstar,\exponent) ) }
		\pgfmathsetmacro{\AEll}{ 
			\BEll * H2(\pstar,\exponent) / H1(\pstar,\exponent) };

	\else

		\pgfmathsetmacro{\AEll}{ \AElllogamma };
		\pgfmathsetmacro{\BEll}{ \BElllogamma };
		\pgfmathsetmacro{\AZero}{
			\AEll + ( \l - \BEll * H2(\pstar,\exponent) ) / H1(\pstar,\exponent) };

	\fi


	\draw [domain=0:\pstar, variable=\p,          samples=2, very thick]
		plot ( {\p}, { \yscale * 0 } );
	\draw [domain=\pstar:\pstarstar, variable=\p, samples=2, very thick]
		plot ( {\p}, { \yscale * \l } );
	\draw [domain=\pstarstar:1, variable=\p,      samples=2, very thick]
		plot ( {\p}, { \yscale * \h } );
	\draw ( { (\pstarstar+1)/2 }, { \yscale * \h } )
		node[anchor=south] {$\boldsymbol{u}$};

	\draw [domain=0:\pstar, variable=\p, smooth] 
		plot ( {\p}, 
		{ \yscale * valuefn(\p,\exponent,0,\AZero,0) } );
	\draw [domain=\pstar:\pstarstar*1.001, variable=\p, smooth]
		plot ( {\p}, 
		{ \yscale * min( valuefn(\p,\exponent,\l,\AEll,\BEll), \h ) } );
	\draw ( { 0.2*\pstar + 0.8*\pstarstar }, 
		{ \yscale * valuefn(0.2*\pstar + 0.8*\pstarstar,\exponent,\l,\AEll,\BEll) } )
		node[anchor=north west] {$v$};

	\draw [domain=\philop:\phihip, variable=\p, smooth, dotted] 
		plot ( {\p}, 
		{ \yscale * ( \phia*\p^2 + \phib*\p + \phiintercept ) } );
	\draw ( { \philop }, 
		{ \yscale * ( \phia*( \philop )^2 
		+ \phib*( \philop ) 
		+ \phiintercept ) } )
		node[anchor=north east] {$\phi$};



	\draw[-]  (0,0) -- ( 1, \yscale * 0       ); 
	\draw[->] (0,0) -- ( 0, \yscale * \h*1.05 ); 
	
	\draw ( \pstar,     \yscale * 0 ) node[anchor=north] {$1/2$};
	\draw ( \pstarstar, \yscale * 0 ) node[anchor=north] {$3/4$};
	\draw ( 1,          \yscale * 0 ) node[anchor=north] {$1$};

	\draw ( 1, \yscale * 0 ) node[anchor=west] {$p$};

	\draw[-] ( \pstar,     - \ticklength ) -- ( \pstar,     \ticklength );
	\draw[-] ( \pstarstar, - \ticklength ) -- ( \pstarstar, \ticklength );
	\draw[-] ( 1,          - \ticklength ) -- ( 1,          \ticklength );

	\draw ( 0, \yscale * \l ) node[anchor=east] {$\l$};
	\draw ( 0, \yscale * \h ) node[anchor=east] {$\h$};

	\draw[-] ( - \ticklength, \yscale * \l ) -- ( \ticklength, \yscale * \l );
	\draw[-] ( - \ticklength, \yscale * \h ) -- ( \ticklength, \yscale * \h );

	\fill[white] ( \pstar, \yscale * 0 ) circle[radius=\radius pt];
	\draw[thick] ( \pstar, \yscale * 0 ) circle[radius=\radius pt];
	\fill[white] ( \pstarstar, \yscale * \l ) circle[radius=\radius pt];
	\draw[thick] ( \pstarstar, \yscale * \l ) circle[radius=\radius pt];

\end{tikzpicture}


%% file: tikz/3act2_12_subsoln2.tex


\begin{tikzpicture}[scale=5.2, line cap=round,
	declare function={
		H1(\p,\exponent) 
			= \p^\exponent * (1-\p)^(1-\exponent);
		H2(\p,\exponent)
			= H1( 1-\p, \exponent );
		valuefn(\p,\exponent,\k,\A,\B)
			= \k + \A * H1(\p,\exponent) - \B * H2(\p,\exponent);
	}]

	\pgfmathsetmacro{\yscale}{1/3}; 
	\pgfmathsetmacro{\ticklength}{\yscale/20}; 
	\pgfmathsetmacro{\samples}{200}; 
	\pgfmathsetmacro{\radius}{\pgflinewidth/2}; 

	\pgfmathsetmacro{\pstar}{1/2}; 
	\pgfmathsetmacro{\pstarstar}{3/4}; 
	\pgfmathsetmacro{\l}{1}; 
	\pgfmathsetmacro{\h}{3}; 
	\pgfmathsetmacro{\prior}{1/4}; 
	\pgfmathsetmacro{\vol}{2}; 
	\pgfmathsetmacro{\r}{1}; 

	\pgfmathsetmacro{\phislope}{5};
	\pgfmathsetmacro{\philop}{ 3/8 };
	\pgfmathsetmacro{\phihip}{ 5/8 };

	\pgfmathsetmacro{\exponent}{ (1/2) + sqrt( (1/4) + 2*\vol*\vol*\r ) }; 
	\pgfmathsetmacro{\prod}{ 0.5 * ( \exponent - 0.5 )^2 - 0.125 }; 
	\pgfmathsetmacro{\BElllogamma}{ 
		\l * ( \exponent - \pstar ) / ( H2(\pstar,\exponent) * ( 2*\exponent - 1 ) ) };
	\pgfmathsetmacro{\AElllogamma}{ 
		( \h-\l + \BElllogamma * H2(\pstarstar,\exponent) ) / H1(\pstarstar,\exponent) };
	\pgfmathsetmacro{\higammatest}{ 
		\AElllogamma * H1(\pstar,\exponent) - \BElllogamma * H2(\pstar,\exponent) };

	\pgfmathsetmacro{\phia}{ \phislope };
	\pgfmathsetmacro{\phib}{ 1 - \phia/2 };

	\ifdim \higammatest pt < 0.0 pt \relax

		\pgfmathsetmacro{\AZero}{ 
			\l / H1(\pstar,\exponent) };
		\pgfmathsetmacro{\BEll}{ 
			( (\h-\l) * H1(\pstar,\exponent) ) /
			( H1(\pstarstar,\exponent) * H2(\pstar,\exponent) 
			- H1(\pstar,\exponent) * H2(\pstarstar,\exponent) ) }
		\pgfmathsetmacro{\AEll}{ 
			\BEll * H2(\pstar,\exponent) / H1(\pstar,\exponent) };

	\else

		\pgfmathsetmacro{\AEll}{ \AElllogamma };
		\pgfmathsetmacro{\BEll}{ \BElllogamma };
		\pgfmathsetmacro{\AZero}{
			\AEll + ( \l - \BEll * H2(\pstar,\exponent) ) / H1(\pstar,\exponent) };

	\fi


	\draw [domain=0:\pstar, variable=\p,          samples=2, very thick]
		plot ( {\p}, { \yscale * 0 } );
	\draw [domain=\pstar:\pstarstar, variable=\p, samples=2, very thick]
		plot ( {\p}, { \yscale * \l } );
	\draw [domain=\pstarstar:1, variable=\p,      samples=2, very thick]
		plot ( {\p}, { \yscale * \h } );
	\draw ( { (\pstarstar+1)/2 }, { \yscale * \h } )
		node[anchor=south] {$\boldsymbol{u}$};

	\draw [domain=0:\pstar, variable=\p, smooth] 
		plot ( {\p}, 
		{ \yscale * valuefn(\p,\exponent,0,\AZero,0) } );
	\draw [domain=\pstar:\pstarstar*1.001, variable=\p, smooth]
		plot ( {\p}, 
		{ \yscale * min( valuefn(\p,\exponent,\l,\AEll,\BEll), \h ) } );
	\draw ( { 0.2*\pstar + 0.8*\pstarstar }, 
		{ \yscale * valuefn(0.2*\pstar + 0.8*\pstarstar,\exponent,\l,\AEll,\BEll) } )
		node[anchor=north west] {$v$};

	\draw [domain=\philop:\phihip, variable=\p, smooth, dotted] 
		plot ( {\p}, 
		{ \yscale * ( \phia*\p + \phib ) } );
	\draw ( { \philop }, 
		{ \yscale * ( \phia*( \philop )
		+ \phib ) } )
		node[anchor=south east] {$\phi$};



	\draw[-]  (0,0) -- ( 1, \yscale * 0       ); 
	\draw[->] (0,0) -- ( 0, \yscale * \h*1.05 ); 
	
	\draw ( \pstar,     \yscale * 0 ) node[anchor=north] {$1/2$};
	\draw ( \pstarstar, \yscale * 0 ) node[anchor=north] {$3/4$};
	\draw ( 1,          \yscale * 0 ) node[anchor=north] {$1$};

	\draw ( 1, \yscale * 0 ) node[anchor=west] {$p$};

	\draw[-] ( \pstar,     - \ticklength ) -- ( \pstar,     \ticklength );
	\draw[-] ( \pstarstar, - \ticklength ) -- ( \pstarstar, \ticklength );
	\draw[-] ( 1,          - \ticklength ) -- ( 1,          \ticklength );

	\draw ( 0, \yscale * \l ) node[anchor=east] {$\l$};
	\draw ( 0, \yscale * \h ) node[anchor=east] {$\h$};

	\draw[-] ( - \ticklength, \yscale * \l ) -- ( \ticklength, \yscale * \l );
	\draw[-] ( - \ticklength, \yscale * \h ) -- ( \ticklength, \yscale * \h );

	\fill[white] ( \pstar, \yscale * 0 ) circle[radius=\radius pt];
	\draw[thick] ( \pstar, \yscale * 0 ) circle[radius=\radius pt];
	\fill[white] ( \pstarstar, \yscale * \l ) circle[radius=\radius pt];
	\draw[thick] ( \pstarstar, \yscale * \l ) circle[radius=\radius pt];

\end{tikzpicture}


%% file: tikz/3act2_12_supersoln.tex


\begin{tikzpicture}[scale=5.2, line cap=round,
	declare function={
		H1(\p,\exponent) 
			= \p^\exponent * (1-\p)^(1-\exponent);
		H2(\p,\exponent)
			= H1( 1-\p, \exponent );
		valuefn(\p,\exponent,\k,\A,\B)
			= \k + \A * H1(\p,\exponent) - \B * H2(\p,\exponent);
	}]

	\pgfmathsetmacro{\yscale}{1/3}; 
	\pgfmathsetmacro{\ticklength}{\yscale/20}; 
	\pgfmathsetmacro{\samples}{200}; 
	\pgfmathsetmacro{\radius}{\pgflinewidth/2}; 

	\pgfmathsetmacro{\pstar}{1/2}; 
	\pgfmathsetmacro{\pstarstar}{3/4}; 
	\pgfmathsetmacro{\l}{1}; 
	\pgfmathsetmacro{\h}{3}; 
	\pgfmathsetmacro{\prior}{1/4}; 
	\pgfmathsetmacro{\vol}{2}; 
	\pgfmathsetmacro{\r}{1}; 

	\pgfmathsetmacro{\phislope}{4};
	\pgfmathsetmacro{\phiintercept}{-10};

	\pgfmathsetmacro{\exponent}{ (1/2) + sqrt( (1/4) + 2*\vol*\vol*\r ) }; 
	\pgfmathsetmacro{\prod}{ 0.5 * ( \exponent - 0.5 )^2 - 0.125 }; 
	\pgfmathsetmacro{\BElllogamma}{ 
		\l * ( \exponent - \pstar ) / ( H2(\pstar,\exponent) * ( 2*\exponent - 1 ) ) };
	\pgfmathsetmacro{\AElllogamma}{ 
		( \h-\l + \BElllogamma * H2(\pstarstar,\exponent) ) / H1(\pstarstar,\exponent) };
	\pgfmathsetmacro{\higammatest}{ 
		\AElllogamma * H1(\pstar,\exponent) - \BElllogamma * H2(\pstar,\exponent) };

	\pgfmathsetmacro{\phia}{ 4*(\phiintercept-1) + 2*\phislope };
	\pgfmathsetmacro{\phib}{ \phislope - \phia };
	\pgfmathsetmacro{\philop}{ 
		( -\phib - sqrt( (\phib)^2 - 4*\phia*(\phiintercept) ) ) 
		/ (2*\phia) };
	\pgfmathsetmacro{\phihip}{ 
		( -\phib + sqrt( (\phib)^2 - 4*\phia*(\phiintercept) ) ) 
		/ (2*\phia) };

	\ifdim \higammatest pt < 0.0 pt \relax

		\pgfmathsetmacro{\AZero}{ 
			\l / H1(\pstar,\exponent) };
		\pgfmathsetmacro{\BEll}{ 
			( (\h-\l) * H1(\pstar,\exponent) ) /
			( H1(\pstarstar,\exponent) * H2(\pstar,\exponent) 
			- H1(\pstar,\exponent) * H2(\pstarstar,\exponent) ) }
		\pgfmathsetmacro{\AEll}{ 
			\BEll * H2(\pstar,\exponent) / H1(\pstar,\exponent) };

	\else

		\pgfmathsetmacro{\AEll}{ \AElllogamma };
		\pgfmathsetmacro{\BEll}{ \BElllogamma };
		\pgfmathsetmacro{\AZero}{
			\AEll + ( \l - \BEll * H2(\pstar,\exponent) ) / H1(\pstar,\exponent) };

	\fi


	\draw [domain=0:\pstar, variable=\p,          samples=2, very thick]
		plot ( {\p}, { \yscale * 0 } );
	\draw [domain=\pstar:\pstarstar, variable=\p, samples=2, very thick]
		plot ( {\p}, { \yscale * \l } );
	\draw [domain=\pstarstar:1, variable=\p,      samples=2, very thick]
		plot ( {\p}, { \yscale * \h } );
	\draw ( { (\pstarstar+1)/2 }, { \yscale * \h } )
		node[anchor=south] {$\boldsymbol{u}$};

	\draw [domain=0:\pstar, variable=\p, smooth] 
		plot ( {\p}, 
		{ \yscale * valuefn(\p,\exponent,0,\AZero,0) } );
	\draw [domain=\pstar:\pstarstar*1.001, variable=\p, smooth]
		plot ( {\p}, 
		{ \yscale * min( valuefn(\p,\exponent,\l,\AEll,\BEll), \h ) } );
	\draw ( { 0.2*\pstar + 0.8*\pstarstar }, 
		{ \yscale * valuefn(0.2*\pstar + 0.8*\pstarstar,\exponent,\l,\AEll,\BEll) } )
		node[anchor=north west] {$v$};

	\draw [domain=\philop:\phihip, variable=\p, smooth, dotted] 
		plot ( {\p}, 
		{ \yscale * ( \phia*\p^2 + \phib*\p + \phiintercept ) } );
	\draw ( { \philop }, 
		{ \yscale * ( \phia*( \philop )^2 
		+ \phib*( \philop ) 
		+ \phiintercept ) } )
		node[anchor=south west] {$\phi$};



	\draw[-]  (0,0) -- ( 1, \yscale * 0       ); 
	\draw[->] (0,0) -- ( 0, \yscale * \h*1.05 ); 
	
	\draw ( \pstar,     \yscale * 0 ) node[anchor=north] {$1/2$};
	\draw ( \pstarstar, \yscale * 0 ) node[anchor=north] {$3/4$};
	\draw ( 1,          \yscale * 0 ) node[anchor=north] {$1$};

	\draw ( 1, \yscale * 0 ) node[anchor=west] {$p$};

	\draw[-] ( \pstar,     - \ticklength ) -- ( \pstar,     \ticklength );
	\draw[-] ( \pstarstar, - \ticklength ) -- ( \pstarstar, \ticklength );
	\draw[-] ( 1,          - \ticklength ) -- ( 1,          \ticklength );

	\draw ( 0, \yscale * \l ) node[anchor=east] {$\l$};
	\draw ( 0, \yscale * \h ) node[anchor=east] {$\h$};

	\draw[-] ( - \ticklength, \yscale * \l ) -- ( \ticklength, \yscale * \l );
	\draw[-] ( - \ticklength, \yscale * \h ) -- ( \ticklength, \yscale * \h );

	\fill[white] ( \pstar, \yscale * 0 ) circle[radius=\radius pt];
	\draw[thick] ( \pstar, \yscale * 0 ) circle[radius=\radius pt];
	\fill[white] ( \pstarstar, \yscale * \l ) circle[radius=\radius pt];
	\draw[thick] ( \pstarstar, \yscale * \l ) circle[radius=\radius pt];

\end{tikzpicture}
